\documentclass{aa}  

\usepackage{graphicx}
\usepackage{txfonts}
\usepackage{xcolor}
\usepackage{placeins}
\begin{document}

   \title{QSO MUSEUM. II. Search for extended Ly$\alpha$ emission around eight $z\sim3$ quasar pairs}

   \author{Eileen Herwig\inst{1}
   \and
          Fabrizio Arrigoni Battaia\inst{1}
          \and
          Jay Gonz\'alez Lobos\inst{1}
          \and
          Emanuele P. Farina\inst{2}
          \and
          Allison W. S. Man\inst{3}
          \and
          Eduardo Ba\~nados\inst{4}
          \and
          Guinevere Kauffmann\inst{1}
          \and
          Zheng Cai\inst{5}
          \and
          Aura Obreja\inst{1,6,7}
          \and
          J. Xavier Prochaska\inst{8,9}
          }

   \institute{Max-Planck-Institut f\"ur Astrophysik, Karl-Schwarzschild-Straße 1, D-85748 Garching bei M\"unchen, Germany\\
              \email{eherwig@mpa-garching.mpg.de}
             \and
             International Gemini Observatory, NSF’s NOIRLab, 670 N A’ohoku Place, Hilo, Hawai’i 96720, USA
             \and
             Department of Physics \& Astronomy, University of British Columbia, 6224 Agricultural Road, Vancouver, BC V6T 1Z1, Canada
             \and
             Max Planck Institut f\"ur Astronomie, K\"onigstuhl 17, D-69117, Heidelberg, Germany
             \and
             Department of Astronomy, Tsinghua University, Beijing 100084, People’s Republic of China
             \and
             Interdisziplin\"ares Zentrum f\"ur Wissenschaftliches Rechnen, Universit\"at Heidelberg, Im Neuenheimer Feld 205, D-69120 Heidelberg, Germany
             \and
             Zentrum für Astronomie, Institut f\"ur Theoretische Astrophysik, Universit\"at Heidelberg, Albert-Ueberle-Straße 2, D-69120 Heidelberg, Germany
             \and
             Department of Astronomy and Astrophysics, University of California, Santa Cruz, CA 95064, USA
             \and
             Kavli Institute for the Physics and Mathematics of the Universe, 5-1-5 Kashiwanoha, Kashiwa 277-8583, Japan
             }

   \date{}
 
  \abstract{Extended Ly$\alpha$ emission is routinely found around single quasars across cosmic time. However, few studies have investigated how such emission changes in fields with physically associated quasar pairs, which should reside in dense environments and are predicted to be linked through intergalactic filaments. We present VLT/MUSE snapshot observations (45 minutes/source) to unveil extended Ly$\alpha$ emission on scales of the circumgalactic medium (CGM) around the largest sample of physically associated quasar pairs to date, encompassing eight pairs (14 observed quasars) at $z\sim 3$ with an $i$-band magnitude between 18 and 22.75, corresponding to absolute magnitudes $M_i(z=2)$ between -29.6 and -24.9. The pairs are either at close ($\sim$50-100~kpc, five pairs) or wide ($\sim$450-500~kpc, three pairs) angular separation and have velocity differences of $\Delta v\leq2000$~km~s$^{-1}$. We detected extended emission around 12 of the 14 targeted quasars and investigated the luminosity, size, kinematics, and morphology of these Ly$\alpha$ nebulae. On average, they span about 90~kpc and are 2.8~$\times 10^{43}$~erg~s$^{-1}$ bright. Irrespective of the quasars' projected distance, the nebulae often ($\sim$45~\%) extend toward the other quasar in the pair, leading to asymmetric emission whose flux-weighted centroid is at an offset position from any quasar location. We show that large nebulae are preferentially aligned with the large-scale structure, as traced by the direction between the two quasars, and conclude that the cool gas (10$^4$~K) in the CGM traces well the direction of cosmic web filaments. Additionally, the radial profile of the Ly$\alpha$ surface brightness around quasar pairs can be described by a power law with a shallower slope ($\sim-1.6$) with respect to single quasars ($\sim-2$), indicative of increased CGM densities out to large radii and/or an enhanced contribution from the intergalactic medium (IGM) due to the dense environments expected around quasar pairs. The sample presented in this study contains excellent targets for ultra-deep observations to directly study filamentary IGM structures in emission. This work demonstrates that a large snapshot survey of quasar pairs will pave the way to direct statistical study of the IGM.
  }


   \maketitle
%

\section{Introduction}

In the current paradigm of galaxy evolution, the circumgalactic medium (CGM) plays a major role in regulating star formation and the activity of active galactic nuclei (AGNs). Loosely defined to be the gas virially bound to a galaxy but beyond the visible stellar disk \citep{Tumlinson17}, the CGM is believed to be heavily influenced by galactic outflows, heating and chemically enriching the gas reservoir, as well as inflows of cool gas ($\approx$10$^4$ K) from the intergalactic medium (IGM). While this cold accretion from the IGM filamentary gas structures, also known as the cosmic web, onto the CGM is ubiquitously seen in state-of-the-art cosmological simulations (e.g., \citealp{Dubois14, Schaye15, Springel2018}), it is difficult to directly observe due to the expected low densities.

Constraints on CGM properties (e.g., gas density, temperature, metallicity, and kinematics) have been achieved through line-of-sight absorption studies (e.g., \citealp{Hennawi2006, Prochaska2013, Rudie19, Peroux20, Donahue22, Faucher23}) or by directly observing CGM emission with narrowband filters or long-slit spectra (e.g., \citealt{HuCowie1987,Heckman1991,Weidinger2005,Hennawi2013}). However, the new generation of sensitive integral field unit (IFU) spectrographs such as the Multi Unit Spectroscopic Explorer (MUSE; \citealt{Bacon2010}) on the Very Large Telescope (VLT) and the Keck Cosmic Web Imager (KCWI; \citealt{Morrissey2012}) have made it possible to routinely observe the emission of the cool CGM gas around individual galaxies by utilizing, for example, bright nebular lines such as Ly$\alpha$. Most commonly, high-$z$ quasars ($2<z<6$) are targeted for such studies (e.g., \citealp{Borisova16, FAB19a, Farina19, Cai19, OSullivan20, Mackenzie21, Fossati21}), often unveiling large portions of the host galaxies' CGM and with the bulk of the Ly$\alpha$ emission extending for $\sim80$~kpc. 
Indeed, quasars are estimated to reside in relatively massive halos across cosmic times ($\sim10^{12.5}$ M$_{\odot}$; \citealt{Shen07, White12, Eftek15, Fossati21, Pizzati24, Costa23}), resulting in a virial radius of ${R_{\rm vir} \sim 100}$ kpc at $z\sim3$, close to their cosmic noon.

While these AGNs produce copious amounts of ionizing photons, greatly outshining their host galaxies, the exact balance between the powering mechanisms of their Ly$\alpha$ nebulae remains disputed. The suggested possibilities include the combination of recombination radiation after the ionization of the material by the quasar ionizing photons \citep{Cantalupo05, Kollmeier10, Costa22}, resonant scattering of Ly$\alpha$ photons from the quasar \citep{Cantalupo14, Costa22}, shocks \citep{Taniguchi00}, gravitational cooling radiation \citep{Haiman00, Dijkstra06}. Additionally, in situ scattering of Ly$\alpha$ photons is likely important in shaping the nebulae morphology \citep{Costa22, FAB2023b}. Notwithstanding these uncertainties, both current observations and simulations emphasize the main role of the quasar radiation in powering the extended emission \citep{Costa22, Jay23, Langen23, Obreja24}.

The most extreme Ly$\alpha$ nebulae known, commonly referred to Enormous Ly$\alpha$ nebulae (ELANe), extend up to $\approx$~500 projected kiloparsec and therefore exceed the predicted virial radius for quasar halos. They have been detected around systems hosting multiple quasars: the Slug nebula, situated between a radio-loud and a radio-quiet quasar at redshift $z=2.3$, which extends for 460 kpc \citep{Cantalupo14}; \cite{Hennawi15} reported 
the detection of four quasars at $z \sim 2$ embedded in a 310 kpc long nebula; \cite{FAB18} presented a Ly$\alpha$ nebula extending over 297 kpc and associated with one bright and two faint quasars at $z \sim 3$; and \cite{Cai18} detected a 232 kpc wide nebula surrounding a $z \sim 2.45$ quasar pair. This phenomenon might be explained by the overdensity of UV luminous galaxies able to power nebular emission and the surplus of gas fed into the system through tidal interactions during close encounters, which in turn could also fuel AGN activity.

Subsequently, targeted quasar pair fields often show remarkable extended emission. \cite{FAB19b} discovered a Ly$\alpha$ nebula stretching between two $z\sim3$ quasars with a projected separation of $\sim 100$ kpc. \cite{Li23} targeted four pairs consisting of one bright ($g<18$ mag) and one faint quasar at $z \sim 2.4$ with the Palomar Cosmic Web Imager (PCWI; \citealt{Matuszewski2010}) and 
found a maximum projected extent upward of 200 kpc in three of them, indicative of cool gas outside the CGM being illuminated. A quasar pair at ${z = 3.23}$ with a projected separation of 500 kpc observed for 40 hours as part of the MUSE Ultra Deep Field (MUDF) program hosts multiple Ly$\alpha$ nebulae extending up to 100 kpc and preferentially along the connecting line between the quasars \citep{Lusso19}. This is interpreted as possible evidence of a large-scale gaseous filament connecting the two galaxies.

   \begin{figure*}[h]
   \centering         {\includegraphics[width=0.9\textwidth]{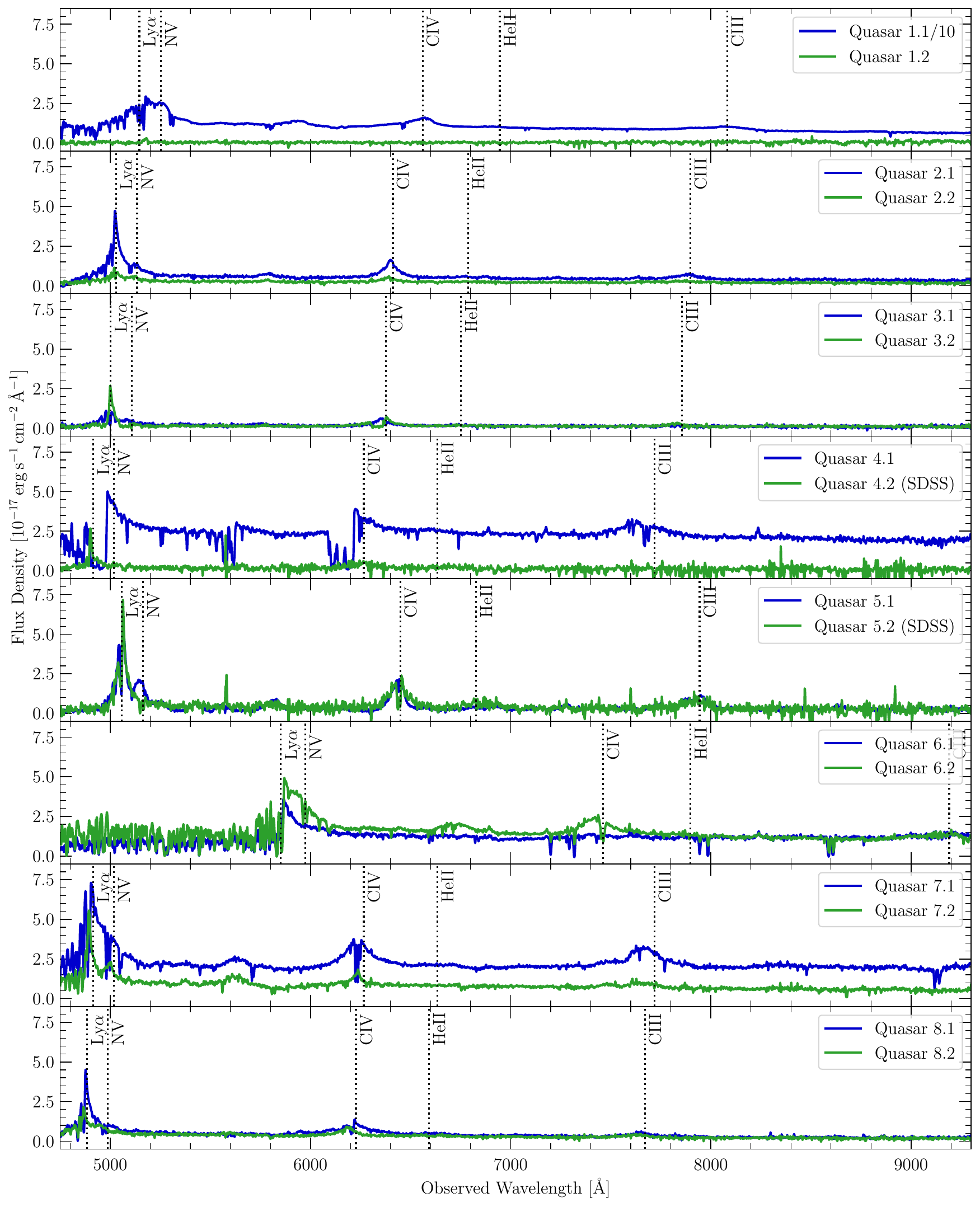}}
      \caption{Eight targeted quasar pairs. Each panel shows the spectra for the two AGNs extracted either from the final MUSE data cubes in an aperture of 3\arcsec\ or available from SDSS. Vertical dotted lines indicate the location of key emission lines at the SDSS redshift of the brighter quasar. The spectrum of quasar 1.1 is scaled down by a factor of 10. The mean noise per channel in the MUSE data at the respective wavelength of Ly$\alpha$ is ${\rm 0.08 \cdot 10^{-17}\ erg\, s^{-1}\, cm^{-2}\, \AA^{-1}}$.
              }
         \label{fig:qsospectra}
         
   \end{figure*}

Given the aforementioned expected quasar halo mass of $\sim$10$^{12.5}$ M$_{\odot}$, quasar pairs could indeed be suitable tracers of filament orientations.
In the MilleniumTNG simulation (\citealt{Pakmor2023}), the mean halo mass at the nodes of the cosmic web is 10$^{12.47}$ M$_{\odot}$ at redshift three, with almost all halos above 10$^{13}$ M$_{\odot}$ being identified as a node \citep{Galarraga24}. Per definition, nodes indicate the endpoint of cosmic web filaments, and therefore, two neighboring nodes are expected to be connected to each other by a gaseous filament.

In this paper, we extend the research of previous works by searching for extended nebular emission around a sample of eight quasar pairs observed with VLT/MUSE. Our work has been conducted in the framework of the survey {Quasar
Snapshot Observations with MUse: Search for Extended Ultraviolet
eMission} (QSO MUSEUM; \citealt{FAB19a}), using a very similar strategy. The paper
is structured as follows. Section~\ref{sec:obs} describes the observations and the data reduction. In Sect.~\ref{sec:data_analysis}, we explain how the data are analyzed to unveil the extended Ly$\alpha$ emission. Sections~\ref{sec:results} and \ref{sec:discussion} present the results of this work and discuss them in comparison to previous studies. Finally, Sect.~\ref{sec:summary} summarizes our findings.
Throughout this paper, we assume a flat ${\Lambda}$CDM cosmology with ${H_0\ =\ 67.7}$~km~s$^{-1}$~Mpc$^{-1}$, ${\Omega_m\ =\ 0.31}$, and ${\Omega_{\Lambda}\ =\ 0.69}$. At the mean redshift of our sample, ${z \approx 3.2}$, one arcsecond corresponds to roughly 7.7 kpc. Reported quasar magnitudes are in the AB system, and absolute magnitudes are K-corrected relative to redshift two \citep{Ross13}.

\section{Observations and data reduction}
\label{sec:obs}

\begin{figure}
   \centering
   \includegraphics[width=\hsize]{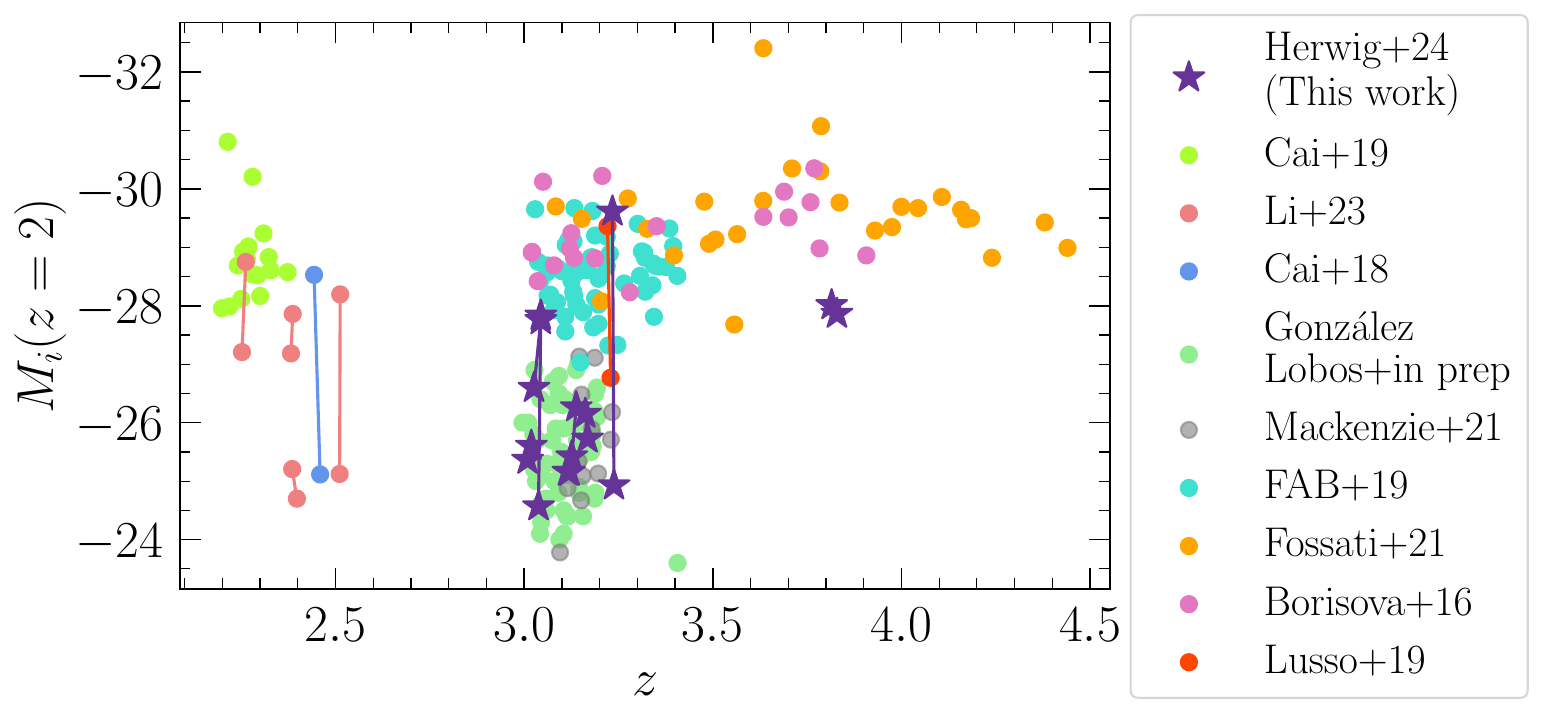}
      \caption{Quasar magnitudes and redshift covered by different studies of Ly$\alpha$ nebulae. Single quasars are shown as dots (\citealt{Borisova16, FAB19a, Cai19, Mackenzie21, Fossati21}; Gonz\'alez Lobos et al. in prep.), while physically associated quasar pairs are connected with a line (our work as purple stars; \citealt{Cai18, Lusso19, Li23}).
              }
         \label{fig:miz}
   \end{figure}

The sample was selected from the SDSS DR12 quasar catalog \citep{Paris17} complemented by AGN searches at high-$z$ (e.g., \citealt{Bielby13}) as described in \cite{FAB19b}. In brief, to ensure the physical association of the quasar pairs, targets were selected to be within $\Delta z \leq 0.03$ of each other (or about $\leq 2000$~km~s$^{-1}$) based on their systemic redshift provided by the aforementioned works and to have an angular separation of $\leq$ 1 arcmin ($\sim450-500$~kpc). Moreover, the redshift of the quasar pairs was restricted to the range $3 \leq z<3.9$. The lower limit ensures that the Ly$\alpha$ line is within the wavelength range of MUSE, while the upper limit prevents it from being heavily contaminated by sky emission lines. Although the sample selected in this way comprised 17 pairs covering a wide range of projected distances between the two quasars, only close ($\sim$ 50-100 kpc) and wide ($\sim$ 450 kpc) pairs were observed due to weather conditions. In the two widest pairs, only one quasar was targeted in each system due to the large angular separation.
One quasar pair from the selected and observed sample (ID 8.1 and 8.2) has already been published in \cite{FAB19b}. 
We included it here to increase the sample size, and re-analyzed it with the same procedures for comparison.
The full spectra of the quasars are shown in Fig.~\ref{fig:qsospectra}. If the quasar is not in the field of view (FoV) of the MUSE observations, its SDSS spectra is instead shown.
In the VLA FIRST survey \citep{Becker94}, a radio source is detected 0.8$\arcsec$ away from quasar 5.1 with $f_{1.4 {\rm GHz}} = 1.4\ {\rm mJy\ beam^{-1}}$. No other quasar from this sample is detected in VLA FIRST.
Figure~\ref{fig:miz} displays the redshift and magnitude of quasars in our sample in comparison to previous surveys of both single quasar nebulae and quasar pairs.

The sample presented in this paper was observed from November 2017 to March 2018 as part of the program 0100.A-0045(A) (PI: F. Arrigoni Battaia) in service mode with the MUSE instrument in Wide Field Mode on the VLT telescope YEPUN. Quasar properties and information about the observations are summarized in Table~\ref{tab:1}. The observations were conducted at an average seeing of 1.16" (or $\sim 8.9$~kpc at the mean redshift of the sample) with clear weather conditions or thin cloud coverage. For each target, three exposures with 880 seconds of on source time each were taken, rotated by 90 degrees with respect to each other and a few arcseconds dithering pattern. The only exception is quasar pair 7, for which only two exposures are available.

Each cube covers a wavelength range of 4750 $\AA$ - 9350 $\AA$ with a spectral resolution at the mean wavelength of Ly$\alpha$ of $R = 1815$ (corresponding to ${\rm FWHM}=2.81~\AA$ or 165~km~s$^{-1}$) and at a spectral sampling of 1.25 $\AA$ and a field of view of approximately 1$\arcmin$x1$\arcmin$ with a spatial sampling of 0.2$\arcsec$. 
The data was reduced using the MUSE pipeline v2.8.7 \citep{Weilbacher14, Weilbacher16, Weilbacher20} as described in \cite{Farina19} and \cite{Jay23}, applying bias-subtraction, flat-fields, twilight and illumination correction, sky-subtraction, wavelength and flux calibration.
After running the pipeline, there are still residual sky emission lines affecting the data. Therefore, we applied an additional skyline subtraction using the Zurich Atmospheric Purge tool (ZAP, \citealt{SOTO16}), minimizing both the residuals of sky emission lines and the influence of this procedure on the flux of emission at the expected wavelength and position of the Ly$\alpha$ glow.
Subsequently, we masked artifacts due to the edges of the MUSE integral field units via visual inspection and determined the offset between individual exposures by centroid fitting of a bright point source in the FoV. Then, we median combined the science exposures and variance cubes for each object. 
To determine a more accurate estimate of the noise by taking into account the correlated noise introduced by the pipeline interpolations, we rescaled the associated variance cubes layer by layer to the RMS spectrum of a sigma-clipped background region of the science cubes as usually done in the literature (e.g., \citealt{Borisova16, Bacon2017}). This final variance cube was subsequently used to calculate the surface brightness limit (SBL) and errors. The average SBL in a 30 $\AA$ pseudo narrowband (PNB) image centered on Ly$\alpha$ and after masking continuum sources via sigma-clipping is ${\rm 1.54 \cdot 10^{-18}\ erg\, s^{-1}\, cm^{-2}\, arcsec^{-2}}$ in a 1 arcsec$^2$ aperture, while the average SBL in the same aperture and in the respective central slice is ${\rm 3.15 \cdot 10^{-19}\ erg\, s^{-1}\, cm^{-2}\, arcsec^{-2}}$. This is comparable, but lower than the layerwise SBL reported in \cite{FAB19a} (${\rm 4.4 \cdot 10^{-19}\ erg\, s^{-1}\, cm^{-2}\, arcsec^{-2}}$) and \cite{Borisova16} (${\rm 5 \cdot 10^{-19}\ erg\, s^{-1}\, cm^{-2}\, arcsec^{-2}}$). Both surveys have comparable observational setups, but include more targets at $z \sim 3.3$, where sky emission lines can lower the sensitivity at the Ly$\alpha$ line.

\section{Data analysis}
\label{sec:data_analysis}

To reveal the extended emission, the quasar point spread function (PSF) has to be subtracted from the science data cube as it outshines the extended emission \citep{Moller00, Christensen06, Husemann14}. Here we used the same empirical method described extensively in \cite{Jay23}.
Specifically, the size of the PSF and therefore the required subtraction window are dependent on the seeing, which we estimated from the quasar continuum emission. We created narrowbands of 25 $\AA$ width and fitted a 2D Moffat function onto the quasar profile while conservatively excluding spectral regions where extended emission might occur, that is, slices around the quasar emission lines typically spanning around 500 \AA. Extrapolated to the Ly$\alpha$ wavelength at the systemic redshift of the respective quasar, the Moffat FWHM provides a good estimate of the seeing of the observations. In order to obtain more robust results, we performed this estimation on the brighter quasar in the field, although the results are consistent for all pairs if performed individually. Additionally, we determined the seeing at the wavelength of Ly$\alpha$ using a star in the FoV, if available. Especially for faint quasars, the host galaxy contribution can be high enough that the quasar no longer appears as a point source, leading to an overestimation of the seeing. However, the seeing determined from a star is on average only 0.3 \% smaller than the one determined from the respective quasar, with the biggest deviation being 6\%.

In previous works the PSF subtraction was performed in circular windows with a diameter ranging from 6 to 10 times the seeing depending on the luminosity of the quasar (\citealt{Borisova16,Mackenzie21}), with brighter quasars requiring a larger subtraction region. 
Given the diversity of luminosities in our sample, we accordingly varied the window size between objects in a similar range as previous studies (circular aperture with diameter 4 to 9 times the seeing), aiming to have virtually no contribution from the quasar PSF to the flux value at the edge of the circular region. If the chosen window size is too big, oversubtraction can occur, and sometimes lead to spurious emission detection in the PSF and the continuum subtracted data cube if there are absorption features in the Ly$\alpha$ line of the quasar.

We constructed an empirical PSF 
layerwise by creating narrowbands in the subtraction window and scaling it to the flux value in the central 1$\arcsec$x1$\arcsec$. In this normalization area, residual values are very error-prone and we therefore masked the pixels for subsequent analysis (as usually done in similar analyses, e.g., \citealt{Borisova16}). To avoid including possible extended nebular emission in the PSF construction, wavelength slices containing quasar emission lines were skipped when constructing narrowbands and instead, the next narrowband redward of the line was used to create the PSF. This is especially important for the Ly$\alpha$ line as the narrowband at shorter wavelengths than the line can be heavily affected by absorption. The width of narrowbands has to be varied depending on the magnitude of the quasar in order to achieve sufficient signal-to-noise ratios to constrain the PSF shape, with a typical width of 350 channels or 437.5 $\AA$ for this sample.
We subtracted the empirical PSF constructed in this way from the respective quasar in the data cube to reveal the large scale emission around it. Subtraction of two PSFs was not necessary in every field: In the quasar pairs 4 and 5, only the brighter quasar is within the MUSE FoV due to the large angular separation, and quasar 1.2 is very dim with an $r$-band magnitude of 24.6 at the time of  discovery with VLT/VIMOS \citep{Bielby13} and is detected as a faint continuum source in the MUSE observation ($i$-band 22.75). Therefore, we subtracted only one PSF in these fields. 

Also, two systems (quasar 5.1 and 3.2) required ``non-standard'' PSF subtractions.
In quasar 5.1, continuum sources very close to the PSF disturb the subtraction and we therefore masked them before constructing the empirical PSF. Examining the nature of these sources will be the focus of a companion paper \citealt{Herwig2024b}.
The PSF of quasar 3.2 has a very asymmetric spectral shape, leading to severe oversubtraction in one half of the subtraction window when using our standard approach, with a mean value within a 30 $\AA$ narrowband around the Ly$\alpha$ peak equivalent to -2.3$\sigma$, and undersubtraction in the other half. This is likely due to a very bright narrow nebular line compared to the fainter and broader Ly$\alpha$ from the quasar. Therefore, we masked the quasar spectrum in the slices where nebular emission is suspected and replaced the values with an interpolation. During the construction, we then scaled the PSF by the masked spectrum. The mask encompasses 12 slices, or 15 $\AA$, and was chosen to be the narrowest mask capable of removing the oversubtraction.

Lastly, we utilized the ZAP package to estimate the continuum using the function {\tt contsubfits}, which applies a median filter. The obtained continuum cube was then subtracted from the PSF-subtracted data cube, yielding a data cube that should only contain nebular emission lines and noise. 
In the subsequent analysis, continuum sources, excluding the quasars, were masked.

\begin{table}
\caption{Properties of the PNBs used in the analysis.}             
\label{tab:2}      
\centering                          
\begin{tabular}{c c c c c}        
\hline\hline                 
ID & ${\rm \lambda_{Neb}}$\tablefootmark{a} & ${z_{\rm Neb}}$\tablefootmark{b} & SBL$_{\rm PNB}$ \tablefootmark{c} & SBL$_{\rm Slice}$ \tablefootmark{d} \\    
\hline                        
    1 & 5182.1 & 3.264 & 1.30 & 2.50 \\      
    2 & 5019.7 & 3.131 & 1.24 & 2.47 \\
    3 & 4997.2 & 3.112 & 1.28 & 2.66 \\
    4 & 4894.6 & 3.028 & 2.29 & 4.59 \\
    5 & 5067.3 & 3.170 & 1.48 & 2.78 \\
    6 & 5867.3 & 3.828 & 1.35 & 3.28 \\
    7 & 4894.8 & 3.028 & 1.88 & 3.74 \\
    8 & 4876.5 & 3.013 & 1.52 & 3.18 \\
\hline                                   
\end{tabular}
\tablefoot{
\tablefoottext{a}{Wavelength of the peak of the nebular Ly$\alpha$ line, determined from a Gaussian fit to the spectrum.}
\tablefoottext{b}{Redshift of the peak of the nebular Ly$\alpha$ line, determined from a Gaussian fit to the spectrum.}
\tablefoottext{c}{Surface brightness limit in [${\rm 10^{-18}\ erg\, s^{-1}\, cm^{-2}\, arcsec^{-2}}$], calculated in the 30 $\AA$ narrowband encompassing 24 slices.}
\tablefoottext{d}{Surface brightness limit in [${\rm 10^{-19}\ erg\, s^{-1}\, cm^{-2}\, arcsec^{-2}}$], calculated in one slice of width 1.25 $\AA$ at $\lambda_{\rm Neb}$.}
}
\end{table}

\section{Results}
\label{sec:results}

In this section we report all the observational results of our analysis. We discuss the most important implications in Sect.~\ref{sec:discussion}. 

   \begin{figure*}
   \centering
   \includegraphics[height=4cm]{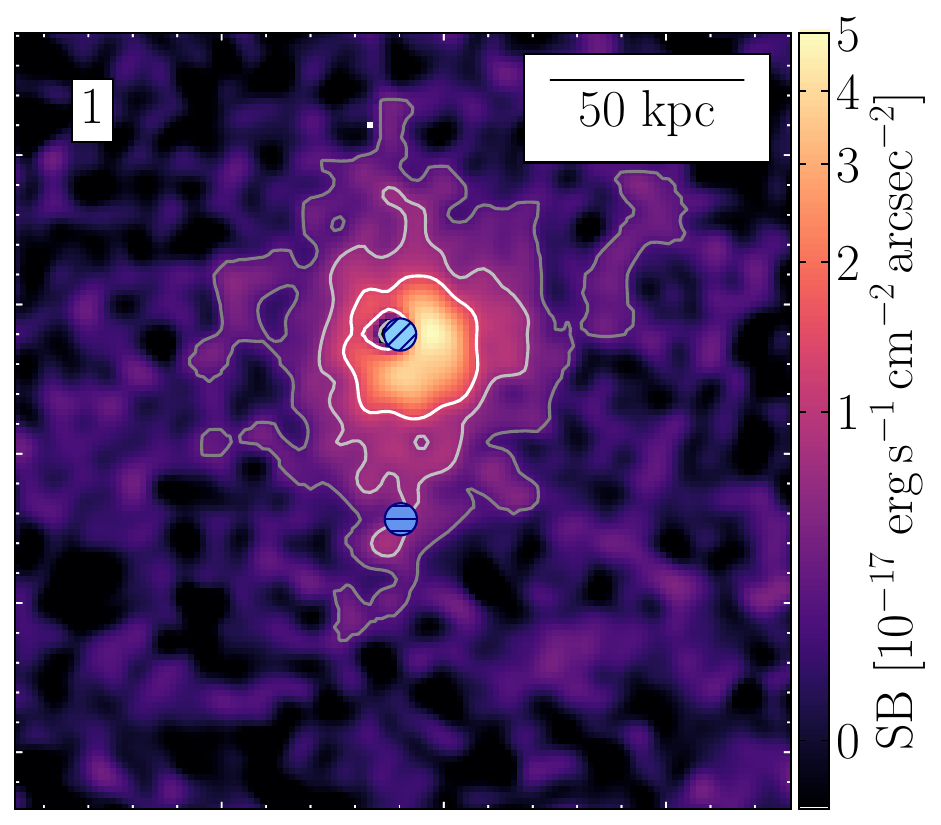}
   \includegraphics[height=4cm]{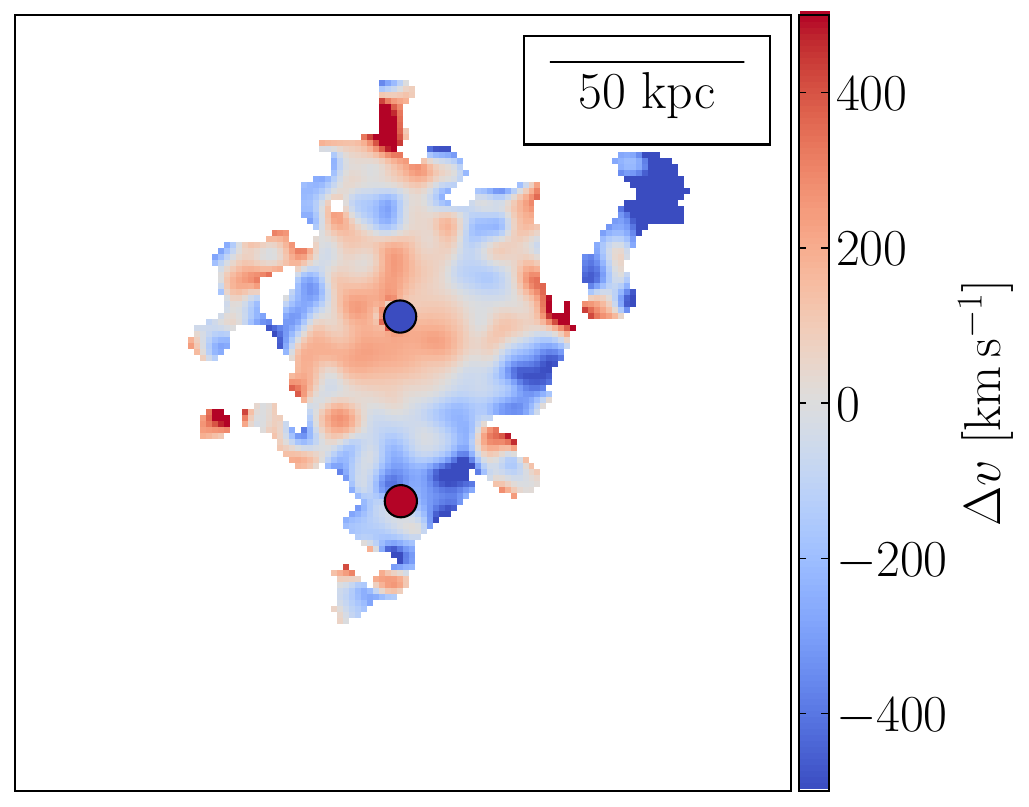}
   \includegraphics[height=4cm]{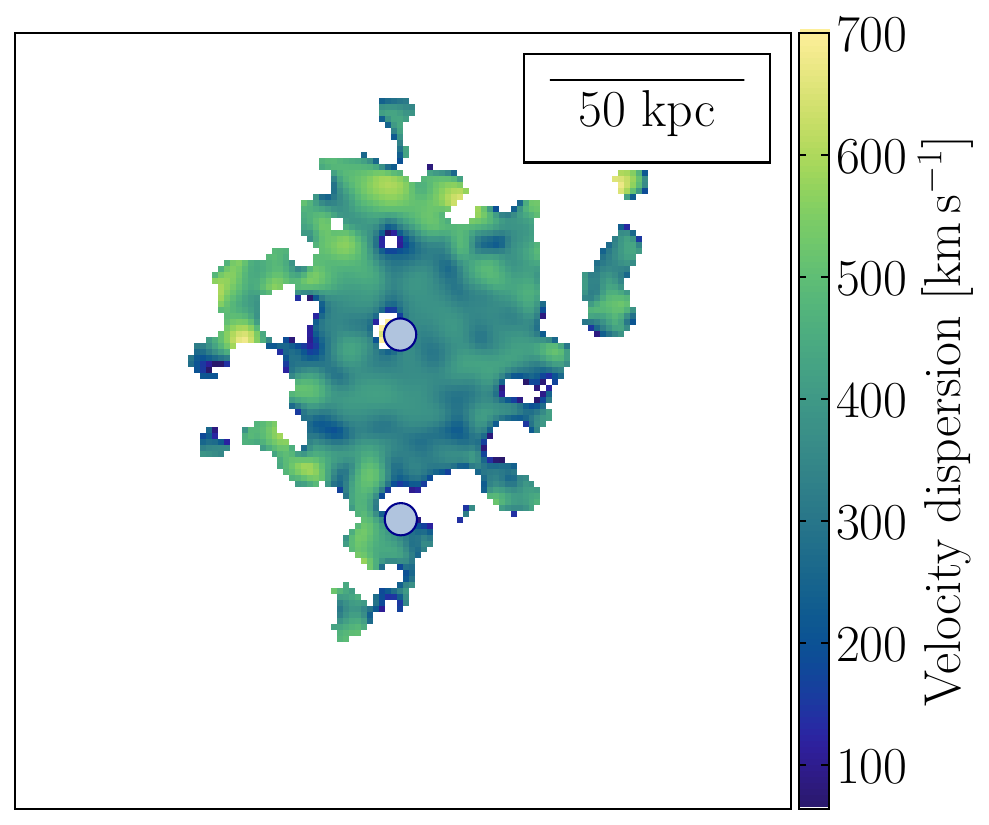}
   \includegraphics[height=4cm]{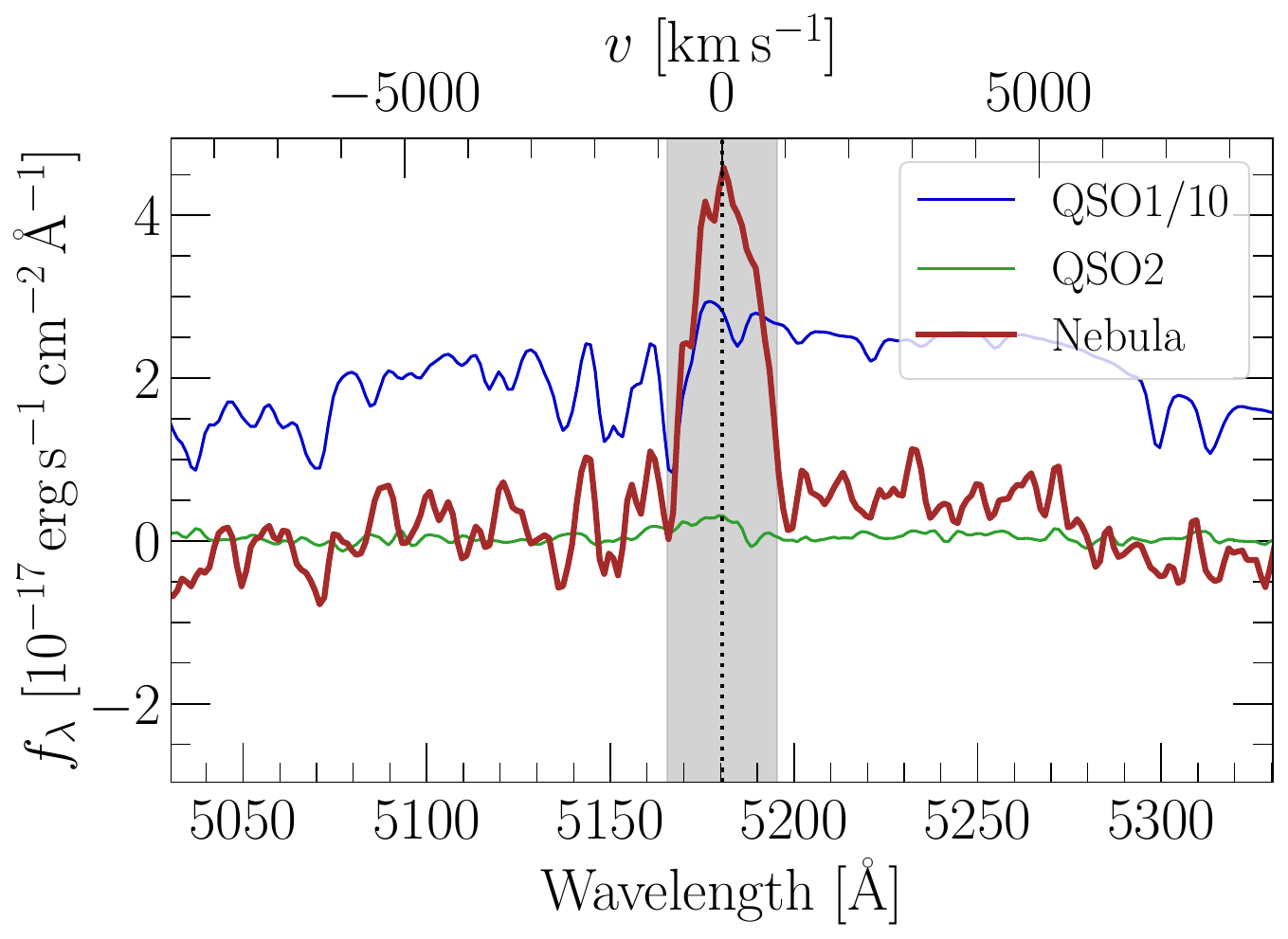}
   \includegraphics[height=4cm]{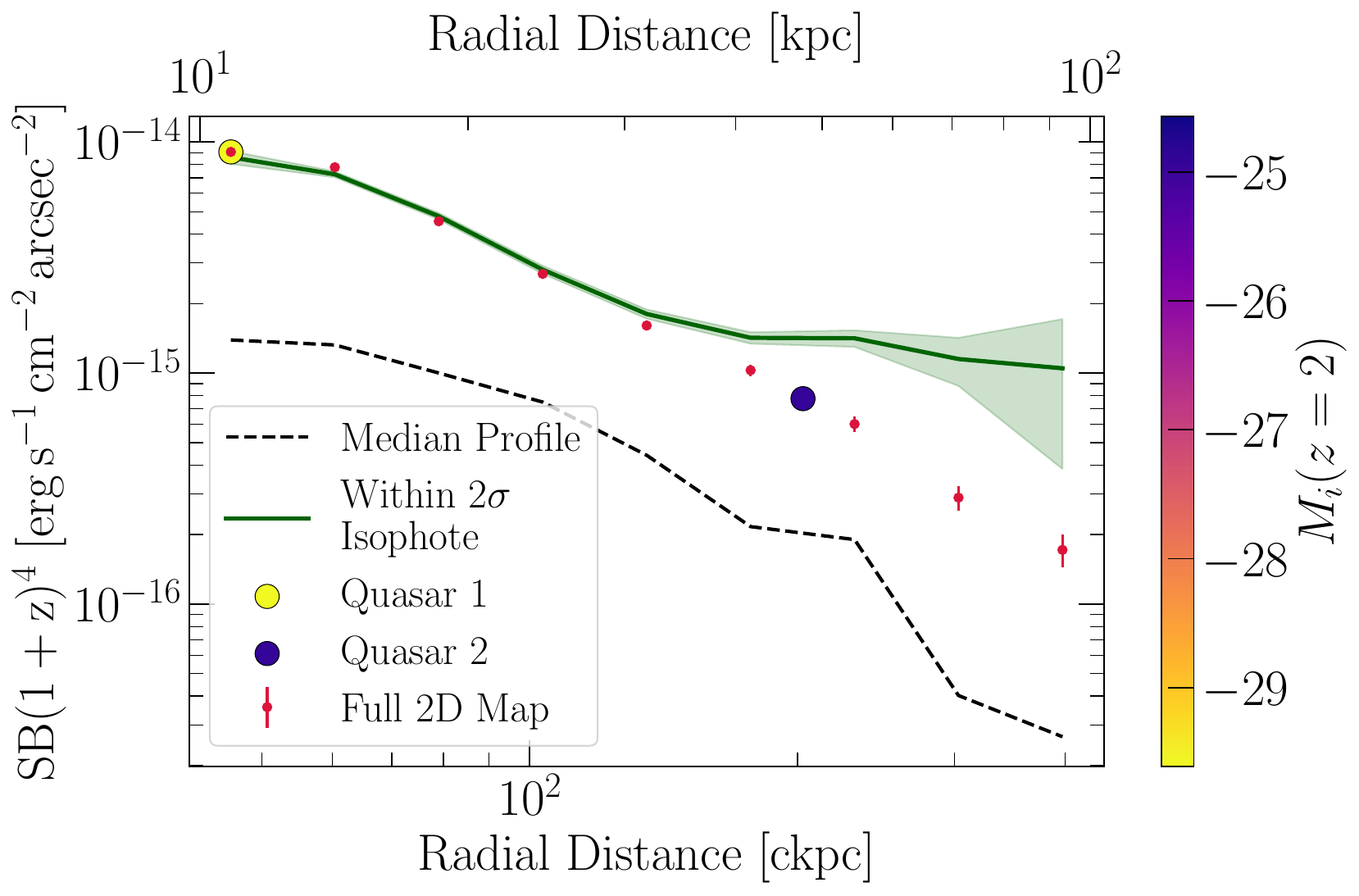}
      \caption{Ly$\alpha$ maps and profiles for quasar pair 1 (ID 1.1 and ID 1.2). All maps have been smoothed by a 2D Gaussian kernel with a standard deviation of 2 pixels or 0.4$\arcsec$ and were obtained from the PSF-subtracted optimized PNBs. Shown is a $26\arcsec \times 26\arcsec$ (or about $200$~kpc~$\times$~$200$~kpc) cut-out centered between quasar 1.1 and quasar 1.2. \textbf{Top left:} Ly$\alpha$ surface brightness map. The position of the brighter quasar is indicated with a light blue, diagonally dashed circle with a diameter of 1 arcsec, the dimmer quasar is indicated by the darker blue, horizontally dashed circle with the same diameter. Contour lines indicate a detection of 2$\sigma$ (dark gray), 4$\sigma$ (light gray) and 10$\sigma$ (white). The surface brightness maps of all pairs are shown with the same colorbar range to ease comparison.
      \textbf{Top middle:} Velocity offset with respect to the PNB center of the Ly$\alpha$ line within the 2$\sigma$ isophote, calculated within the PNB wavelength range. Quasar positions are marked with circles color-coded by their systemic redshift on the same colorbar scale as the map. \textbf{Top right:} Velocity dispersion of the Ly$\alpha$ line within the 2$\sigma$ isophote. \textbf{Bottom left:} Spectrum of the brighter (QSO1) and dimmer (QSO2) quasar integrated within an aperture of 3$\arcsec$ and spectrum of the nebula integrated within the 2$\sigma$ isophote indicated in the surface brightness map. The spectra are centered on the Ly$\alpha$ peak of the nebula chosen as the PNB center (black dashed line). The slices used for the PNB are marked in gray. All spectra are smoothed with a Gaussian kernel with a standard deviation of 1 pixel or 1.25 $\AA$. The spectrum of quasar 1.1 is scaled down by a factor of 10.
      \textbf{Bottom right:} Cosmologically corrected surface brightness profiles calculated in logarithmically spaced annuli around the brighter quasar. Quasar positions are indicated by the dots, color coded by $M_i(z=2)$.
      The profile represented by red dots is calculated in the full PNB after masking continuum sources and subtracting the background, the errorbars in y-direction show the 1$\sigma$ error. For comparison, the median stacked profile of all pairs is plotted as black dashed line. The green curve is calculated within the 2$\sigma$ isophote indicated in the surface brightness map, with shaded regions showing the 1$\sigma$ error calculated in the same annuli in the variance cube.}
         \label{fig:qso1}
   \end{figure*}

\subsection{Surface brightness levels of extended Ly$\alpha$ emission}

As described in the previous section, the PSF and continuum subtracted data cubes should only contain nebular emission lines and noise. 
In this work we only focus on the extended Ly$\alpha$ emission surrounding the quasar pairs. To facilitate comparisons with previous studies, for the subsequent analysis we chose PNBs with a width of 30 $\AA$ in observed frame around the Ly$\alpha$ nebula line. To select the central wavelength of the PNBs, we proceeded as follows. After constructing a PNB centered on the peak of the nebula emission line determined by visual inspection, we fitted a Gaussian function to the spectrum extracted within the largest 2$\sigma$ isophote close to the respective quasar, determined from the SBL in this PNB. We then constructed a new PNB centered on the peak of the fitted line and iterated the process. 
We selected those that maximize the flux of the fitted line as optimized PNBs. 
Table~\ref{tab:2} summarizes the properties of such optimized PNBs for each pair, and Appendix~\ref{sec:appPSF} provides an example of a PNB before PSF subtraction as well as the corresponding example of SB profile extracted from the PNB before PSF subtraction, after PSF subtraction, and of the empirical PSF.

To account for narrower or wider lines and different nebula redshifts in cubes with two quasars in the FoV, we performed the same iteration individually for each quasar while also changing the width of the PNB to two times the FWHM of the Gaussian fit. This does not significantly change the resulting PNB or sensitivity, as the narrowbands optimized in this way are mostly already 30 $\AA$ wide, encompassing both nebulae. The only exception is quasar pair 2 (Fig.~\ref{fig:qso2}), where the determined peak wavelengths ${\rm\lambda_{Neb}}$ are offset from each other by 7.5 $\AA$. 
The obtained PNBs for all quasar pairs are shown as surface brightness maps in the first panel of Figs.~\ref{fig:qso1} and ~\ref{fig:qso2} to \ref{fig:qso8}.

Additionally, we computed the mean surface brightness in logarithmically spaced annuli around the brighter quasar in each pair, extending out to 450 comoving kiloparsecs or 110 physical kiloparsecs at the typical redshift of our sample. During this profile calculation, residual background emission was subtracted by calculating the mean surface brightness within a sigma-clipped background region of each image.
This calculation was performed both for the full PNB and for the PNB after masking regions outside the 2$\sigma$ isophote shown in Figs.~\ref{fig:qso1} and ~\ref{fig:qso2} to \ref{fig:qso8}. The second set of profiles are typically brighter, as they do not include regions with no signal in the calculation.

\begin{figure}
   \centering
   \includegraphics[width=\hsize]{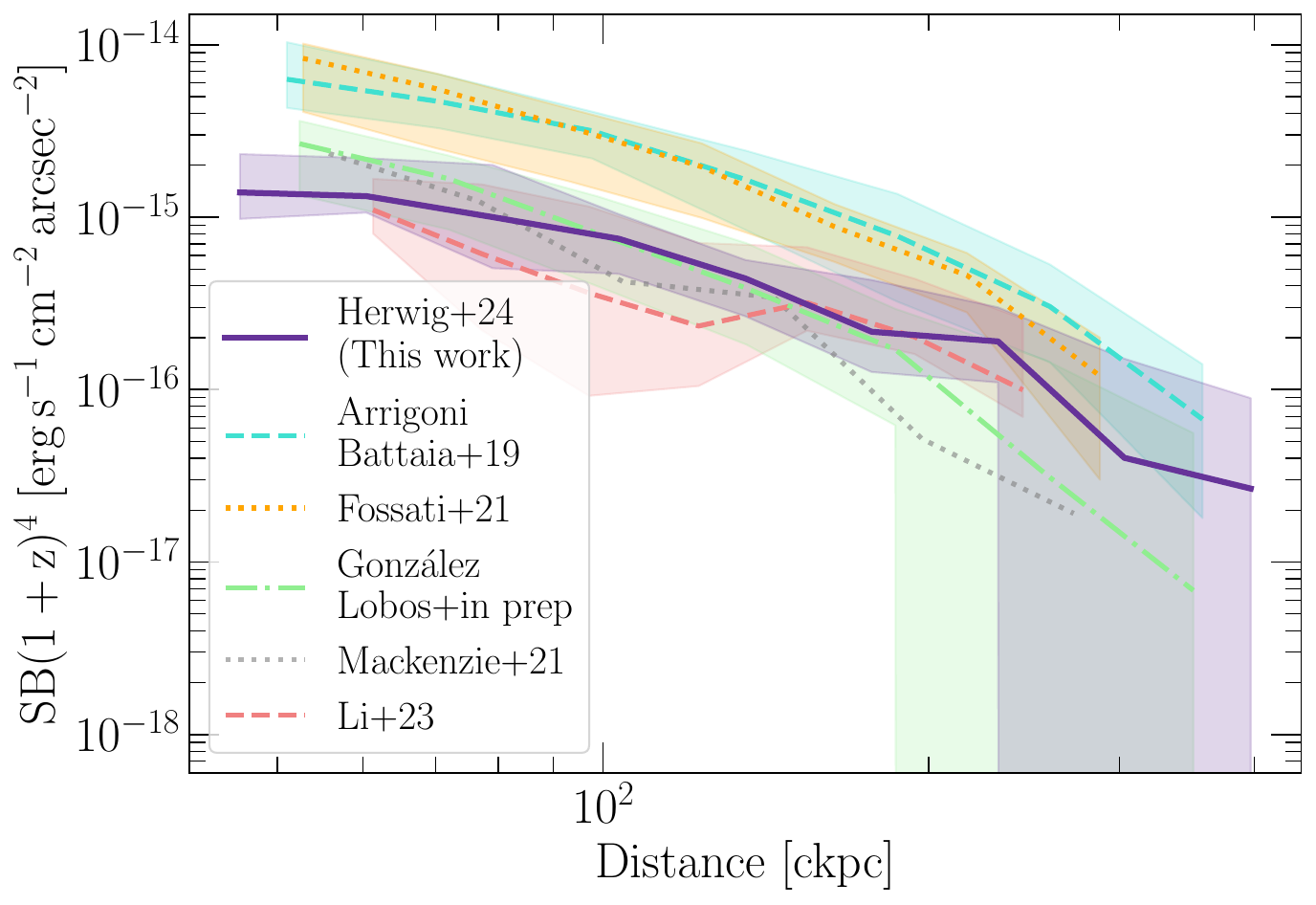}
      \caption{Cosmology corrected, circularly averaged Ly$\alpha$ surface brightness profiles of different quasar samples. The obtained average Ly$\alpha$ SB profile for $z\sim3$ quasar pairs is compared with (i) samples of individual $z\sim3$ (bright quasars, \citealt{FAB19a,Fossati21}; faint quasars, \citealt{Mackenzie21}; Gonz\'alez Lobos et al. in prep.) and (ii) with $z\sim2$ quasar pairs (\citealt{Li23}).  
              }
         \label{fig:SBstacked}
   \end{figure}

To get the typical surface brightness profile of physical quasar pairs, we stacked the individual profiles, putting the brighter quasar in the pair to 0 ckpc and taking the median value for each radius (Fig.~\ref{fig:SBstacked}, Table~\ref{tab:sbprofile}). We made this choice because we always find brighter emission closer to the brighter quasar of the pair even though in quasar pair 3 the two AGNs have similar magnitudes and one does not show associated extended emission. For comparison, the median surface brightness profiles of other $z\sim2-3$ quasar samples are shown. These can typically be described with a power law ${\rm SB_{Ly \alpha}} \propto r^{\alpha}$ with $\alpha \approx -2$ (e.g., \citealt{Borisova16,FAB19a}). However, in the case of quasar pairs, the surface brightness starts out with relatively low values close to the brighter quasar and falls off more gradually. A fitted power law to the profile weighted by the symmetrized 25th and 75th percentiles yields $\alpha = -1.57 \pm 0.17$. We performed fitting using the nonlinear least squares function \texttt{curve\_fit} in \texttt{scipy} \citep{2020SciPy-NMeth}.
As the slope might be influenced by the presence of dense gas in the host halo of the companion quasar, we repeated this analysis by only stacking the three pairs at wide separation. The obtained profile is not contaminated by a second quasar halo and is best fitted with a power law index of $\alpha = -1.17 \pm 0.28$ for the median stacked profile and $\alpha = -1.33 \pm 0.16$ for the mean profile.

\begin{table}
\caption{Median Ly$\alpha$ SB profile for $z\sim3$ quasar pairs.}             
\label{tab:sbprofile}      
\centering                          
\begin{tabular}{c c c c}        
\hline\hline                 
$R$ [ckpc]\tablefootmark{a} & Ly$\alpha$ SB\tablefootmark{b}\tablefootmark{c} & 25$^{\rm th}$ percentile\tablefootmark{b}\tablefootmark{c} & 75$^{\rm th}$ percentile\tablefootmark{b}\tablefootmark{c}\\    
\hline                        
40 - 52  & 139.1 & 97.7  & 232.2 \\ 
52 - 68  & 132.4 & 106.2 & 220.0 \\ 
68 - 90  & 100.1 & 50.6  & 200.8 \\ 
90 - 117 & 74.9  & 46.9  & 104.0 \\ 
117 - 153 & 44.1  & 26.6  & 56.5  \\ 
153 - 201 & 21.6  & 12.6  & 43.2  \\ 
201 - 263 & 19.0  & 11.0  & 30.0  \\ 
263 - 344 & 4.0   & -0.7  & 15.1  \\ 
344 - 450 & 2.7   & -0.9  & 8.9   \\ 

\hline                                   
\end{tabular}
\tablefoot{
\tablefoottext{a}{Inner and outer radius of annulus around brighter quasar in comoving units.}
\tablefoottext{b}{in units of 10$^{-17}$~erg~s$^{-1}$~cm$^{-2}$~arcsec$^{-2}$.}
\tablefoottext{c}{corrected for cosmological dimming by multiplying the individual profiles by $(1+z)^{4}$.}
}
\end{table}

\subsection{Ly$\alpha$ nebulae spectra and kinematics}

We obtained a global nebula spectrum for each detection by adding the emission from spaxels within the largest 2$\sigma$ isophote in the PSF-subtracted surface brightness maps close to the respective quasars.
These spectra, together with the corresponding quasar spectra extracted within a $3\arcsec$ aperture centered on each quasar, are displayed in the fourth panel of Figs.~\ref{fig:qso1} and ~\ref{fig:qso2} to \ref{fig:qso8}.
As the area over which the nebula spectra are integrated can be quite large and the targeted quasars are relatively faint, the peak of the nebular emission line is in some cases brighter than the peak of one of the quasar's emission at the nebula peak wavelength (i.e., quasar pair 1, 3, 4, and 8). 

From these Ly$\alpha$ nebula spectra we can obtain a first probe of the cool gas kinematics, even though such integrated spectra encompass large portions of the quasars' surroundings, mixing kinematics on different scales. We calculated the second order of the nebular emission line, $\sigma_{\rm Line}$, in a window of 2 $\times$ FWHM obtained from a Gaussian fit to each spectrum.
Errors were obtained by resampling the spectra with the associated noise spectra 10,000 times and calculating the second moment. The 32nd and 68th percentiles of the obtained distribution were assumed as lower and upper error.
The mean value of $\sigma_{\rm Line}$ of all nebulae, 434~km~s$^{-1}$, is indicative of a relatively quiescent cool CGM gas, as opposed to violent kinematics with typical linewidths and velocity shifts above 1000~km~s${^{-1}}$ \citep{VillarMartin03}.

To further analyze the Ly$\alpha$ kinematics in each nebula, we obtained the line velocity shift and velocity dispersion by calculating the first and second moment from the subcubes (i.e., wavelength range) used to build the PNBs. 
In particular, for each pair, we set the central slice of the PNB wavelength range as reference velocity of the nebula and computed the moments within the 2$\sigma$ isophote. The obtained maps are shown in the second and third panels of Figs.~\ref{fig:qso1} and \ref{fig:qso2} to \ref{fig:qso8}. We find that the detected nebulae have velocity shifts of a few hundred km~s$^{-1}$, and average velocity dispersions of 390~km~s$^{-1}$.

\subsection{Ly$\alpha$ nebulae morphology}
\label{sec:morpho}

To quantify the morphology of the detected nebulae, we determined multiple metrics using the largest 2$\sigma$ isophote associated with the respective quasar: the integrated Ly$\alpha$ nebula luminosity ${L_{\rm Neb}}$, the maximum extent of the nebula ${d_{\rm max}}$ as well as the maximum extent between quasar position and nebula 2$\sigma$ isophote, ${d_{\rm QSO,max}}$ (visualized in Fig.~\ref{fig:qso4}), the offset between quasar and nebula centroid ${d_{\rm QSO-Neb}}$, the enclosed area ${A_{2\sigma}}$, the ratio between major and minor axis ${\alpha}$, and the angular offset between the nebula semi-major axis and the vector connecting the two quasars in the pair, ${\phi}$.
To determine ${\rm \alpha}$ and ${\rm \phi}$, as described in \cite{FAB19a}, we first calculated the 
second order moments of the flux distribution as
\begin{displaymath}
    { M_{xx} = \langle \frac{(x-x_{\rm Cen})^2}{r^2}  \rangle\ , \; M_{yy} = \langle \frac{(y-y_{\rm Cen})^2}{r^2} \rangle \ , }
\end{displaymath}
\begin{displaymath}
    { M_{xy} = \langle \frac{(x-x_{\rm Cen})(y-y_{\rm Cen})}{r^2} \rangle\ , \\}
\end{displaymath}
with the flux-weighted centroid coordinates ${ (x_{\rm Cen}, y_{\rm Cen})}$ and the distance of the point $(x,y)$ to the centroid, $r$.
From this, the Stokes parameters follow as
\begin{displaymath}
Q = M_{xx} - M_{yy}\ ,\ U = 2M_{xy} \\
\end{displaymath}
and were used to calculate the axis ratio, or asymmetry of the nebula,
\begin{displaymath}
    \alpha = \frac{1- \sqrt{Q^2 + U^2}}{1 + \sqrt{Q^2 + U^2}}\\
\end{displaymath}
and the angle of the semi-major axis to the next x- or y-axis, 
\begin{displaymath}
    \gamma = arctan \left( \frac{U}{Q} \right).\\
\end{displaymath}
From $\gamma$, the angle $\phi$ between semi-major axis of the nebula and the connecting line between the quasars can be determined.
The determination of positions and as a result most aforementioned values are seeing-limited and this imprecision accounts for the majority of uncertainties. We therefore assumed an error of 1$\times$seeing on all distance measurements and obtained errors for $\phi$ and $\alpha$ by randomly varying the centroid position 500 times within a circle of radius 1$\times$seeing. The minimum and maximum values of the obtained distribution were taken as upper and lower errors.
Physical area measurements are also influenced by uncertainties on the redshift. A conservative estimate for this error is 10 percent of the measured area \citep{FAB2023b}.
To avoid taking into account the same nebula twice, we only determined morphologies with respect to the brighter quasar in the pair if both of them are associated with the same nebula or only one quasar is in the FoV (ID 1.1, 4.1, 5.1, 6.1 and 8.1). If there are two clearly distinct nebulae or only one nebula, analysis was performed for each of those individually with respect to the quasar they are associated with (ID 2.1, 2.2, 3.2, 7.1). This analysis was not performed for quasar 7.2, as we require a minimum enclosed area of 2 arcsec$^2$ to be able to sensibly constrain the morphology. Table~\ref{tab:3} lists all the aforementioned metrics computed to constrain the morphology of the detected Ly$\alpha$ nebulae. On average, the extended emission spans about 90 kpc with a nebula luminosity of 2.8~$\times 10^{43}$~erg~s$^{-1}$.

   \begin{figure}
   \centering
   \includegraphics[width=\hsize]{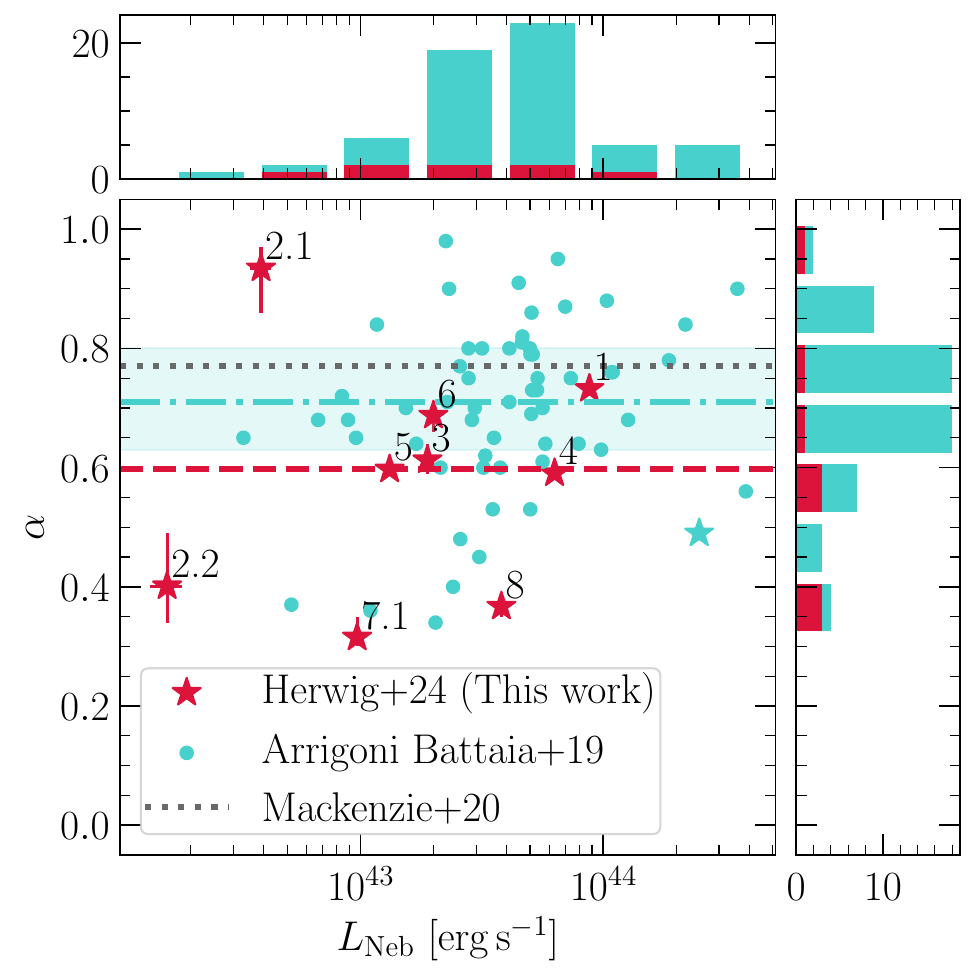}
      \caption{Ly$\alpha$ nebulae axis ratio versus their luminosity. The values for the nebulae discovered around the studied quasar pairs (red stars, accompanied by the quasar ID) are compared with those for nebulae found around bright (cyan dots and cyan dash-dotted line; \citealt{FAB19a}) and faint (dotted gray line; \citealt{Mackenzie21}) $z\sim3$ quasars. The cyan star indicates the value for the Fabulous ELAN in \cite{FAB19a}, which is hosted by an AGN triplet (of which two are type-I quasars as in this work). The cyan shaded region indicates the 25th and 75th percentiles of the distribution of bright quasars.
              }
         \label{fig:lnebvsalpha}
   \end{figure}

\begin{figure}
   \centering
   \includegraphics[width=\hsize]{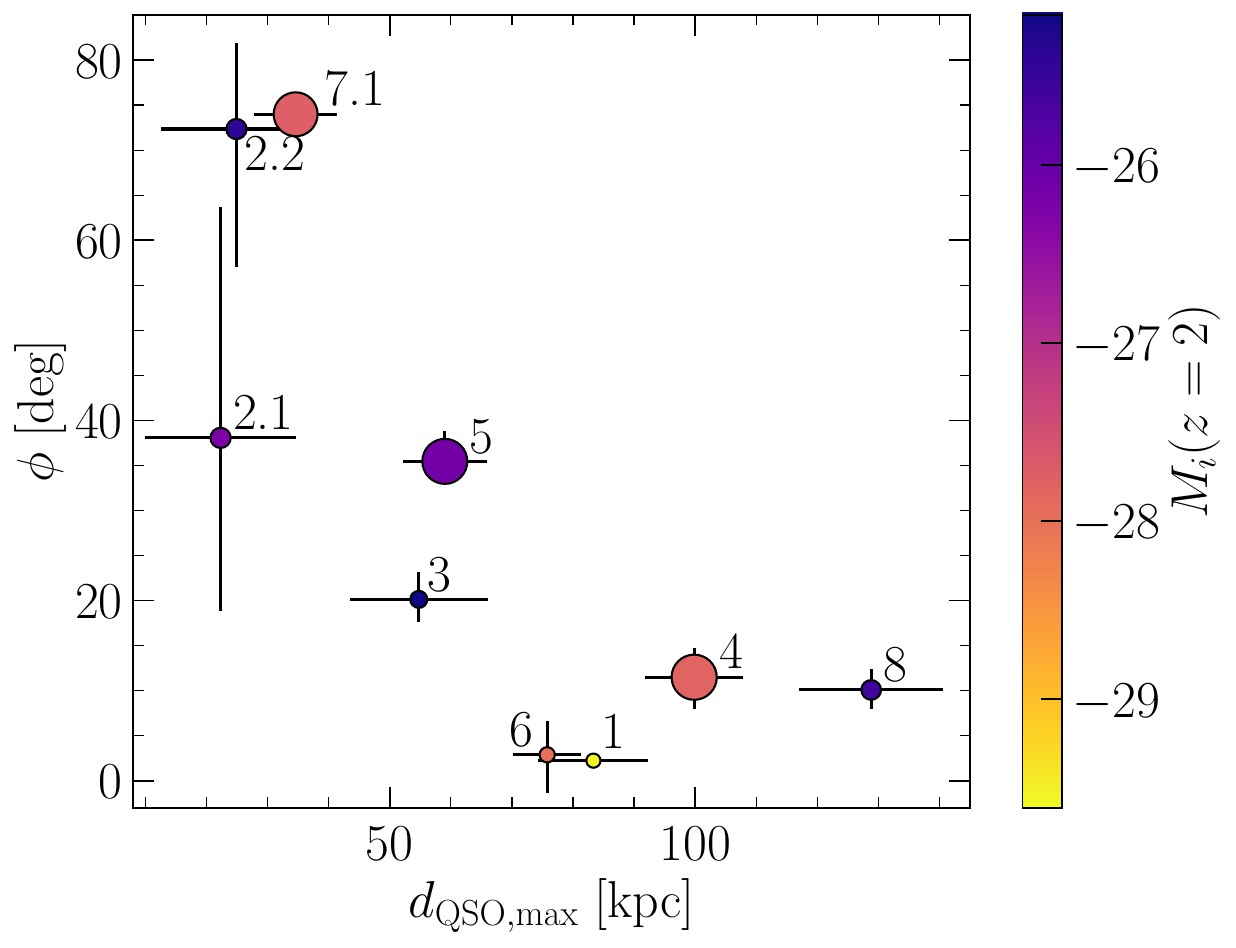}
      \caption{Angle between the Ly$\alpha$ nebula semi-major axis and the line connecting the two quasars, $\phi$ versus the maximum distance between the bright quasar in each pair and its associated Ly$\alpha$ nebula $2\sigma$ isophote, ${d_{\rm QSO,max}}$. The data points are color coded following $M_i(z=2)$ of the associated quasar, while their sizes indicate the projected distance between the quasars in each pair. The quasar IDs are displayed next to the respective data points.
              }
         \label{fig:extendvsangle}
   \end{figure}

We summarize our main findings on nebulae morphologies in Figs.~\ref{fig:lnebvsalpha} and \ref{fig:extendvsangle}.
The first figure shows the axis ratio of each nebula, $\alpha$, against its nebula luminosity ${L_{\rm Neb}}$, with the dashed red line indicating the median value of ${\rm \alpha_{Median} = 0.60}$. For comparison, we also provide the values reported in \cite{FAB19a} for nebulae detected around more luminous single $z\sim3$ quasars (QSO MUSEUM I) together with their median value ${\rm \alpha_{Median} = 0.71}$ (turquoise dash-dotted) and their 25th and 75th percentiles, and the median value of $\alpha$ found for nebulae associated with dimmer $z\sim3$ quasars (dotted gray line, \citealt{Mackenzie21}).
For a better visualization of the data, the side plots of Fig.~\ref{fig:lnebvsalpha} provide the distributions of all the points as histograms. 

We find that the second least luminous nebula considered for the morphological calculations, ID 2.1, has the most circular shape. The rest of the sample can be divided into two groups: The first one falls well within the distribution of QSO MUSEUM~I with moderate asymmetries ($\alpha \sim 0.64$), while the second group shows exceptionally high asymmetries ($\alpha \sim 0.36$). Spanning almost two orders of magnitude in nebula luminosities and encompassing quasars 2.2, 7.1, and 8, and therefore a multitude of projected separations and magnitudes, there is no clear trend in what drives these asymmetries.
Overall, the shape of quasar pair nebulae is more asymmetric than single quasar nebulae with a median value below the 25th percentile of bright quasar nebulae, even though in previous studies, there is a trend of increasing $\alpha$ with lower quasar luminosity. In other words, we find that quasars with luminosities as low as those in \citet{Mackenzie21} but part of a pair are associated with more asymmetric nebulae. The integrated nebula luminosity $L_{\rm{Neb}}$ spans a similar range in both samples, but peaks at a higher value for single quasars, with the brightest nebula having a luminosity of 1.8~$\times 10^{44}$~erg~s$^{-1}$, and a higher average luminosity of 4.5~$\times 10^{43}$~erg~s$^{-1}$.
Such more asymmetric morphologies in quasar pairs could be due to the presence of intergalactic structures connecting the two quasars in each pair. This tentative evidence of more asymmetric morphologies needs to be confirmed by a larger sample of quasar pair observations.

To investigate this further, Fig.~\ref{fig:extendvsangle} displays the relation between ${d_{\rm QSO,max}}$ (the maximum distance between the bright quasar in each pair and its associated nebula $2\sigma$ isophote) and $\phi$ (the angle between the nebula semi-major axis and the line connecting the two quasars). In the case of isolated/unrelated quasars we expect to find a random distribution of $\phi$ angles and no trend with distance. Instead, the two values are correlated with a Spearman correlation coefficient ${r = -0.7833}$ and a chance of coincidental correlation of ${p = 0.0125}$. We further tested the null hypothesis by randomly orienting the nebulae 50,000 times and calculating the Spearman coefficient for the obtained distributions. In 0.7~\% of realizations, the Spearman coefficient is at least as significant as for the data distribution, that is, below -0.7833, and in 0.2~\% of realizations, the 4 most extended nebulae all show alignment angles below 20~degrees.

The correlation is not driven by the magnitude of the quasars (color of the dots) or the projected distance between the pair (size of the dots), although at lower projected distances, the offset angle $\phi$ tends to be smaller. Possibly due to the small sample size, the significance of this trend is low, with a $p$-value of 0.5 determined by performing a 2D Kolmogorov–Smirnov test \citep{Peacock1983, Fasano1987, Press2007} using the public code \textsc{ndtest}\footnote{Written by Zhaozhou Li, \url{https://github.com/syrte/ndtest}}.

In total, 4 out of 9 nebulae ($\sim$45~\%) have $\phi\lesssim20$~degrees, and all of these 4 nebulae extend toward the other quasar in the pair. Due to the small errorbar on large nebulae, this main finding holds true even when considering the upper limit for the alignment angle. 
We therefore expect that the Ly$\alpha$ nebulae with the largest ${d_{\rm QSO,max}}$ and smallest $\phi$ trace intergalactic bridges possibly connecting the two quasars.

\begin{table*}
\caption{Ly$\alpha$ nebulae kinematics and morphologies.}             
\label{tab:3}      
\centering                          
\renewcommand{\arraystretch}{1.3}
\begin{tabular}{c c c c c c c c c c}        
\hline\hline                 
ID & ${L_{\rm Neb}}$\tablefootmark{a} & ${d_{\rm max}}$\tablefootmark{b} & ${d_{\rm QSO,max}}$\tablefootmark{b} & ${d_{\rm QSO-Neb}}$\tablefootmark{b} & $A_{2\sigma} {\rm [kpc^2]}$ & ${\rm \alpha}$ & ${\rm \phi\ [deg]}$ & ${\rm \overline{\sigma}\ [km\,s^{-1}]}$\tablefootmark{c} & ${\rm \sigma_{Line}\ [km\,s^{-1}]}$\tablefootmark{d}\\    
\hline                        
1.1, 1.2 & 8.78 $\pm$ 0.15  & 142 $\pm$ 9 & 83 $\pm$ 9 & 7 $\pm$ 9 & 5891 $\pm$ 589 & 0.73 $^{+0.01}_{-0.00}$ & 2 $\pm$ 1 & 385 $\pm$ 99 & 476 $^{+4}_{-5}$\\ \hline
2.1 & 0.39 $\pm$ 0.04 & 42 $\pm$ 12 & 22 $\pm$ 12 & 0 $\pm$ 12 & 709 $\pm$ 71 & 0.93 $^{+0.04}_{-0.07}$ & 38 $^{+26}_{-19}$ & 433 $\pm$ 136 & 304 $^{+6}_{-8}$\\
2.2 & 0.16 $\pm$ 0.02 & 46 $\pm$ 12 & 25 $\pm$ 12 & 3 $\pm$ 12 & 353 $\pm$ 35 & 0.40 $^{+0.09}_{-0.06}$ & 72 $^{+10}_{-15}$ & 423 $\pm$ 106 & 596 $^{+60}_{-65}$\\ \hline
3.1 & 1.89 $\pm$ 0.08 & 95 $\pm$ 11 & 55 $\pm$ 11 & 10 $\pm$ 11 & 2718 $\pm$ 272 & 0.61 $^{+0.03}_{-0.02}$ & 20  $\pm$ 3 & 388 $\pm$ 128 & 428 $\pm$ 5\\
3.2 & - & - & - & - & - & - & - & - & -\\\hline
4.1 & 6.31 $\pm$ 0.17 & 137 $\pm$ 8 & 100 $\pm$ 8 & 19 $\pm$ 8 & 5366 $\pm$ 537 & 0.59 $^{+0.02}_{-0.01}$ & 11 $^{+3}_{-4}$ & 395 $\pm$ 112 & 497 $^{+6}_{-7}$\\
4.2 & - & - & - & - & - & - & - & - & -\\ \hline
5.1 & 1.32 $\pm$ 0.07 & 66 $\pm$ 7 & 59 $\pm$ 7 & 34 $\pm$ 7 & 1437 $\pm$ 144 & 0.60 $^{+0.01}_{-0.02}$ & 35 $\pm$ 3 & 344 $\pm$ 119 & 312 $^{+6}_{-8}$\\
5.2 & - & - & - & - & - & - & - & - & -\\ \hline
6.1, 6.2 & 2.00 $\pm$ 0.11 & 85 $\pm$ 6 & 76 $\pm$ 6 & 36 $\pm$ 6 & 1848 $\pm$ 185 & 0.69 $^{+0.02}_{-0.03}$ & 3 $\pm$ 4 & 342 $\pm$ 123 & 399 $^{+16}_{-20}$\\ \hline
7.1 & 0.97 $\pm$ 0.07 & 61 $\pm$ 7 & 35 $\pm$ 7 & 7 $\pm$ 7 & 861 $\pm$ 86 & 0.32 $^{+0.03}_{-0.02}$ & 74 $^{+1}_{-2}$ & 351 $\pm$ 122 & 350 $^{+10}_{-12}$\\
7.2 & - & - & - & - & - & - & - & - & -\\\hline
8.1, 8.2 & 3.81 $\pm$ 0.12 & 158 $\pm$ 12 & 129 $\pm$ 12 & 43 $\pm$ 12 & 5361 $\pm$ 536 & 0.37 $^{+0.02}_{-0.02}$ & 10  $\pm$ 2 & 406 $\pm$ 123 & 542 $\pm$ 6 \\ \hline
\hline                                   
\end{tabular}
\tablefoot{
\tablefoottext{a}{Integrated luminosity of the nebula within the 2$\sigma$ isophote in ${\rm 10^{43}\ erg\,s^{-1}}$}
\tablefoottext{b}{In kpc}
\tablefoottext{c}{Mean velocity dispersion within the 2$\sigma$ isophote}
\tablefoottext{d}{Velocity dispersion of the nebula spectrum, integrated within the 2$\sigma$ isophote, determined from the second moment}
}
\end{table*}

\subsection{Notes on each quasar pair}

\subsubsection{Quasars 1.1 and 1.2}

Quasar 1.1 (Fig.~\ref{fig:qso1}) is the brightest object of the sample; it is bright as the quasars targeted in \citet{Borisova16} or \citet{FAB19a}. It is also associated with the brightest Ly$\alpha$ nebula with a luminosity $L_{\rm Neb}\sim 9 \times 10^{43}\ {\rm erg\, s^{-1}}$. Accordingly, its Ly$\alpha$ surface brightness profile is an order of magnitude brighter than the median profile of all quasar pairs. Quasar 1.2, on the other hand, is the faintest targeted object and sits at a projected distance of 47.7~kpc from quasar 1.1.
The Ly$\alpha$ nebula shows a slight elongation between the two quasars, also indicated by the low value of $\phi \approx 2^{\circ}$, and the nebular line velocity shift evolves from ${\rm \sim 220\ km\, s^{-1}}$ around quasar 1.1 to ${\rm \sim -400\ km\, s^{-1}}$ close to quasar 1.2. However, this is not reflected in the systemic redshifts of the AGNs (see top-middle panel in Fig.~\ref{fig:qso1}), likely due to the associated large uncertainties on these values as only broad lines are available for the redshift determination, leading to typical errors of up to ${\rm 400\ km\, s^{-1}}$ \citep{Shen2016}. The extended Ly$\alpha$ emission traces rather quiescent kinematics 
with an average velocity dispersion of 385~km~s$^{-1}$.

\subsubsection{Quasars 2.1 and 2.2}

Quasar pair 2, displayed in Fig.~\ref{fig:qso2}, is associated with the smallest Ly$\alpha$ nebulae in both area and extent. The extended Ly$\alpha$ emission around quasar 2.1 is also the only one in the sample that is almost circularly shaped according to the asymmetry parameter $\alpha$, and therefore, the value of $\phi$ (evaluated here to be about 40$^{\circ}$) is not well constrained. The two nebulae do not seem to be physically connected, supported by their redshift, which is close to the systemic redshift of the respective quasar. This might indicate that, although the projected separation is relatively small ($\sim$100 kpc), the quasars could be at a significant distance from each other. In this framework the pair would be stretched along our line of sight out to a distance of 2.7~Mpc obtained by assuming that the velocity difference between the two quasars ($\sim870$~km~s$^{-1}$) is all due to the Hubble flow. 
Alternatively, the two quasars might not be able to illuminate gas in the transverse direction due to obscuration. In other words, their ionization cones do not illuminate the gas in between the two objects. While quasar 2.1 and 2.2. are relatively faint, there have been observations of individual quasars of similar luminosities that show similar or more extended associated nebulae (e.g., compare ID4, ID5, ID6 and ID7 in \citealt{Mackenzie21}).\\

\subsubsection{Quasar 3.1 and 3.2}

Quasar 3.1 and 3.2 have similar absolute magnitudes $M_i(z=2)=-25.16$ and are the second faintest objects (after ID 1.2) observed in this study, but are associated with the third-biggest nebula, which is displayed in Fig.~\ref{fig:qso3}. We note that this is true even when taking into account the slightly differing PSF-subtraction method used for this source, which impacts the estimated ${\rm A_{2 \sigma}}$. Indeed, the standard subtraction of an empirical PSF leads to artifacts close to the quasar, reducing the area enclosed by the 2$\sigma$ isophote, but does not influence different measures of nebula extent like ${\rm d_{max}}$, which are sensitive to large scale emission.
The discovered extended emission does not stretch between the quasars, akin to other pairs in the sample, but is concentrated around quasar 3.2 and shows relatively quiescent kinematics ($\overline{\sigma}=388$~km~s$^{-1}$) and a shallower surface brightness profile with respect to the median profile of the sample. While the two quasars are similarly luminous, in the spectrum of quasar 3.2, a blueshifted absorption trough is visible in the CIV line (Fig.~\ref{fig:qsospectra}), which may indicate a nuclear outflow (\citealt{Weymann1991}) possibly helping in illuminating the surrounding gas (\citealt{Costa22}).

\subsubsection{Quasar 4.1 and 4.2}

Figure~\ref{fig:qso4} presents the results for quasar 4.1, while quasar 4.2 is not in the FoV of MUSE due to the large projected separation. Its direction is indicated by a green arrow. As can be seen in Fig.~\ref{fig:qsospectra}, the spectrum of quasar 4.1 shows deep absorption troughs blueshifted of all broad lines and has been therefore classified as broad absorption line quasar in past literature (e.g., \citealt{Gibson09}). The extended emission around this object is among the brightest and most extended, spanning almost 100~kpc between the quasar and the 2$\sigma$ isophote. There is a clear asymmetry in the nebula, also evident in the big offset of flux-weighted nebula centroid and quasar position of $\sim$20~kpc and the relatively low value of $\alpha$. A secondary peak of the Ly$\alpha$ surface brightness is visible to the south of the quasar. This substructure could be gas associated with an inflowing companion galaxy, but no continuum source is robustly detected at this position. The $i$-band forced magnitude in an aperture of 3$\arcsec$ for the substructure is 23.7, while the average background magnitude in the same aperture is 24.3. 
The increase in surface brightness is also apparent in the surface brightness profile measured within the 2$\sigma$ isophote, peaking at 300~ckpc. The profile calculated from the full 2D map deviates from the typical quasar pair profile as the central surface brightness is higher, but the emission falls off more steeply than the median profile.
Once again, the kinematics are relatively quiescent throughout the full extent of the extended Ly$\alpha$ structure ($\overline{\sigma}=395$~km~s$^{-1}$).

\subsubsection{Quasar 5.1 and 5.2}

Displayed in Fig.~\ref{fig:qso5} are the results for quasar 5.1, while quasar 5.2 has not been observed due to the high angular separation. Quasar 5.1 is closely accompanied by two continuum sources first detected in the MUSE data cube and examined in more detail in a companion paper \citep{Herwig2024b}. The detected extended emission is found at a significant offset from the quasar, ${\rm d_{QSO-Neb} = 33.6\ kpc}$, and is not directly connected to it. The velocity shift in this nebula spans a large range compared to the rest of the sample, with an evolution from ${\rm 500\ km\, s^{-1}}$ to ${\rm -500\ km\, s^{-1}}$ in the east-west direction. The cool gas traced by the Ly$\alpha$ emission might be expelled from the interstellar medium of the quasar and companion host galaxies due to tidal interactions.

\subsubsection{Quasar 6.1 and 6.2}

Results for the close quasar pair 6 can be found in Fig.~\ref{fig:qso6}. It is the highest redshift pair in the sample and, possibly as a result of this, shows strong IGM absorption blueward of the Ly$\alpha$ peak. The extended Ly$\alpha$ emission may also be affected by the enhanced IGM absorption compared to the rest of the sample.
The quasar pair is associated with a low surface brightness filamentary nebula stretching between the two quasars. The nebula velocity evolves from ${\rm \sim -300\ km\, s^{-1}}$ around the fainter quasar with lower redshift to ${\rm 400\ km\, s^{-1}}$ close to the brighter quasar with higher redshift, indicating that the nebula is indeed associated with both quasars, whose velocity separation is of $870$~km~s$^{-1}$. Therefore, the centroid of the nebula is roughly between the pair, 35~pkpc away from the brighter quasar.
The surface brightness profile within the 2$\sigma$ isophote stays remarkably constant and likewise decreases slowly when calculated in the full surface brightness map.
The kinematics are among the most quiescent in this sample, with $\overline{\sigma}=342$~km~s$^{-1}$.

\subsubsection{Quasar 7.1 and 7.2}

Quasars 7.1 and 7.2 is the only pair at wide (i.e., approximately 500~kpc) separation for which both quasars have been observed (Fig.~\ref{fig:qso7}). The brighter quasar in the pair is associated with a relatively small and faint (${\rm L_{Neb} \sim 10^{43} erg\, s^{-1}}$) nebula with a significant axis asymmetry and a velocity dispersion around ${\rm 350\ km\, s^{-1}}$. Quasar 7.2, on the other hand, is not associated with extended emission exceeding our cut-off area of ${\rm A_{2 \sigma} > 2\ arcsec^2}$ to ensure that the biggest 2$\sigma$ isophote detected close to the quasar is not caused by spurious emission or noise. For this feature, the velocity shift is almost ${\rm 0\ km\, s^{-1}}$ and the velocity dispersion reaches atypically high values above ${\rm 700\ km\, s^{-1}}$, supporting the exclusion of this isophote from the morphological analysis.

\subsubsection{Quasar 8.1 and 8.2}

This object, shown in Figue \ref{fig:qso8}, has already been published in \cite{FAB19b} and is reanalyzed in this work to ensure consistency. While the overall shape of the extended emission stays constant in both works, the method employed here is more conservative (pseudo-NB vs channel-by-channel detection) and consequently does not detect low surface brightness emission features. While the nebulae associated with the two quasars are still connected through a gaseous filament, the emission is much less coherent and smaller in area. However, as the emission appears physically connected, we treat it as one nebula in the analysis.

Quasar pair 8 is embedded in the biggest nebula of the sample with an extent of almost 160 kpc. The velocity shift of the gas evolves from ${\rm -250\ km\, s^{-1}}$ around quasar 8.2 to ${\rm 350\ km\, s^{-1}}$ close to quasar 8.1, and it is therefore comparable with the velocity separation of the two quasars ($896$~km~s$^{-1}$; see discussion in \citealt{FAB19b}). We measure an average velocity dispersion of ${\rm \sim 400\ km\, s^{-1}}$, with higher dispersions around quasar 8.2 (see Sect.~\ref{sec:comp_pairStudies} for further details).
The surface brightness profile of the emission mirrors the median profile of all quasar pairs remarkably well, starting out faint in the center and falling off gently.

\section{Discussion}
\label{sec:discussion}

\subsection{Comparison to previous quasar pair studies}
\label{sec:comp_pairStudies}

To our knowledge, the number of previous works targeting extended emission on CGM/IGM scales around physically associated quasar pairs is limited to seven studies (\citealt{Cai18,FAB19b,Lusso19,Li23}), three of which have been serendipitous detections since the faint companions were not known during observation planning (\citealt{Cantalupo14,Hennawi15,FAB18}).
These studies unanimously reported Ly$\alpha$ nebulae with exceptional extents (up to 460~kpc). In our work, however, extended emission often does not cover similarly large areas (>10$^4$~kpc$^2$) or reach such high luminosities ${\rm L_{Neb}}$ (>$10^{44}$~erg~s$^{-1}$).

\cite{Li23} targeted quasar pairs at $z \sim 2.5$ with one bright ($g$<18 mag) quasar and included a known ELAN first published in \cite{Cai18}, resulting in two out of five quasar pairs meeting their criteria for ELANe.
Compared to that work, nebulae studied here have much smaller extents, even when the integrated luminosity is comparable. This fact is indicative of the differing methods to determine what significant emission means. Here we decided to employ a more conservative approach in order to match previous narrowband study designs, leading to smaller 2$\sigma$ isophotes. A comparison of these results is therefore not straightforward, but we can firmly state that our data do not unveil nebulosities as bright as those in prototypical ELANe, which have SB$_{\rm Ly\alpha}\sim10^{-17}$~erg~s$^{-1}$~cm$^{-2}$~arcsec$^{-2}$ on hundreds of kiloparsecs (e.g., \citealt{Hennawi15,FAB18}). Therefore, the sample presented here does not confirm the overabundance of ELANe around the overall population of quasar pairs as we do not identify any ELAN within our sample of 8 pairs. Importantly, the targeted sample extends the study of large-scale emission around quasar pairs to additional systems in which both quasars are faint (Fig.~\ref{fig:miz}). This fact is also one of the reasons why we do not detect bright and extended nebulae around all pairs (see more details in Sect.~\ref{sec:comp_singleStudies}). In Fig.~\ref{fig:SBstacked} we compare our stacked $z\sim3$ quasar pair Ly$\alpha$ profile to the one obtained using the same method as employed in this work by \cite{Li23} for $z\sim2.5$ quasar pairs after correcting them for cosmological dimming. This exercise shows that overall the SB levels are consistent, contrary to the results found for Ly$\alpha$ emission around single bright quasars, with $z\sim3$ quasars having a factor of $\sim3$ brighter extended Ly$\alpha$ emission (\citealt{Cai19}). In other words, naively, we would have expected quasar pairs at $z\sim3$ to be associated with brighter emission than $z\sim2$ quasar pairs for similar AGN magnitude ranges.  Therefore we inferred that the emission in our sample is modest because of the average lower luminosity of the targeted quasars (Sect.~\ref{sec:comp_singleStudies}). 

Contrarily, the Ly$\alpha$ nebula published by \cite{Lusso19} in the MUSE Ultra Deep Field (MUDF) around a $z\sim3$ quasar pair ($m_r^{\rm QSO1}=17.9$ and  $m_r^{\rm QSO2}=20.5$) shows a similar morphology to some objects studied here, in particular to quasar 4.1. Both pairs have a projected separation of about 500 physical kiloparsecs, with the Ly$\alpha$ emission around the brighter quasar extending toward the fainter second quasar and showing an additional smaller nebula along the same direction (Fig.~\ref{fig:qso4}). While the extremely long integration time of 40 hours in \cite{Lusso19} allows for sensitivity to exceptionally faint emission, the nebula sizes still do not exceed the expected virial radius. The elongation between that quasar pair also appears to fit the relation shown in Fig. \ref{fig:extendvsangle} for our systems. Quasar pairs like the one targeted in \cite{Lusso19} and quasar pair 4 are therefore excellent systems for ultra-deep integrations to unveil intergalactic bridges and directly study the IGM.

Lastly, one object studied in this work, pair 8, has already been published in \cite{FAB19b}. While most of the results are the same, extended emission is detected in that work using a 3D mask, making it more sensitive to faint and narrow emission. Therefore, the area detected above 2$\sigma$ is bigger, and the nebula is more coherent than in this work. Since the velocity information in that work was computed using only the wavelength ranges within the 3D mask and included the nebula outskirts, the average velocity dispersion is much lower (162~km~s$^{-1}$). We caution that the dispersion calculation is highly dependent on the spectral width considered, and the inclusion of spaxels spanning 1.25~$\AA$ to 2.5~$\AA$ in the 3D mask necessarily leads to very low dispersion measures. This comparison highlights the importance of considering the different methods of extended emission extraction when studying Ly$\alpha$ nebula kinematics.

\subsection{Comparison to single quasar samples}
\label{sec:comp_singleStudies}

A large number of single quasar fields around redshift three have already been targeted with IFUs to study extended Ly$\alpha$ emission, although mostly focusing on more luminous quasars than presented in this work (Fig.~\ref{fig:miz}). Multiple surveys comprising more than one hundred bright $z\sim3-4$ quasars detected extended Ly$\alpha$ emission on scales of, on average, 80 to 150~kpc in maximum projected size around 100~\% of the targets \citep{Borisova16,FAB19a, Fossati21}. This is in stark contrast to the quasar pair sample presented here, where two of the 14 observed quasars are not associated with extended emission and the detected nebulae are comparatively small. This difference can be appreciated by comparing the stacked SB profiles presented in Fig.~\ref{fig:SBstacked}. The SB profile of our $z\sim3$ quasar pair sample is 5 times fainter than the SB profile around bright single quasars. This difference is remarkable as some of the targeted quasar pairs could trace ongoing mergers between massive galaxies. In such events several processes (e.g., tides) could boost the gas densities on CGM scales and hence the associated Ly$\alpha$ emission (Sect.~\ref{subsec:gasorigin}). Further, the differences cannot be explained by the observational setup as the same method is adopted for most of the targets in the aforementioned studies (i.e., an exposure time of at least 1 hour in MUSE WFM) but it could possibly arise due to target selection as our quasar pair sample has not been selected from SDSS based on having bright quasars in the pair.

Indeed, a single quasar survey selecting 12 similarly low luminosity quasars at $z\sim3.1$ found extended emission in each targeted field with a minimum extent of 60~pkpc, but did show a connection between quasar Ly$\alpha$ peak luminosity and nebula luminosity \citep{Mackenzie21}. 
We note that \citet{Mackenzie21} adopted an optimal extraction method for constructing a 3D region for the study of the Ly$\alpha$ emission. This technique, applied channel by channel, is more sensitive to the fainter portions of the nebulae with respect to our method. For this reason we tested our analysis by applying it to the two extreme nebula cases in that sample (see Appendix~\ref{sec:appMackenzie}). We find that our analysis is able to reproduce the principal shape and luminosity of the brightest nebula reported in that paper, while our measurement of the extent deviates by 30~\%. The faint nebula is only detected in the peak of the nebula emission and is thus three times smaller in maximum extent than reported in \cite{Mackenzie21}. Therefore, the difference in maximum extent of 15~\% between the faint single-quasar sample and our sample of quasar pairs is likely in part due to the difference in methodology.

The stacked SB profile from that work has roughly similar SB values as the stacked profile for the quasar pairs (Fig.~\ref{fig:SBstacked}), but declines faster at large radii. 
It is therefore likely that the gaseous environments around quasar pairs at $z\sim3$ are not the same as around single quasars at similar redshifts, indicating the presence of more cool gas at large projected distances.

This is further confirmed by comparing our stacked SB profile with the stacked profile of 59 similarly faint ($-27<M_i(z=2)<-24$) $z\sim3.1$ quasars from the QSO MUSEUM III survey (Gonz\'alez Lobos et al. in prep.). The surface brightness is remarkably similar at intermediate distances to the central quasar, but is brighter at close separation and falls off quicker, in agreement with the profile in \citet{Mackenzie21}.

We quantify the differences between SB profiles around single quasars and quasar pairs by fitting them with power laws. 
The SB profile for single $z\sim3$ quasars 
can be described by a power law with an index reported to be between -1.8 \citep{Borisova16} and -1.96 \citep{FAB19a} for bright quasars, and $-2.41 \pm 0.19$ for the faint sample (Gonz\'alez Lobos et al. in prep.). In contrast, we find a slope of $-1.57 \pm 0.17$ for the quasar pair sample. Depending on the Ly$\alpha$ powering mechanism, the shallower slope of the profile could be evidence of 
(i) presence of denser and/or a bigger reservoir of ionized gas at larger radii in the CGM of pairs in comparison to single quasars (if Ly$\alpha$ is dominated by recombination radiation; \citealt{Hennawi2013}) and/or (ii) more neutral hydrogen on large scales in the CGM of pairs (if Ly$\alpha$ is dominated by resonant scattered photons). Intriguingly, \citet{Lusso2018} found evidence for moderate excess of optically thick absorbers (with column densities of HI log$N_{\rm HI}\geq 17.2$) in closely projected $z\sim2$ quasar pairs compared to single quasars, arguing that they are likely to trace mostly structures located in denser (partially neutral)
regions within the CGM or IGM where both quasars reside.

Further evidence of the presence of large-scale structures comes from the morphologies of the discovered nebulae around quasar pairs. In  section~\ref{sec:morpho} we have shown a comparison of their morphologies (described by their axis ratios) with respect to nebulae around single quasars (Fig.~\ref{fig:lnebvsalpha}).
Morphologies of nebulae around single quasars are mostly circular \citep{Borisova16, FAB19a}, with even higher symmetry for faint quasars \citep{Mackenzie21}. ELANe, known to host multiple AGNs, are more asymmetric (\citealt{FAB23}). Similarly, quasar pair nebulae have a median axis ratio lower than nebulae around single quasars (Fig. \ref{fig:lnebvsalpha}). This is akin to redshift $\sim$ 2 single quasar nebulae with a median value of $\alpha = 0.54$ \citep{Cai19} hypothesized to originate in galaxy mergers, which peak at $z=2$. While this interpretation can be supported in the context of quasar pairs, the spread of $\alpha$ in our small sample is substantial and $\alpha$ is not directly correlated with the projected distance. Asymmetries can therefore not only be driven by galaxy mergers, but also by intergalactic structures.

\subsection{Direction of nebulae with respect to the quasar pairs}
\label{subsec:alignment}

Assuming each quasar inhabits a node of the cosmic web, the connecting line between two members of a pair can be used as a proxy for the direction of an intergalactic filament. Following this argument, the value of the angle $\phi$, introduced in Sect.~\ref{sec:morpho}, indicates the level of alignment between the discovered Ly$\alpha$ nebulae and the cosmic web surrounding their host galaxies. A small angle $\phi$, together with a large elongation ${\rm d_{QSO,max}}$ (see Sect.~\ref{sec:morpho} for definition) should correspond to the best candidates for intergalactic bridges as tested in Fig.~\ref{fig:extendvsangle}.

This expectation is based on the assumption that a nebula semi-major axis (used to compute $\phi$) is a good proxy for the direction of the underlying cool gas distribution. Cosmological simulations post-processed with a radiative transfer tool have shown that this is true if the quasar host galaxy is not edge-on \citep{Costa22}. The more edge-on the host galaxy the more lopsided the extended Ly$\alpha$ emission. The fact that the majority of quasar nebulae are symmetric may therefore indicate that their host galaxies are far from being edge-on and are sampling the surrounding hydrogen gas distribution well \citep{FAB23}. 

These simple predictions are further complicated by the fact that the morphology of the Ly$\alpha$ nebulae (and hence ${\rm d_{QSO,max}}$ and the angle $\phi$) can be influenced by a multitude of factors.
Indeed it has been shown that the following factors all play a role in shaping the extents and luminosities of quasar nebulae: dust in the host galaxy or the CGM \citep{Smith22, Jay23} as dust can efficiently absorb Ly$\alpha$ photons; the molecular gas mass of the host galaxy, being a proxy for denser and more dust rich galaxies that may prevent the escape of ionizing and Ly$\alpha$ photons \citep{Nahir22}; the Ly$\alpha$ luminosity and/or magnitude of the powering sources, with the trend of brighter quasars having brighter and often more extended nebulae \citep{Mackenzie21}; the availability of cool, 10$^4$ K gas that has the potential to be Ly$\alpha$ bright; and geometry effects (e.g., orientation of the quasar's ionizing cones with respect to the gas distribution; \citealt{Obreja24}; orientation of cosmic web filaments with respect to the observing angle, which could lead to absorption of Ly$\alpha$). In addition to those, quasar pair nebulae might not trace exclusively inflowing CGM gas (see Sect.~\ref{subsec:gasorigin}), especially in close pairs.

However, an anti-correlation between nebula misalignment and extent can be seen in our quasar pair sample irrespective of quasar magnitude and projected pair distance (Fig.~\ref{fig:extendvsangle}), indicative of a direct link between filaments of the cosmic web and the distribution of cool gas in the CGM. This trend suggests that the extended nebulae discovered in our study with small angle $\phi$ and large ${\rm d_{QSO,max}}$ (e.g., quasar pair 4) probe gas falling into the quasar dark-matter halo after accretion from the cosmic web. Contrarily, less extended nebulae are perfect laboratories to test all the aforementioned processes that could strongly affect the morphology of Ly$\alpha$ nebulae. 
This means that any alignment found between large-scale galaxy distributions around quasars and their Ly$\alpha$ nebulae (\citealt{FAB23}) should break for extremely small nebulae as the smallest discovered in this study. 

In the next section we discuss more about the origin of the cool gas as probed in all the diversity of quasar pair nebulae in this study. 

\subsection{Origin of the cool gas probed by Ly$\alpha$ around quasar pairs}
\label{subsec:gasorigin}

The gas traced by extended Ly$\alpha$ emission around $z\sim3$ quasars is usually assumed to be the cool ($\sim10^4$~K) phase of the CGM, supported by typical Ly$\alpha$ halo radii $< 100$~kpc for the bulk of the emission, just short of the expected virial radius for quasar halos (e.g., \citealt{FAB19a}). This is further confirmed by the levels of the Ly$\alpha$ SB profiles around quasars as a function of redshift. The absence of an evolution from $z\sim6$ down to $z\sim3$ (\citealt{Farina19,Fossati21}) and the subsequent decrease to $z\sim2$ (\citealt{FAB19a,Cai19}) seems in agreement with the theoretical expectation (e.g., \citealt{DekelBirnboim2006}) that high-$z$ ($z\gtrsim3$) massive halos are able to accrete gas in a cool phase keeping the Ly$\alpha$ emission roughly constant, while at lower redshift, the gas mass in the cool phase (and hence the Ly$\alpha$ emission) is reduced by shock heating.  At moderate depths ($\sim10^{-18}$~erg~s$^{-1}$~cm$^{-2}$~arcsec$^{-2}$), a few rare exceptions show very extended emission that can exceed the projected virial radius of the host halo and therefore reach into the IGM (e.g., \citealt{Cantalupo14,FAB19a}). However, most of these outliers have active companions.

Quasar pairs seem to follow the same trends as single quasars, but with, on average, enhanced Ly$\alpha$ emission at larger distances  (Sect.~\ref{sec:comp_singleStudies}). The Ly$\alpha$ emission in quasar pairs can therefore better trace IGM structures given the closer presence of an additional massive halo. Indeed, cosmological simulations report the presence of denser and thicker filaments in denser environments (\citealt{Cautun2014}), similar to the locations quasar pairs are expected to populate in the cosmic web (e.g., \citealt{Boris2007, Onoue18}). Given the presence of the second quasar, they are at least in a group environment. For this reason, processes favored by dense environments could affect the cool gas distribution more than around single quasars. In particular, galaxy interactions may become especially important in pairs at close separation. 
Both tidal stripping and ram pressure stripping could significantly enhance the density of the CGM in a preferential direction aligned with an interacting pair \citep{Salem15, Samuel23} and may therefore boost the observed surface brightness out to large radii. 
Evidence for gas stripping around high-$z$ quasars is indeed mounting (e.g., \citealt{TC2021,Vayner2023}).
The effects of these environmental processes can, in part, explain the alignment of extended nebulae, specifically in close pairs (Fig.~\ref{fig:extendvsangle}), and the comparatively flat surface brightness profile observed in the median stack of quasar pairs (Fig.~\ref{fig:SBstacked}) and especially apparent in the profiles of some individual quasar pairs, for example pair 6 (Fig.~\ref{fig:qso6}). 
However, the properties of the extended emission studied here (Table~\ref{tab:3}) do not correlate with the angular separation between the quasar pairs, as would be expected if the galaxy interaction alters the CGM in an observable way. Therefore, the morphology of extended Ly$\alpha$ emission alone does not provide evidence for the aforementioned environmental effects.
Summarizing, the diversity of extended Ly$\alpha$ emission around quasar pairs probes the combined effects of dense environments and close massive halos, possibly tracing the surrounding large scale structure. 

\section{Summary and conclusions}
\label{sec:summary}

Motivated by previously discovered ELANe around quasar pairs and signs of filamentary gas connecting two quasars in multiple systems, we conducted the largest study of extended Ly$\alpha$ emission around quasar pairs to date. Our sample consists of 14 quasars in eight pairs observed with VLT/MUSE with a snapshot strategy (45 minutes/source). The targets span redshifts of $z\sim 3-4$ and $i$-band magnitudes between 18 and $\sim$ 22.8, dimmer than most single-quasar samples. Our key findings are as follows:

\begin{itemize}
\item Twelve out of the 14 observed quasars are associated with extended Ly$\alpha$ emission spanning 40 to 160 kpc, and three of the five close pairs are embedded in the same Ly$\alpha$ nebula (Table~\ref{tab:3}, Figs.~\ref{fig:qso1}, ~\ref{fig:qso2} - \ref{fig:qso8}). In contrast to previously discovered quasar pairs, none of these nebulosities meet the criteria for ELANe in either $L_{\rm Neb}$ or extent, lowering the fraction of ELANe around known quasar pairs significantly (Sect.~\ref{sec:comp_pairStudies}).
On average, the nebulae in our sample are 15 \% smaller in maximum projected extent than nebulae around similar single-quasar samples, but they have comparably quiescent kinematics (Section~\ref{sec:comp_singleStudies}).
\item Larger nebulae are preferentially aligned with the expected direction of the large-scale structure between the associated quasar pair, while smaller nebulae tend to be misaligned (Fig.~\ref{fig:extendvsangle}). This fact is indicative of a direct connection between filaments and the cool CGM (Sect.~\ref{subsec:alignment}). 
Wide pairs highly aligned with their filaments, such as quasar pair 4, are excellent targets for ultra-deep follow-up observations to detect these faint IGM structures.
\item The SB profile of quasar pairs is comparable in brightness to single-quasar profiles at similar redshift but is fainter close to the quasar and falls off more gradually with a power law slope of $-1.57 \pm 0.17$ (Fig.~\ref{fig:SBstacked}). Due to galaxy interactions important in close quasar pairs, the higher surface brightness at larger radii can in part be explained by stripped gas increasing the densities in the CGM. The presence of a second large halo can also add to the Ly$\alpha$ emission through increased contribution from IGM structures (Sect.~\ref{subsec:gasorigin}). 
\item Extended emission around this sample of quasar pairs is morphologically diverse, but on average, it is more asymmetric than single-quasar nebulae with large offsets between the quasar position and the flux-weighted nebula centroid (Table~\ref{tab:3}, Fig.~\ref{fig:lnebvsalpha}). This can be explained by the presence of more substructures in the overdense regions likely populated by quasar pairs and by close pairs sometimes sharing the same nebula.
\item We did not find a trend in brightness or morphology of the extended Ly$\alpha$ emission with the projected distance between the associated quasar pair, as would be expected if the host galaxies' interaction influences the CGM in a way that is observable through the extended Ly$\alpha$ emission (Sect.~\ref{subsec:gasorigin}). While the sample studied in this work only encompasses close ($\sim50-100$~pkpc) and wide ($\sim450-500$~pkpc) pairs, extending the sample of observed quasar pairs to intermediate projected distances has the potential to reveal the full picture.

\end{itemize}

This study highlights once again the promise of Ly$\alpha$ nebulae around quasars tracing the large-scale structure in dense environments. As quasar pair nebulae show a higher surface brightness level at larger radii than individual quasars, and they provide the most accessible laboratory to directly study the IGM and its connection to the associated galaxies. Indeed, our sample contains excellent targets for future deep and ultra-deep observations to study the large-scale gaseous structures of the IGM (e.g., \citealt{Tornotti2024}). Additional snapshot observations of quasar pairs are needed to increase the number of such suitable targets and therefore cover the diversity of IGM filament structures and active galaxy interactions at different separation stages. 

\begin{acknowledgements}
We thank the referee for their useful comments that helped us improve the manuscript.
Based on observations collected at the European Southern Observatory under ESO programme 0100.A-0045(A). AWSM acknowledges the support of the Natural Sciences and Engineering Research Council of Canada (NSERC) through grant reference number RGPIN-2021-03046. 
E.P.F. is supported by the international Gemini Observatory, a program of NSF NOIRLab, which is managed by the Association of Universities for Research in Astronomy (AURA) under a cooperative agreement with the U.S. National Science Foundation, on behalf of the Gemini partnership of Argentina, Brazil, Canada, Chile, the Republic of Korea, and the United States of America.
All figures have been made with \texttt{matplotlib} (\citealt{Hunter2007}) and \texttt{APLpy} (\citealt{aplpy2012,aplpy2019}). We have also used the Python libraries \texttt{numpy} (\citealt{Walt2011}), \texttt{scipy} (\citealt{2020SciPy-NMeth}), \texttt{spectral-cube} (\citealt{adam_ginsburg_2015_11485}),
and \texttt{astropy} (\citealt{astropy:2013, astropy:2018, astropy:2022}).
\end{acknowledgements}

\bibliographystyle{aa} 
\bibliography{lit.bib}

\begin{appendix}

\section{Additional material}
\label{sec:addmaterial}

\begin{figure*}
   \centering
   \includegraphics[height=4cm]{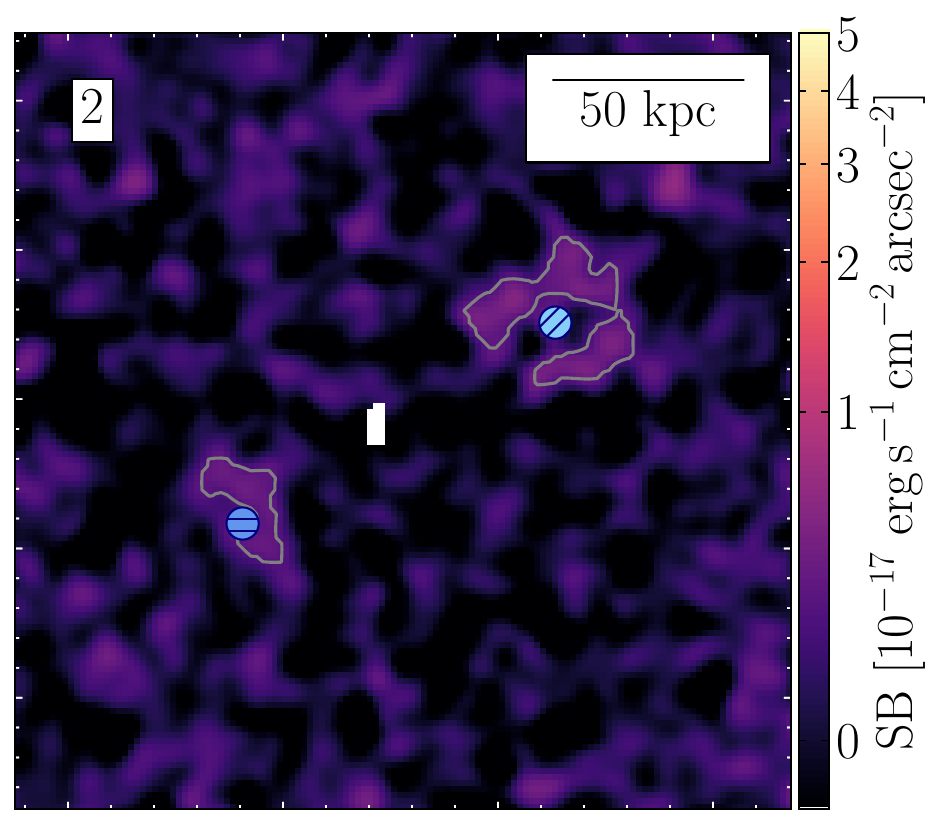}
   \includegraphics[height=4cm]{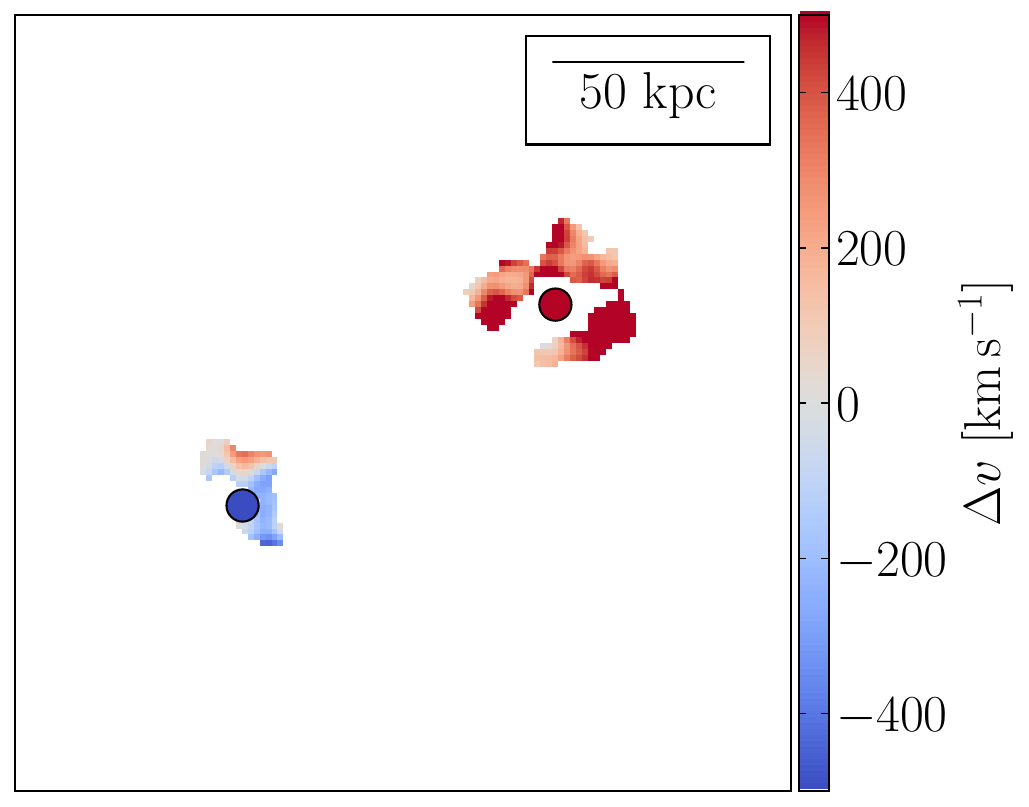}
   \includegraphics[height=4cm]{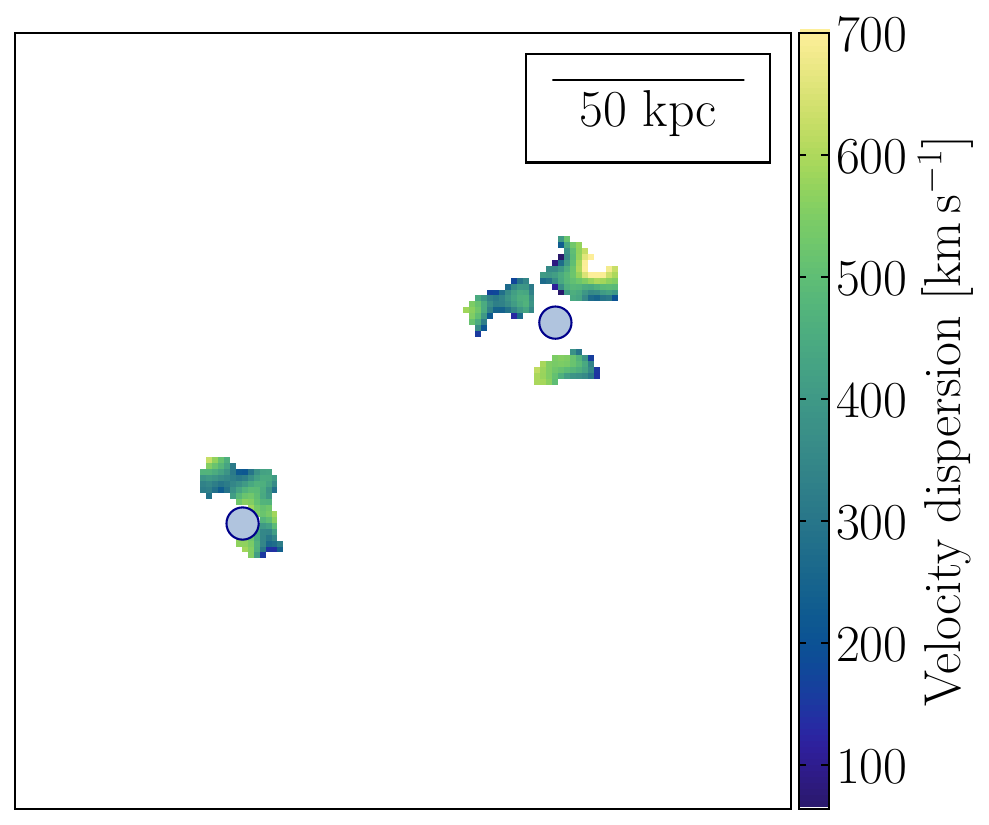}
   \includegraphics[height=5cm]{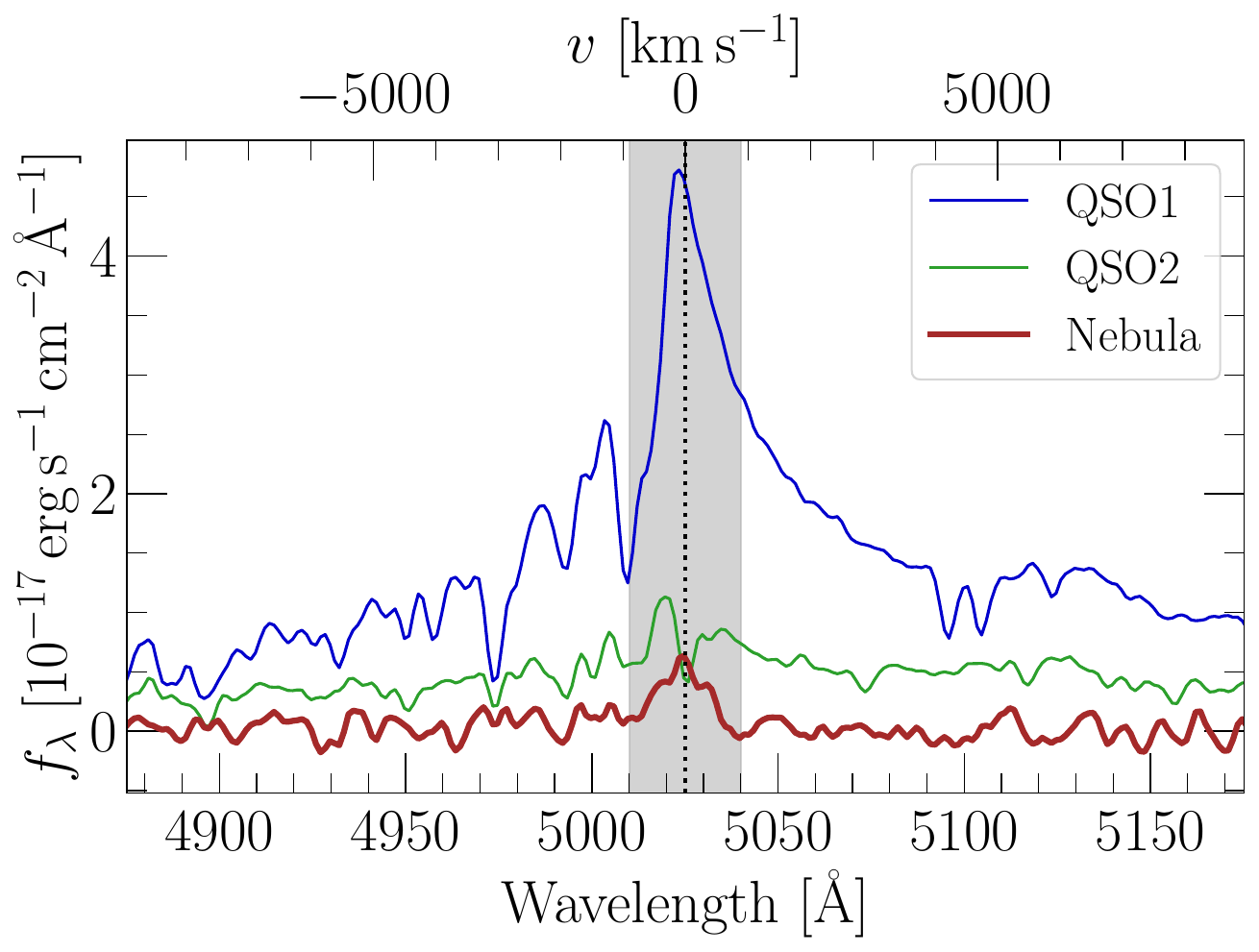}
   \includegraphics[height=5cm]{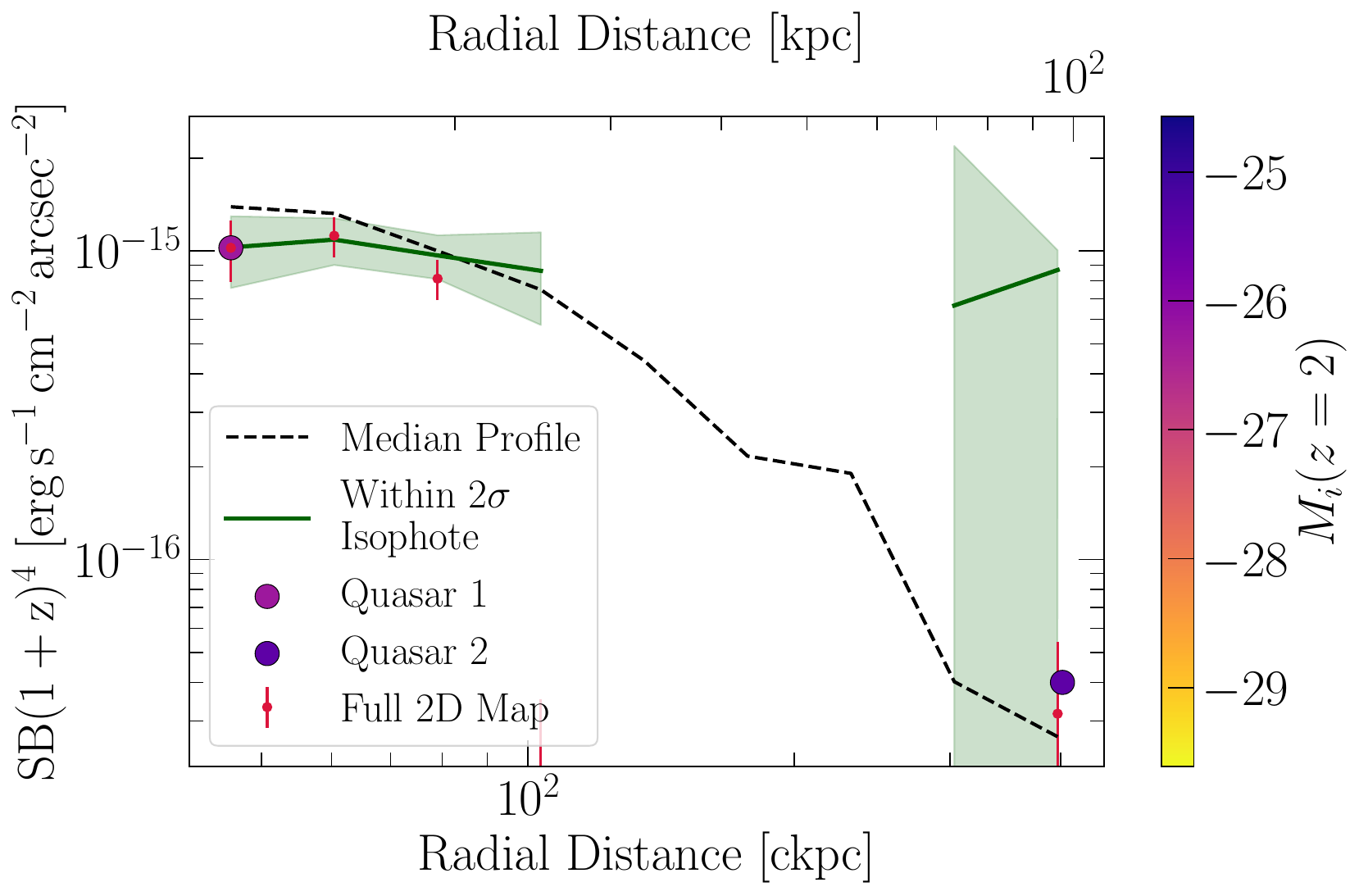}
      \caption{Same as Fig.~\ref{fig:qso1} but for quasar pair 2 (ID 2.1 and ID 2.2).}
         \label{fig:qso2}
   \end{figure*}

\begin{figure*}
   \centering
   \includegraphics[height=4cm]{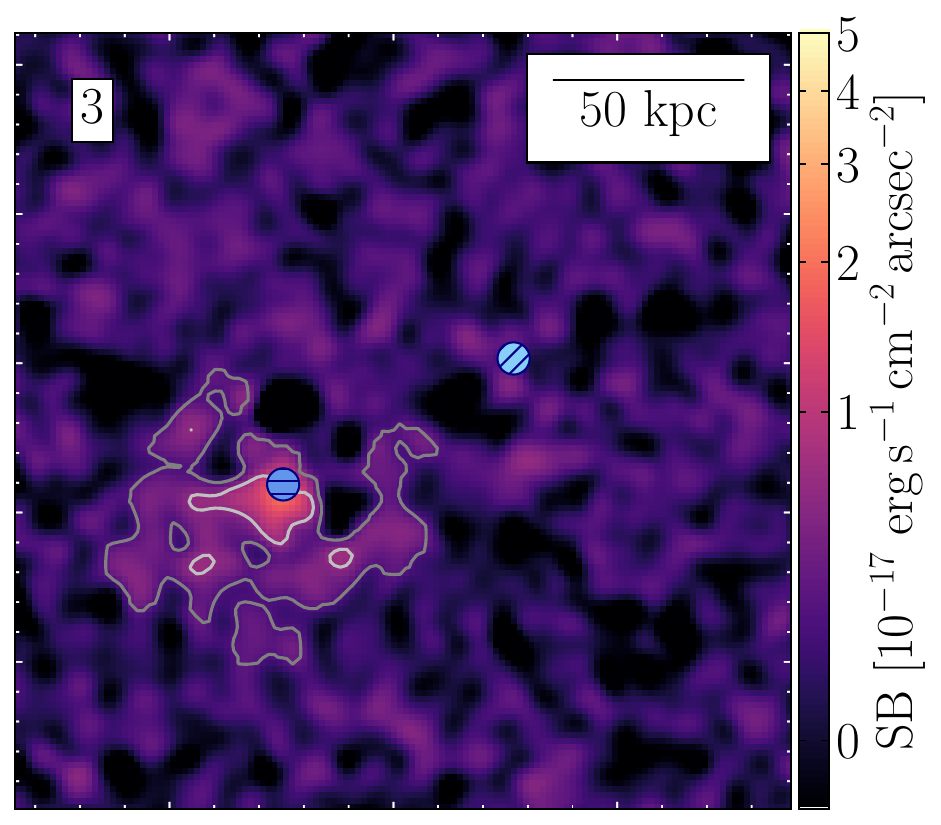}
   \includegraphics[height=4cm]{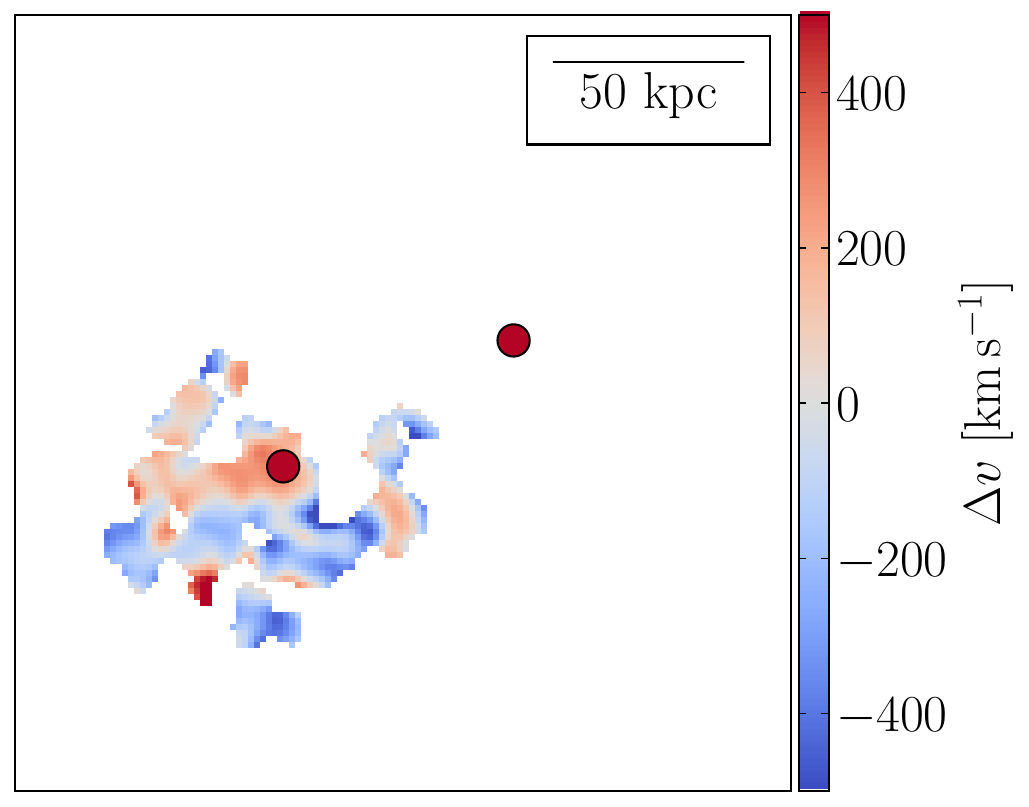}
   \includegraphics[height=4cm]{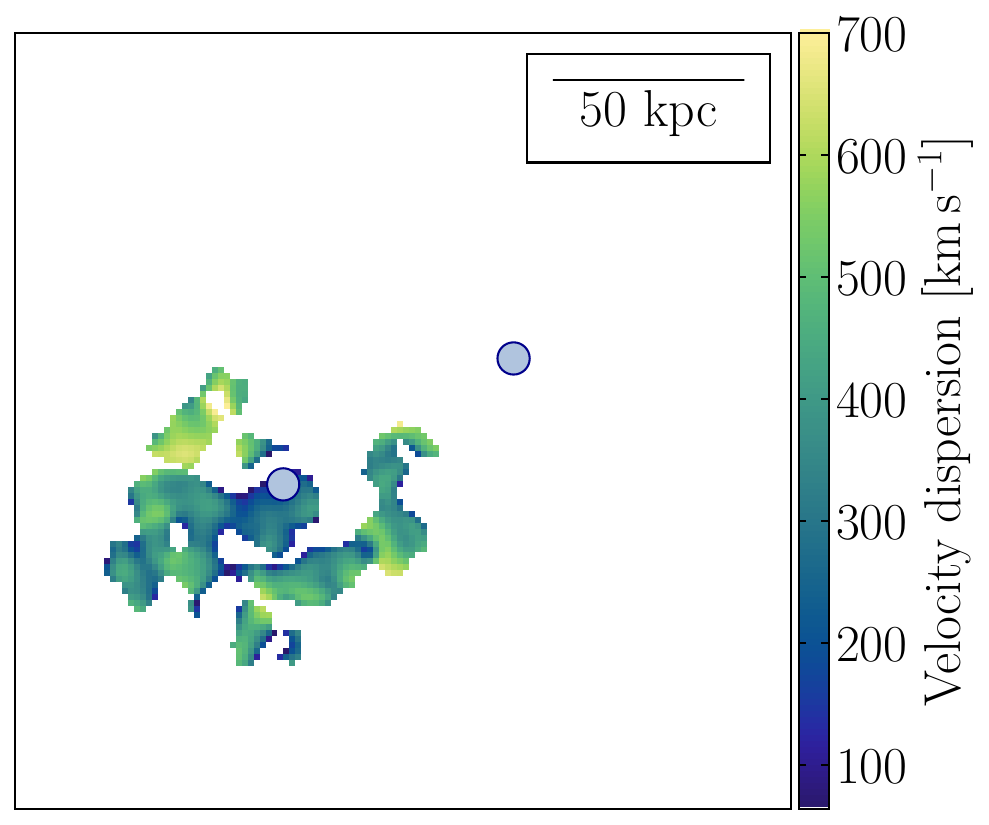}
   \includegraphics[height=5cm]{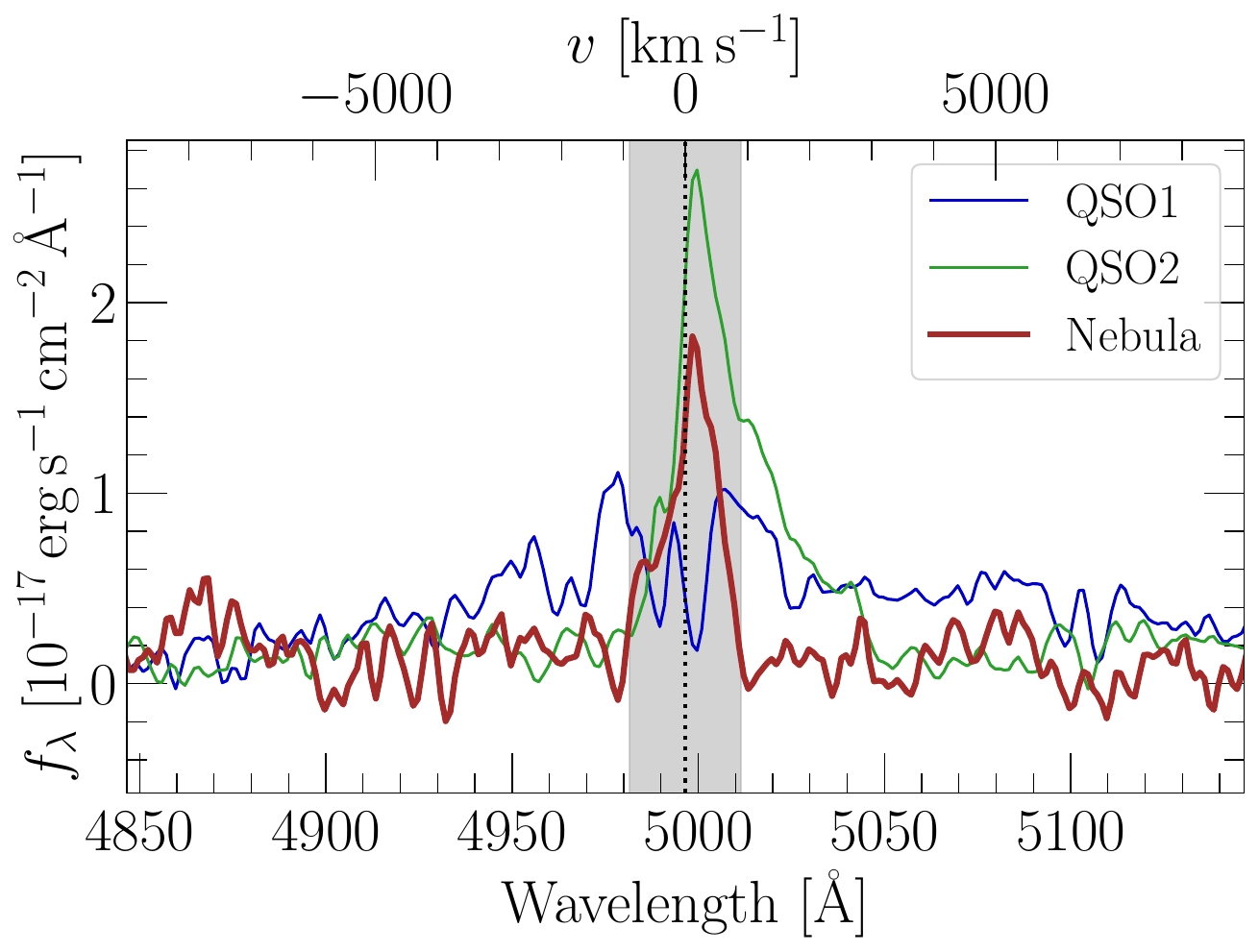}
   \includegraphics[height=5cm]{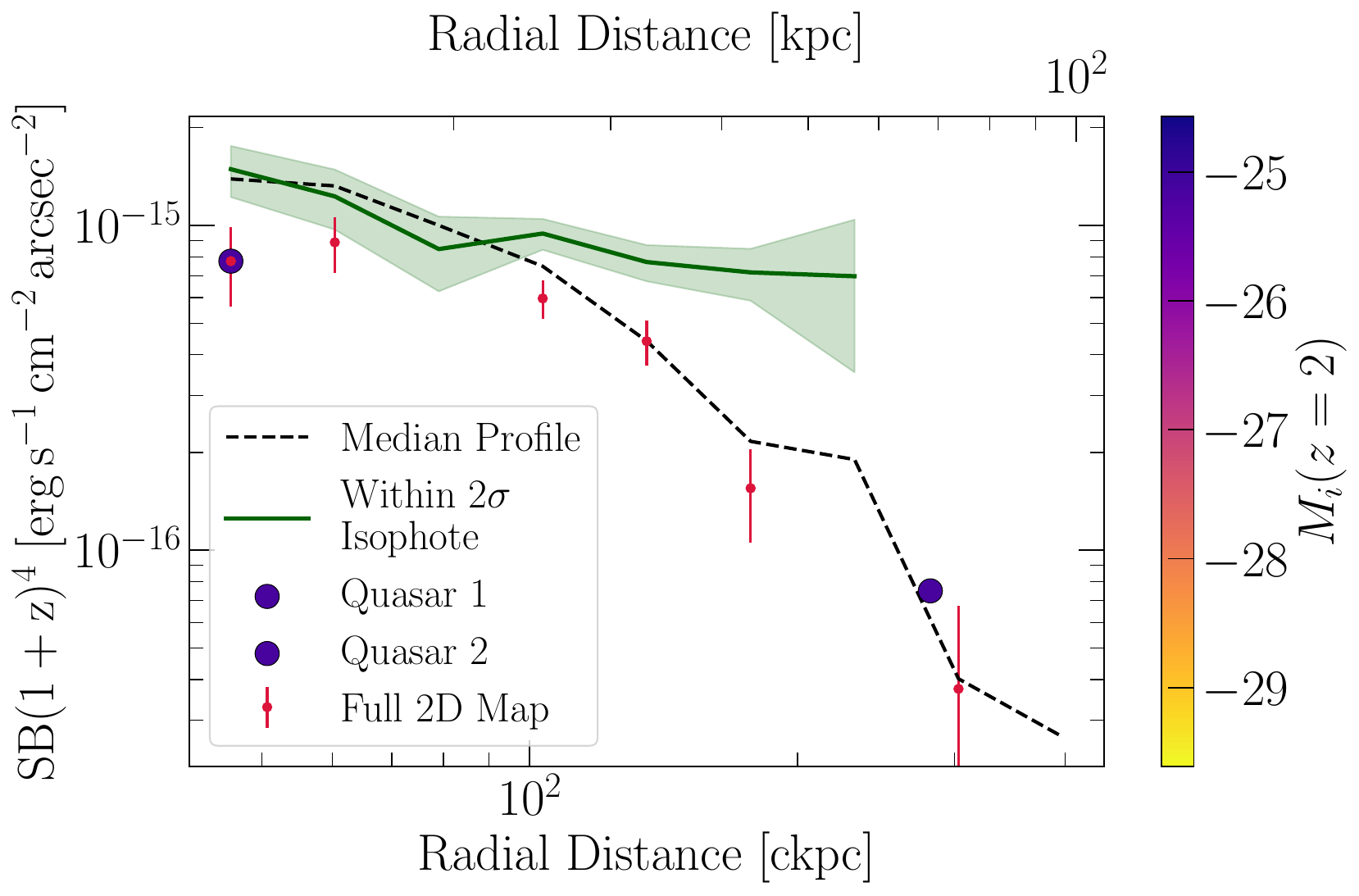}
      \caption{Same as Figure \ref{fig:qso1} but for quasar pair 3 (ID 3.1 and ID 3.2).}
         \label{fig:qso3}
   \end{figure*}

\begin{figure*}
   \centering
   \includegraphics[height=4cm]{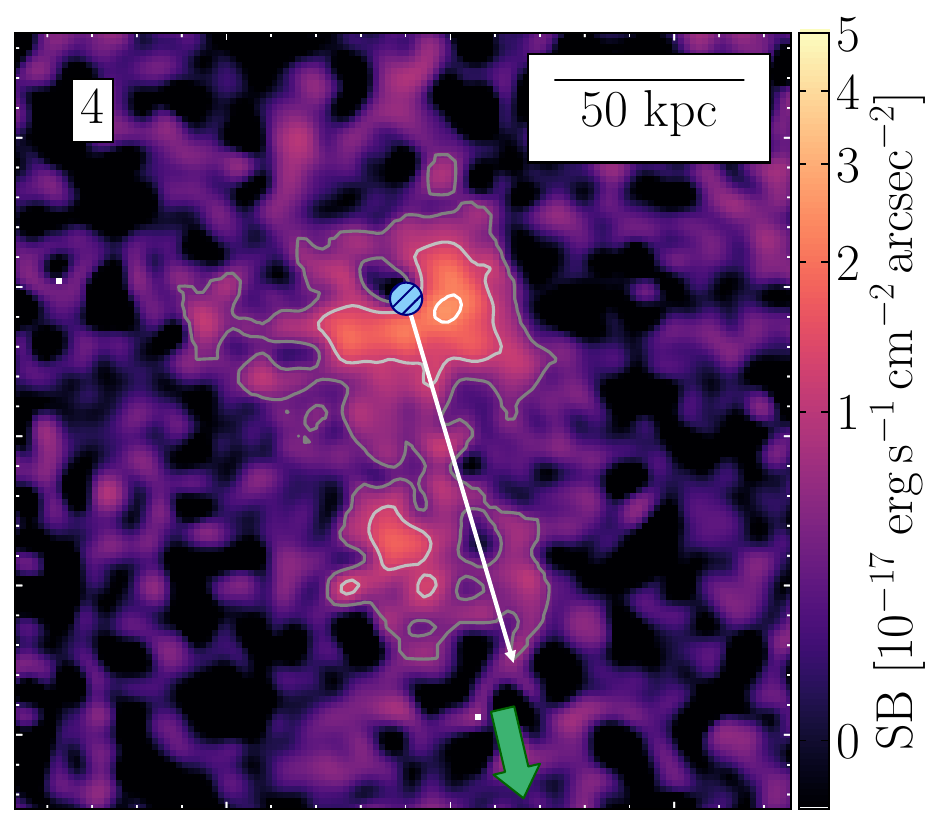}
   \includegraphics[height=4cm]{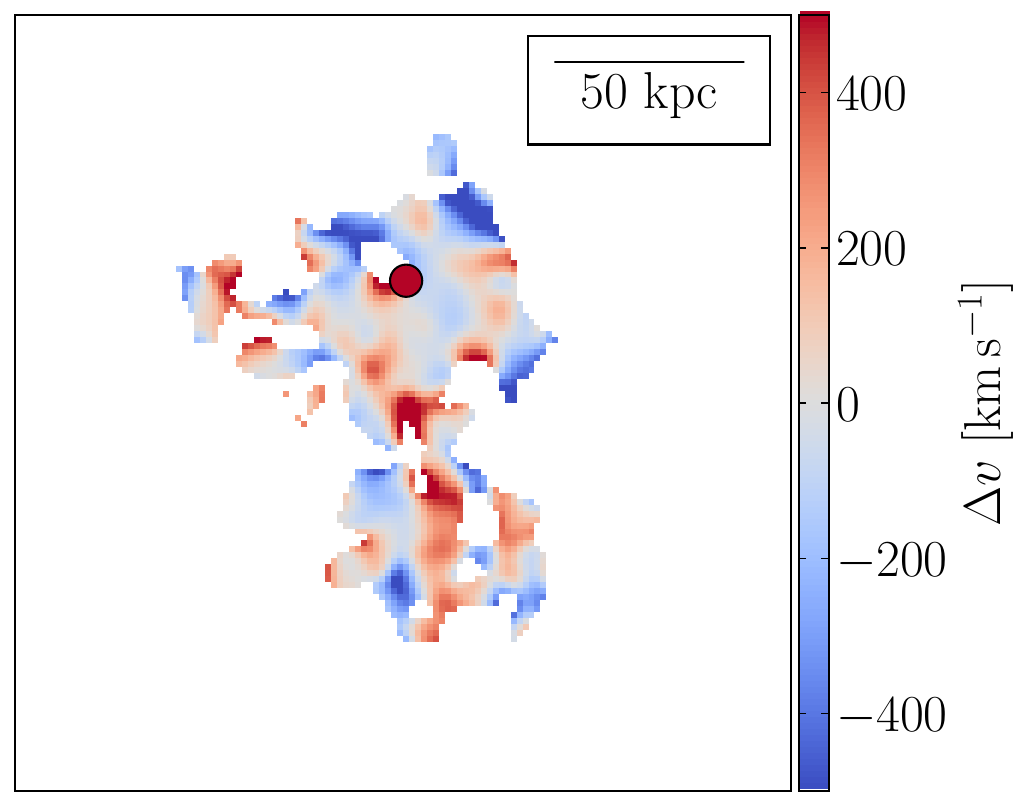}
   \includegraphics[height=4cm]{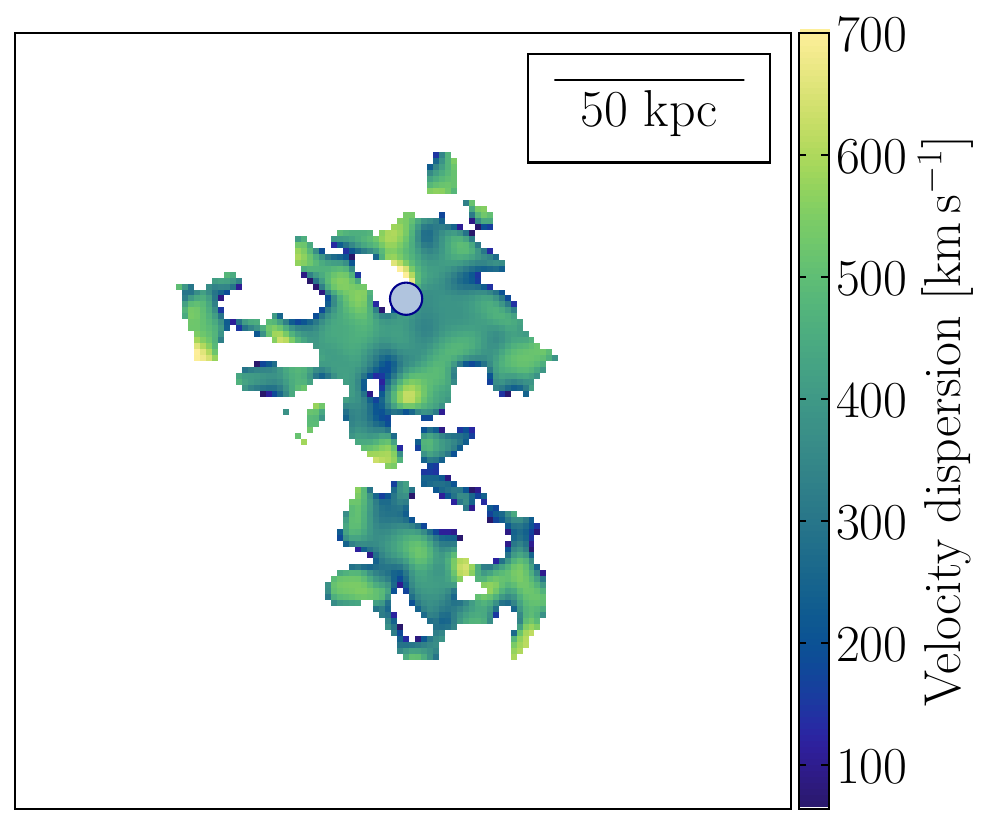}
   \includegraphics[height=5cm]{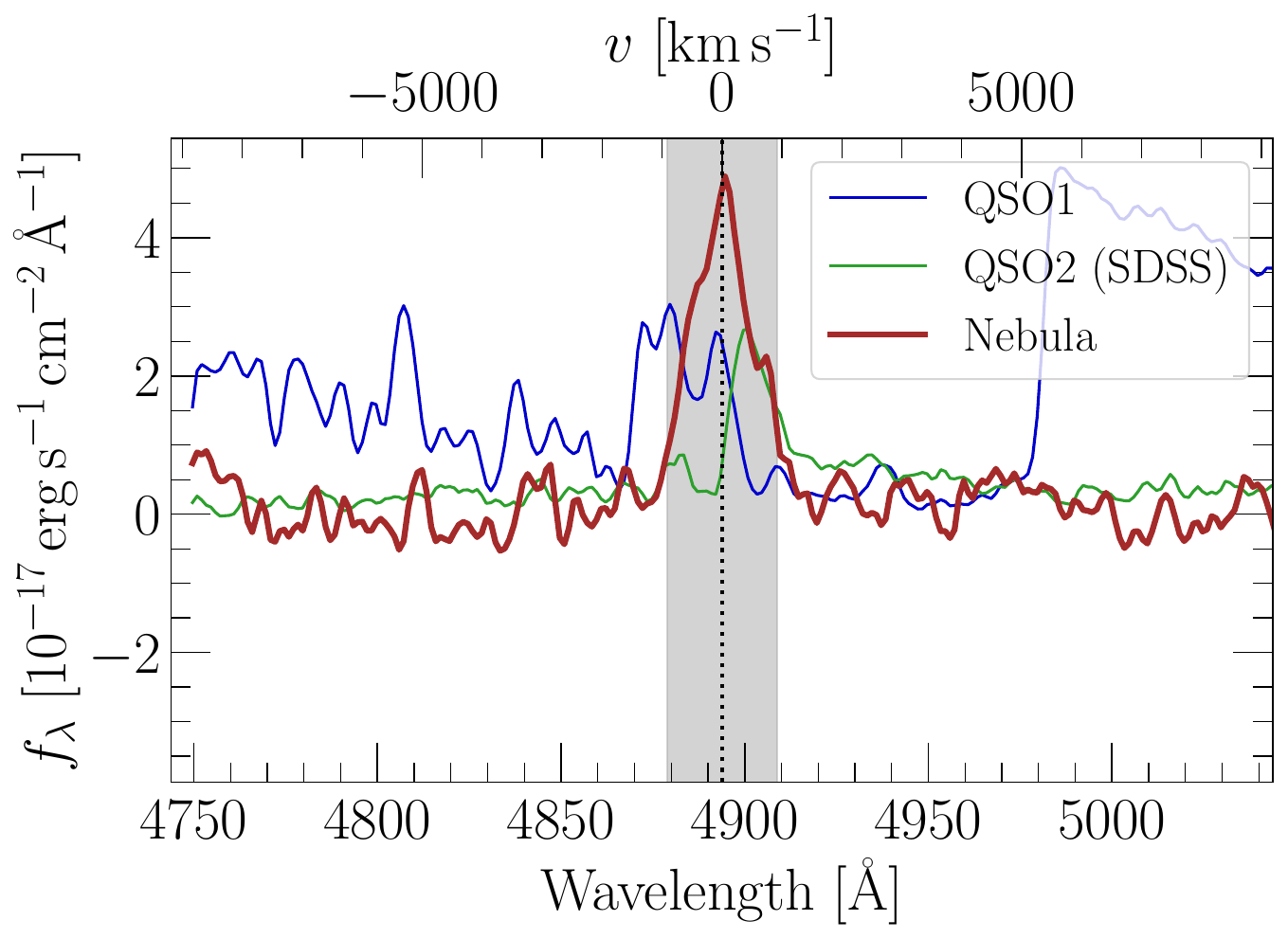}
   \includegraphics[height=5cm]{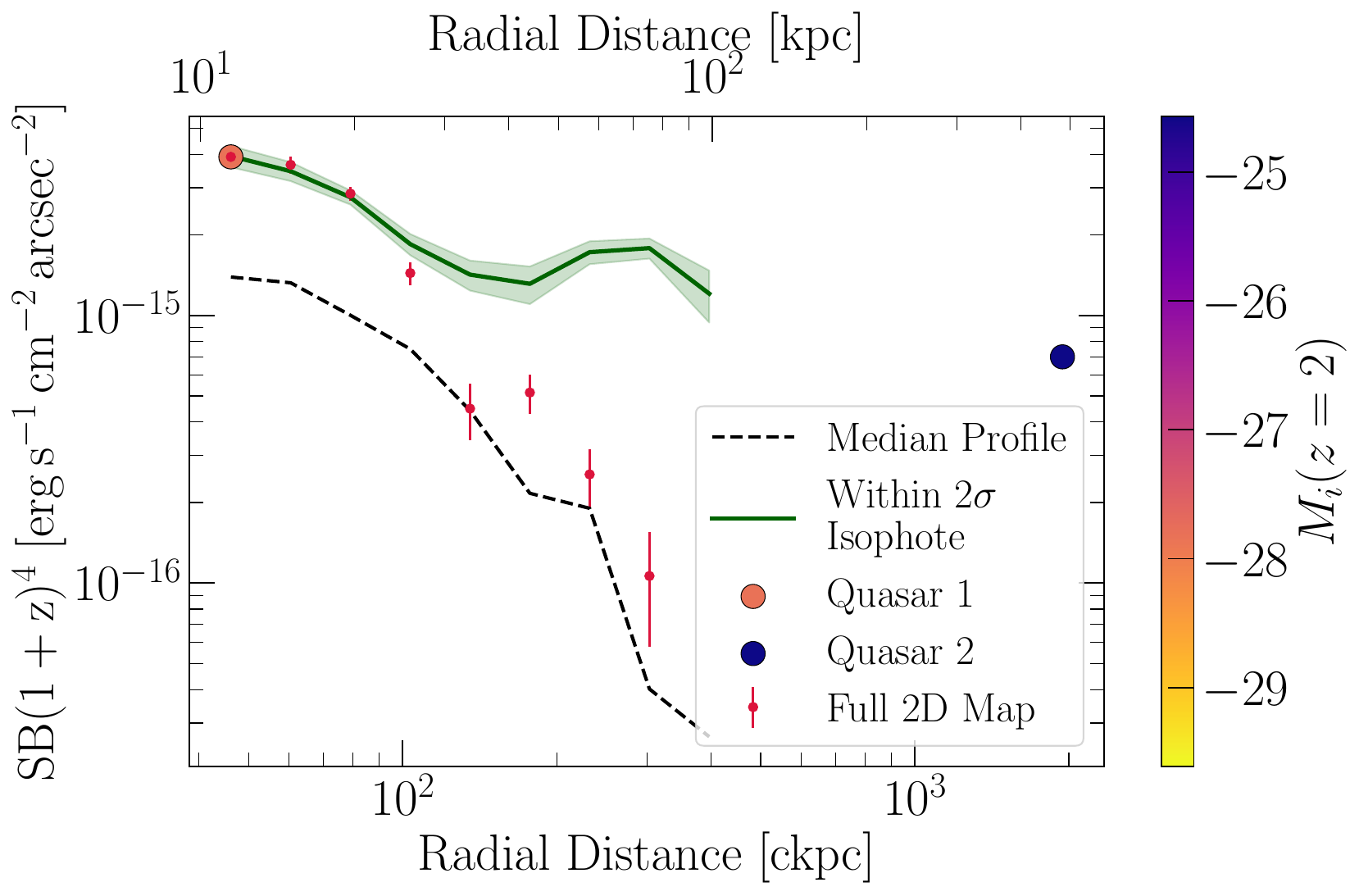}
      \caption{Same as Fig.~\ref{fig:qso1} but for quasar pair 4 (ID 4.1). Quasar 4.2 is not within the MUSE FoV, but the direction toward it is indicated with a green arrow in the Ly$\alpha$ surface brightness map. The white arrow indicates the line corresponding to ${d_{\rm QSO,max}}$ in this object.}
         \label{fig:qso4}
   \end{figure*}

\begin{figure*}
   \centering
   \includegraphics[height=4cm]{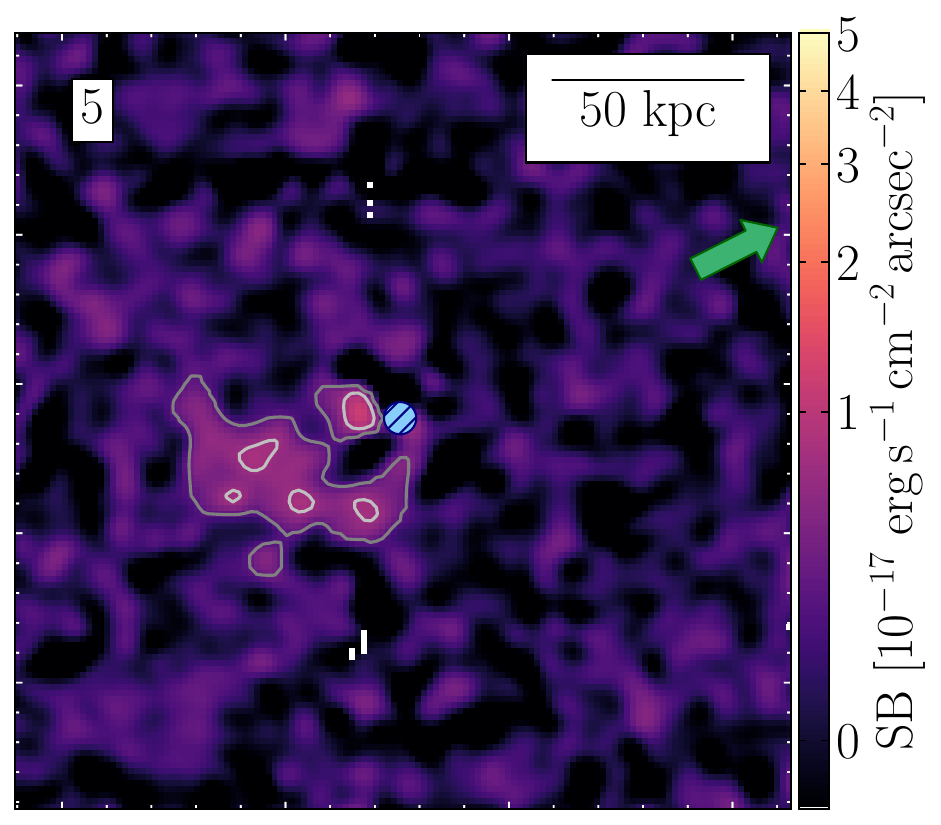}
   \includegraphics[height=4cm]{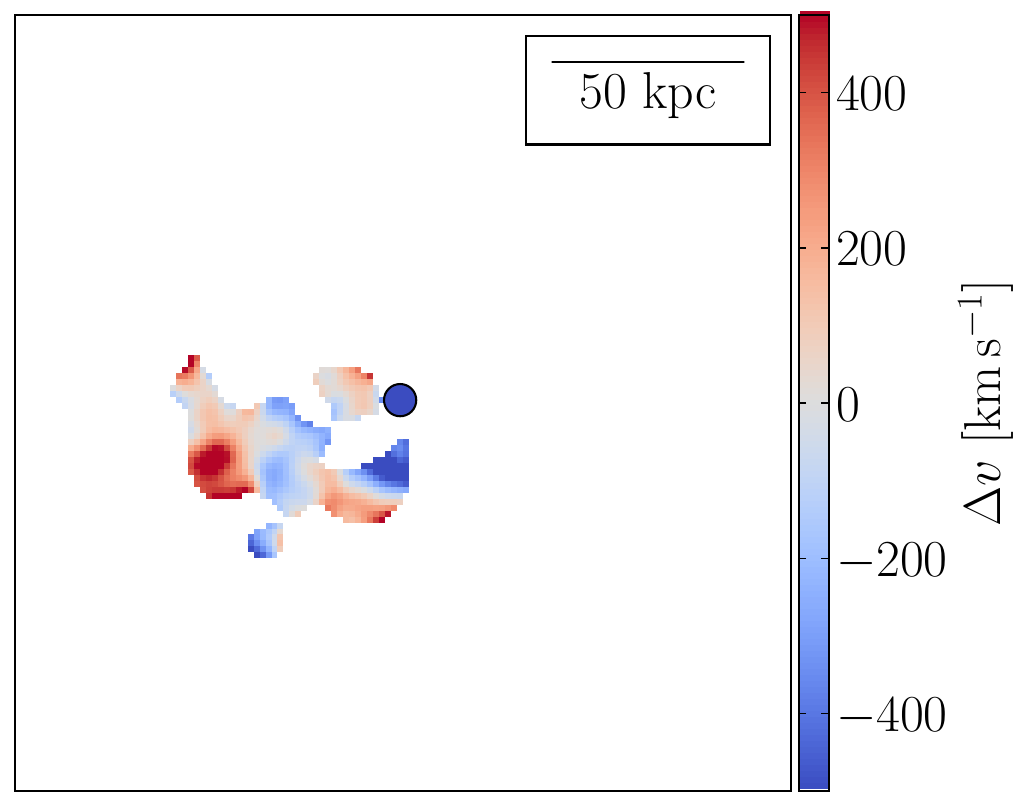}
   \includegraphics[height=4cm]{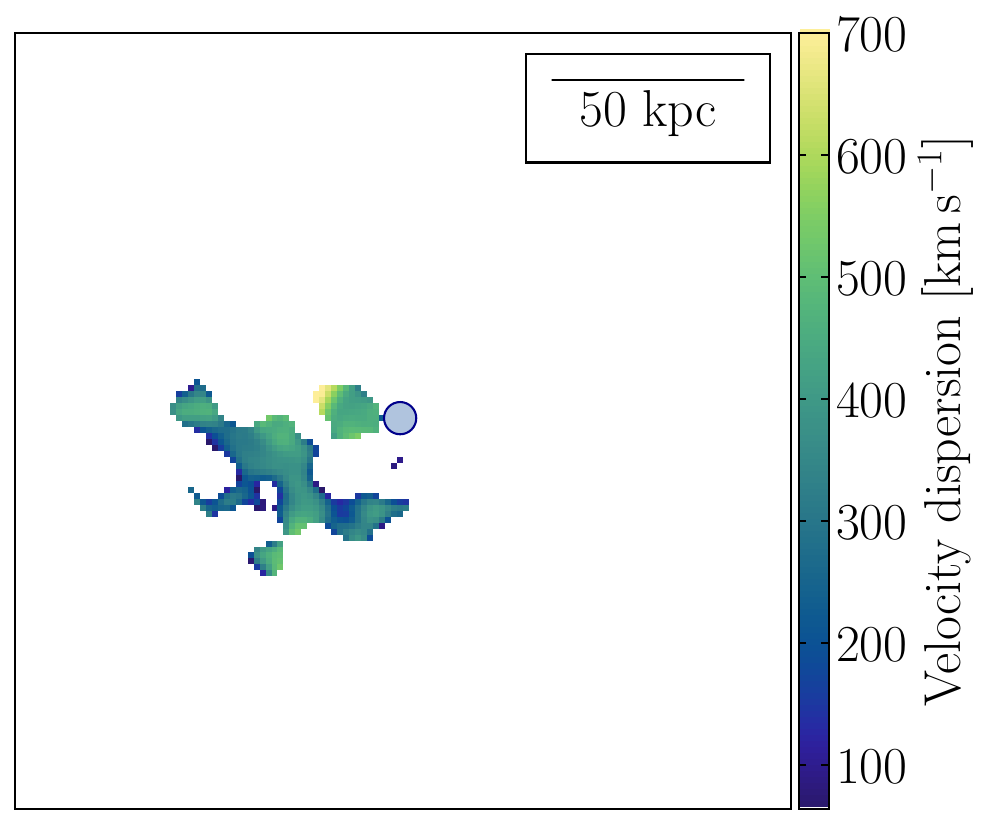}
   \includegraphics[height=5cm]{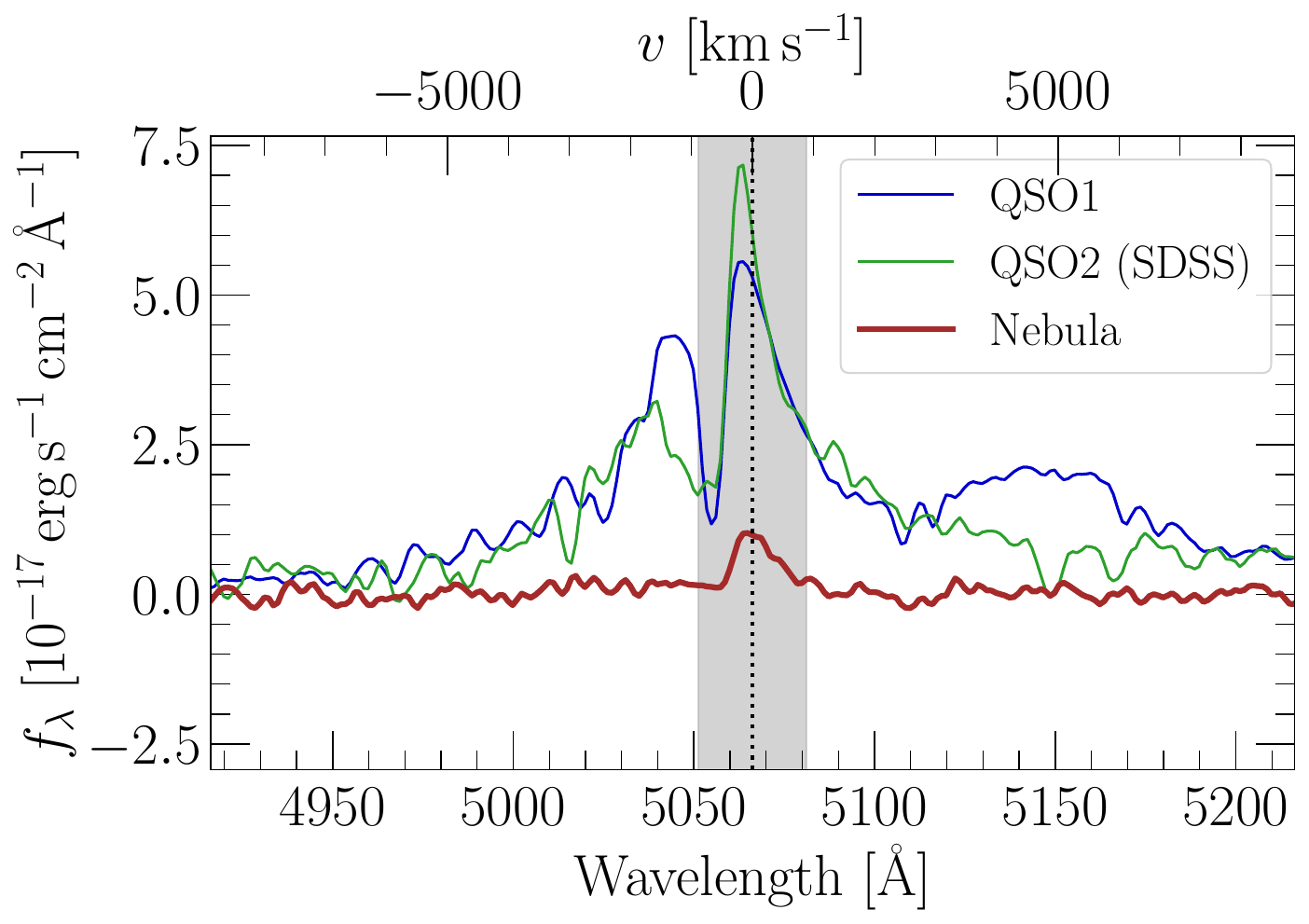}
   \includegraphics[height=5cm]{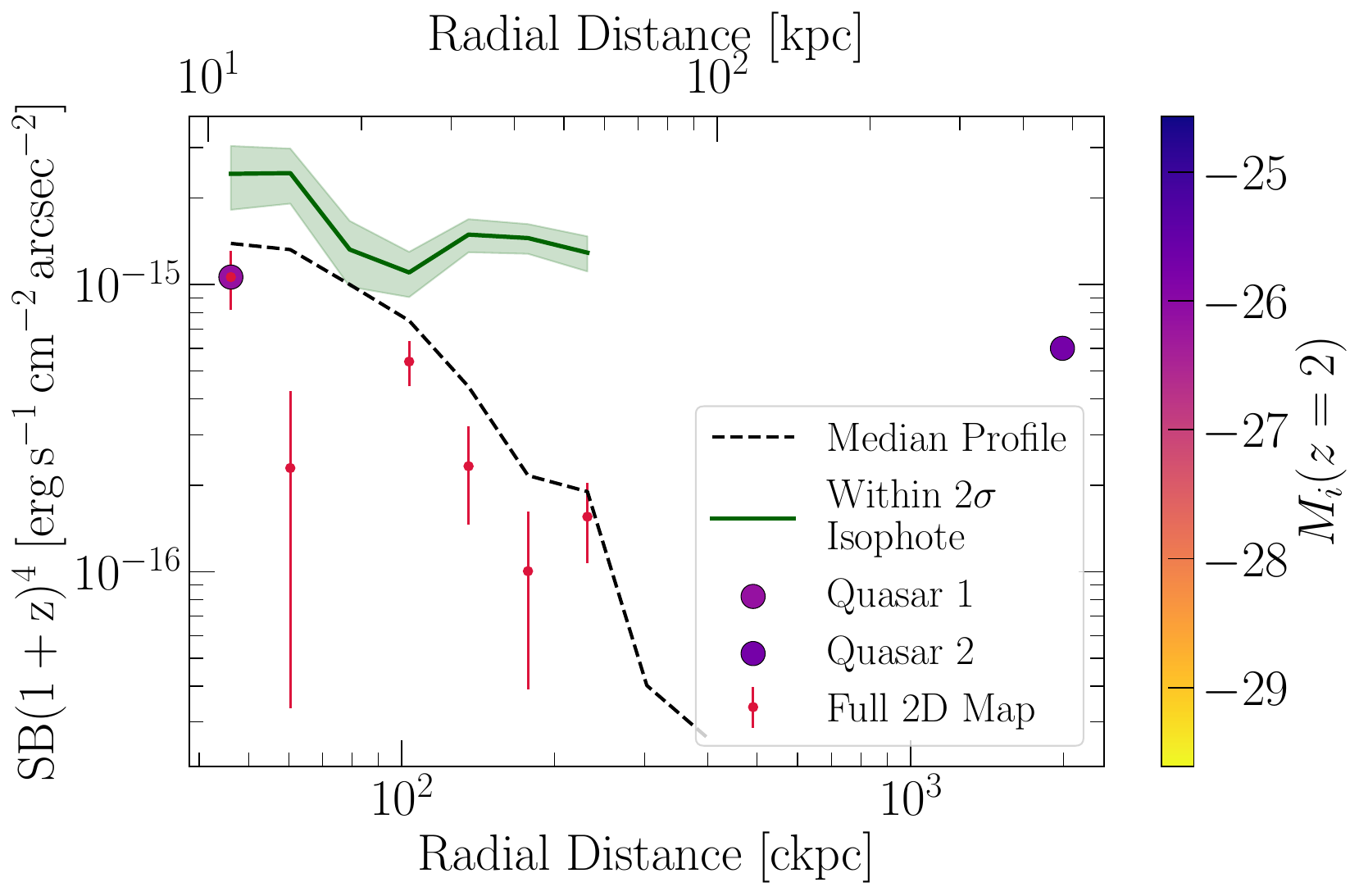}
      \caption{Same as Fig.~\ref{fig:qso1} but for quasar pair 5 (ID 5.1). Quasar 5.2 is not within the MUSE FoV, but the direction toward it is indicated with a green arrow in the Ly$\alpha$ surface brightness map.}
         \label{fig:qso5}
   \end{figure*}

\begin{figure*}
   \centering
   \includegraphics[height=4cm]{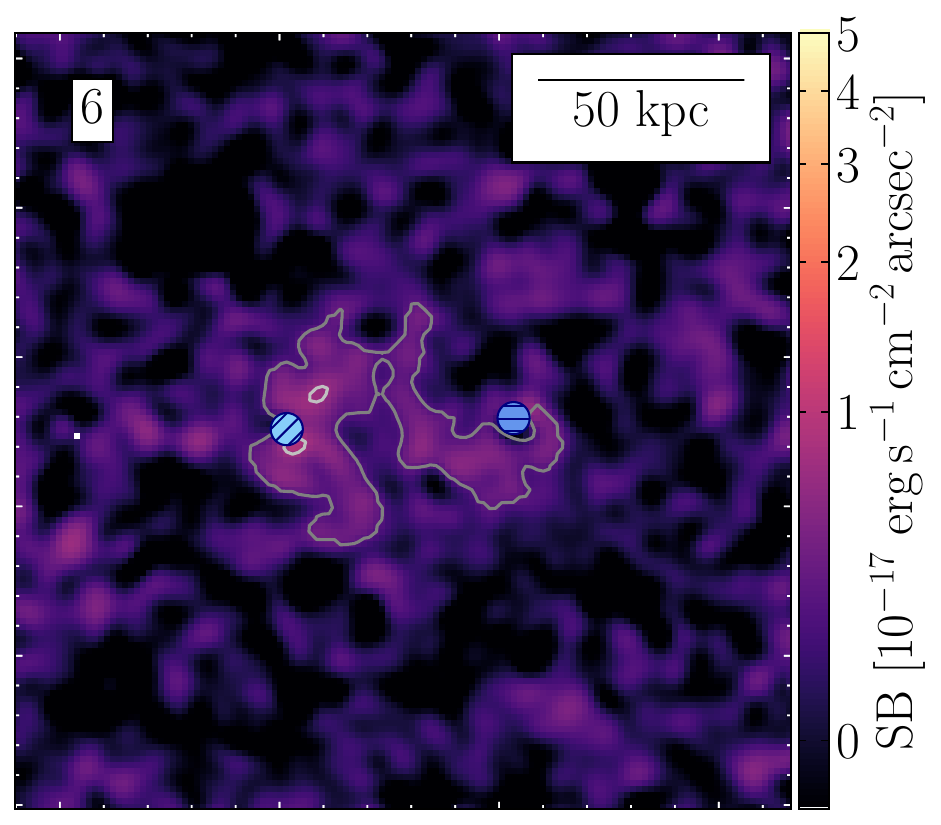}
   \includegraphics[height=4cm]{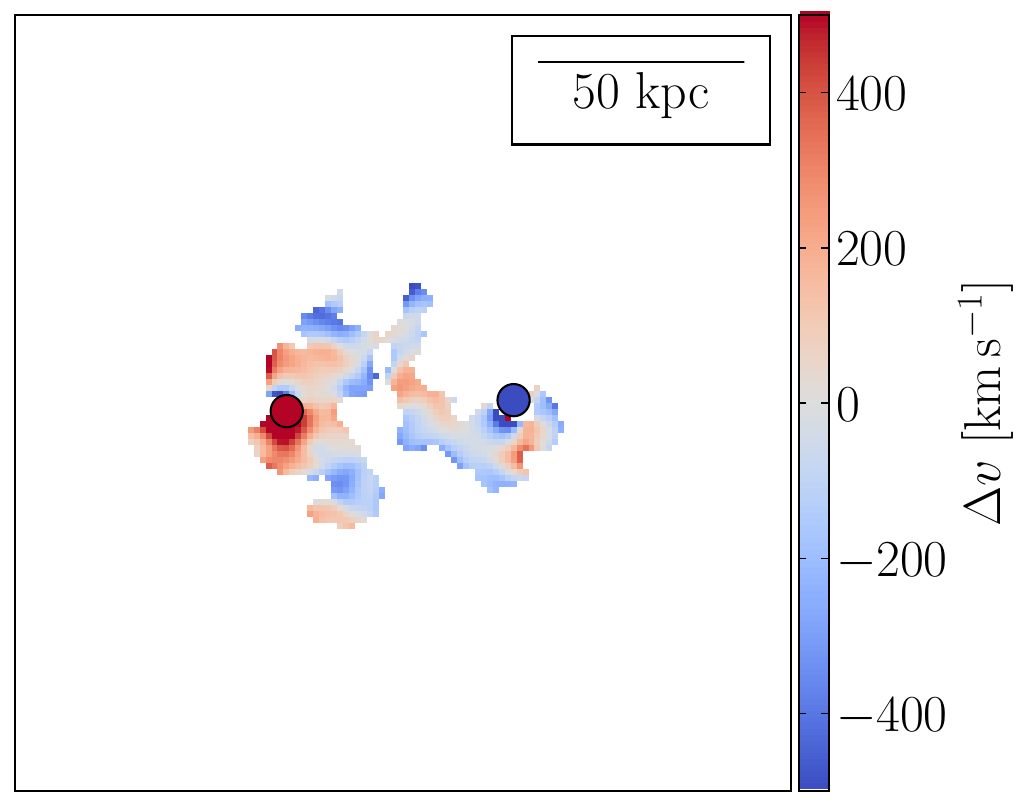}
   \includegraphics[height=4cm]{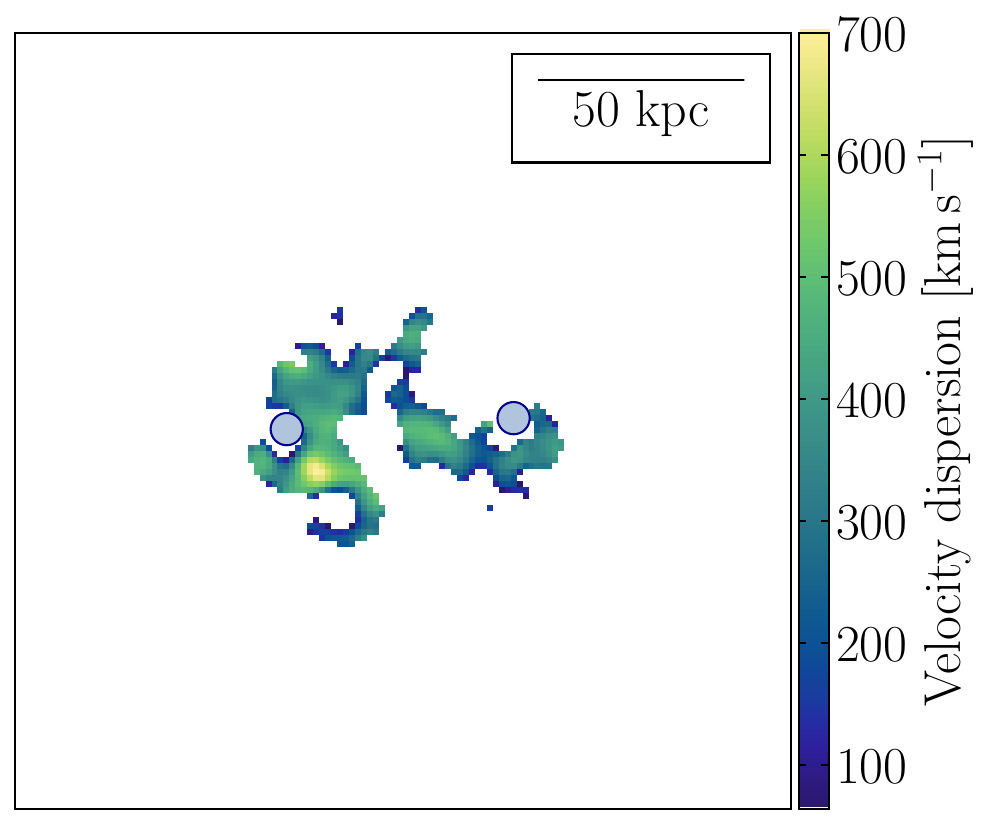}
   \includegraphics[height=5cm]{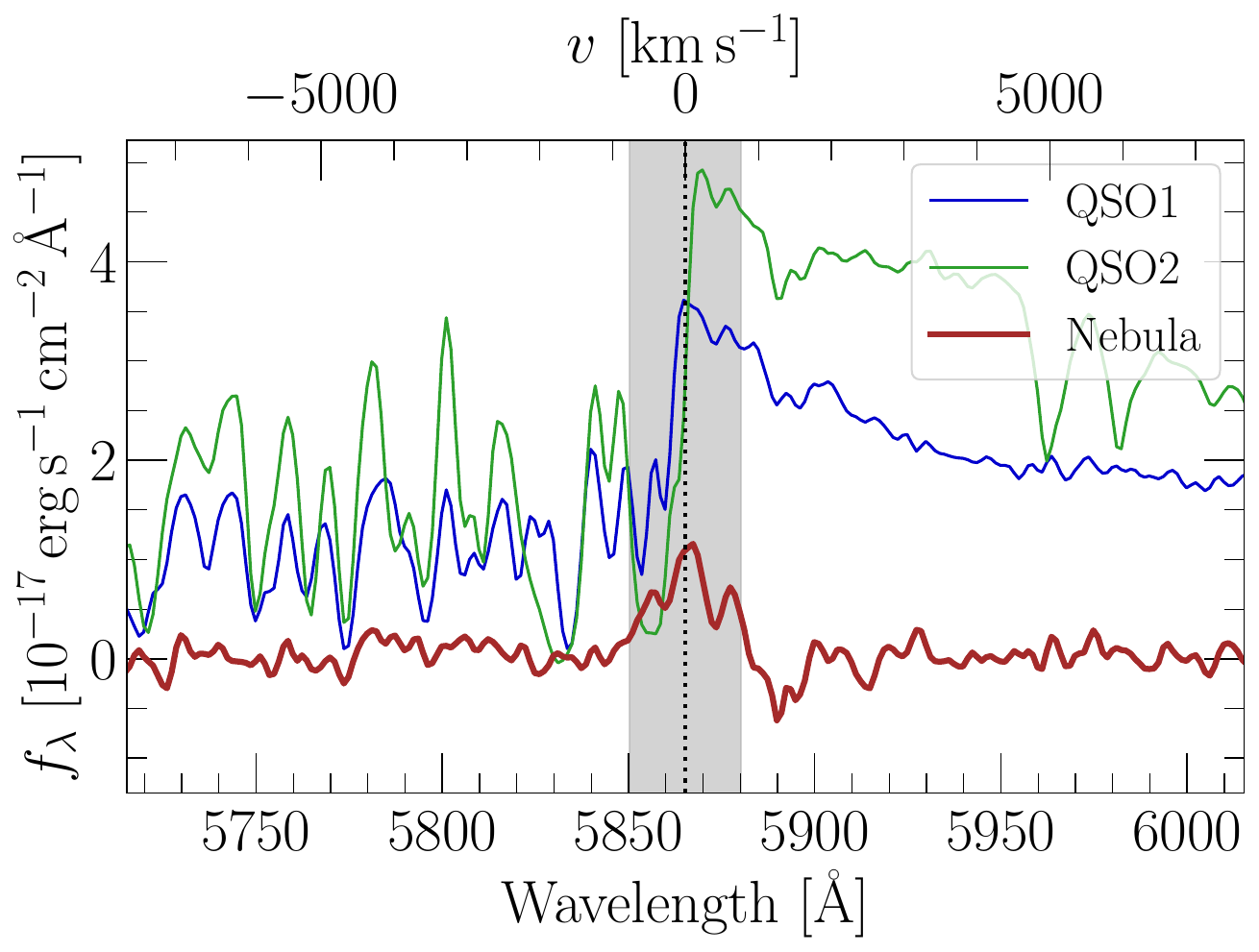}
   \includegraphics[height=5cm]{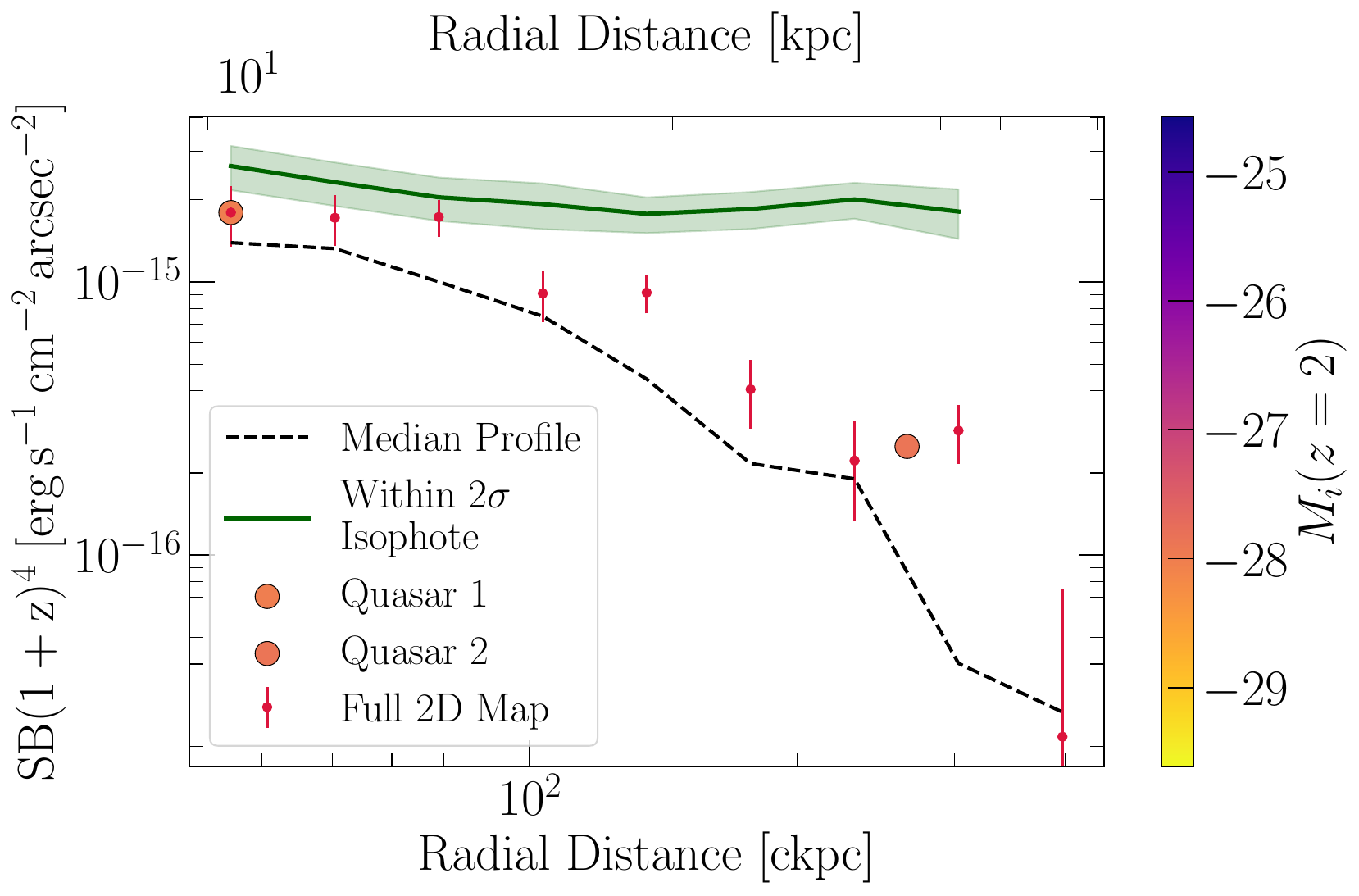}
      \caption{Same as Fig.~\ref{fig:qso1} but for quasar pair 6 (ID 6.1 and ID 6.2).}
         \label{fig:qso6}
   \end{figure*}

\begin{figure*}
   \centering
   \includegraphics[height=4cm]{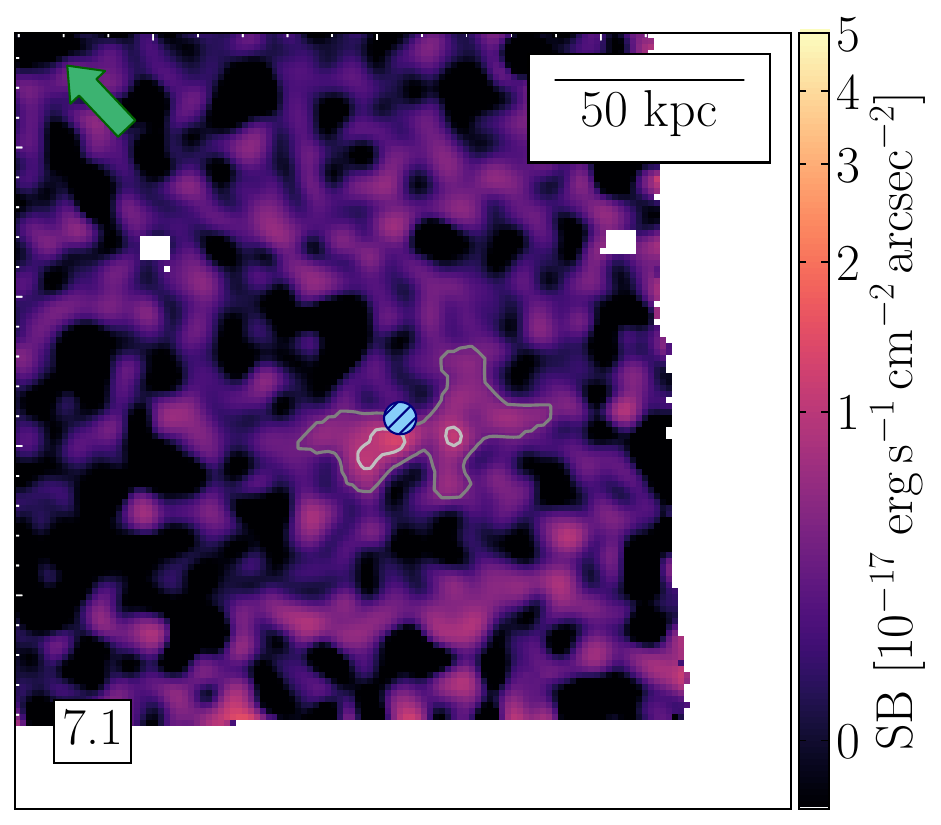}
   \includegraphics[height=4cm]{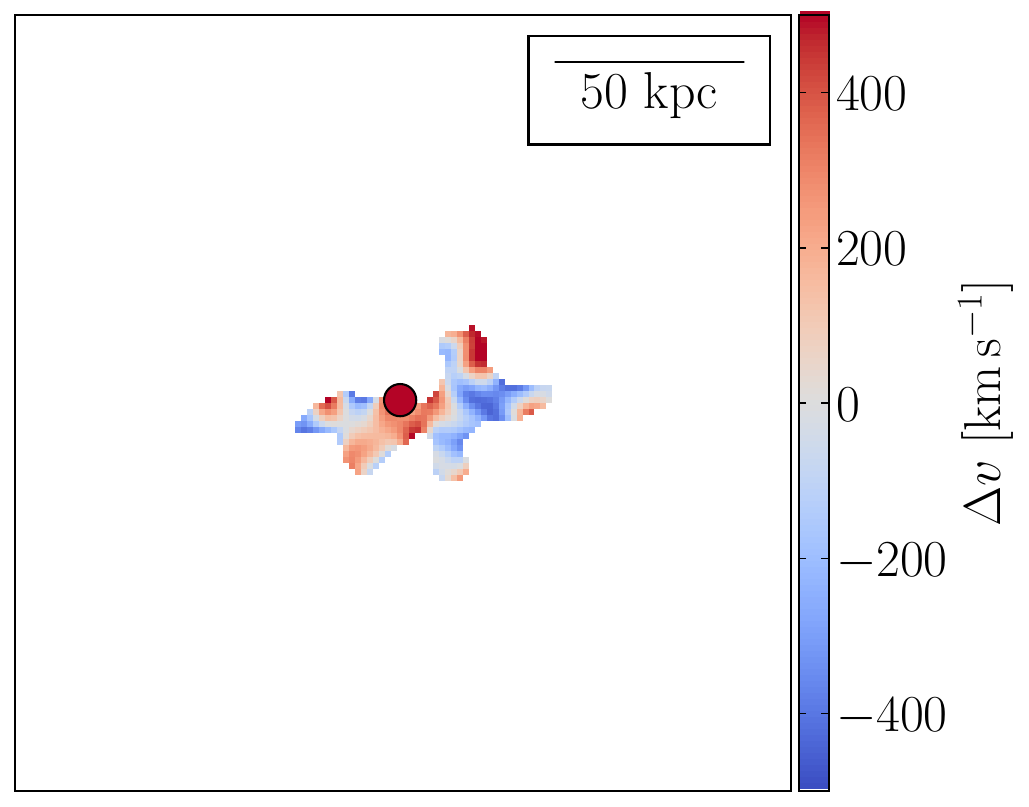}
   \includegraphics[height=4cm]{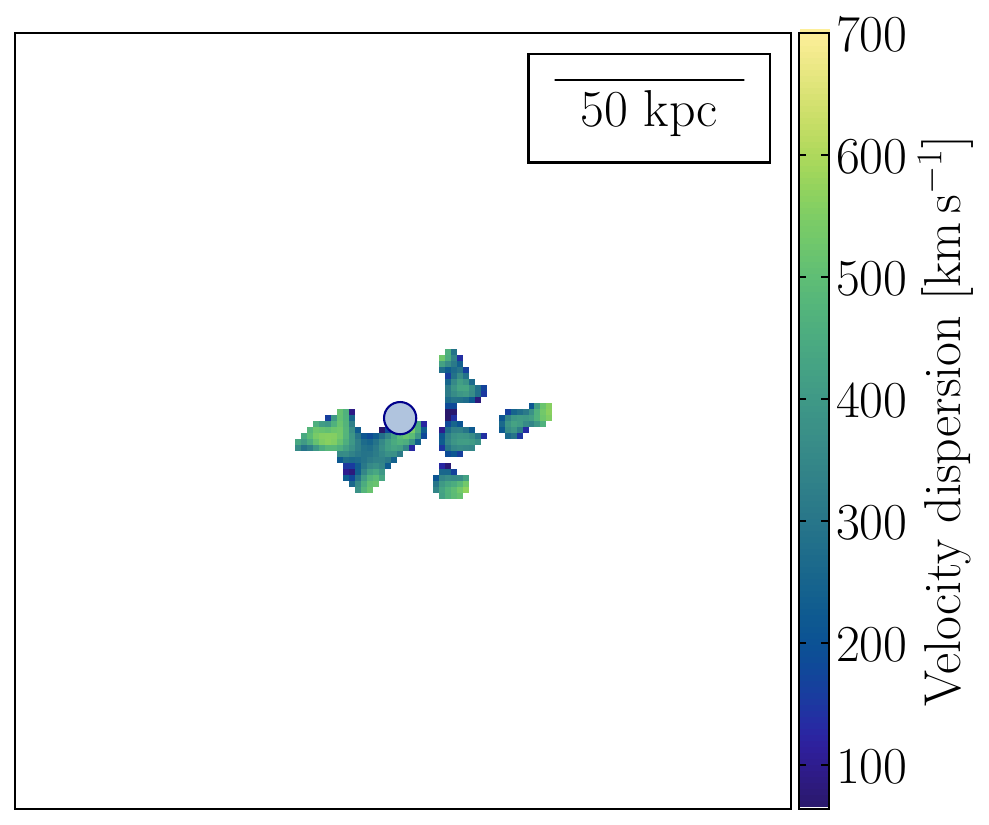}
   \includegraphics[height=4cm]{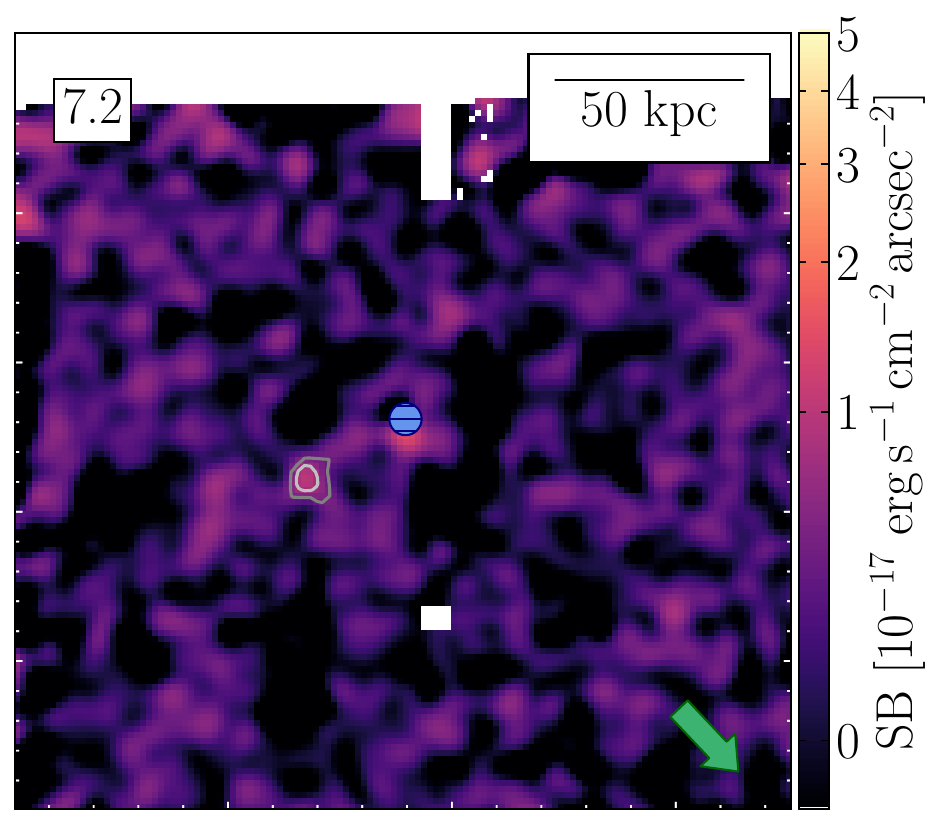}
   \includegraphics[height=4cm]{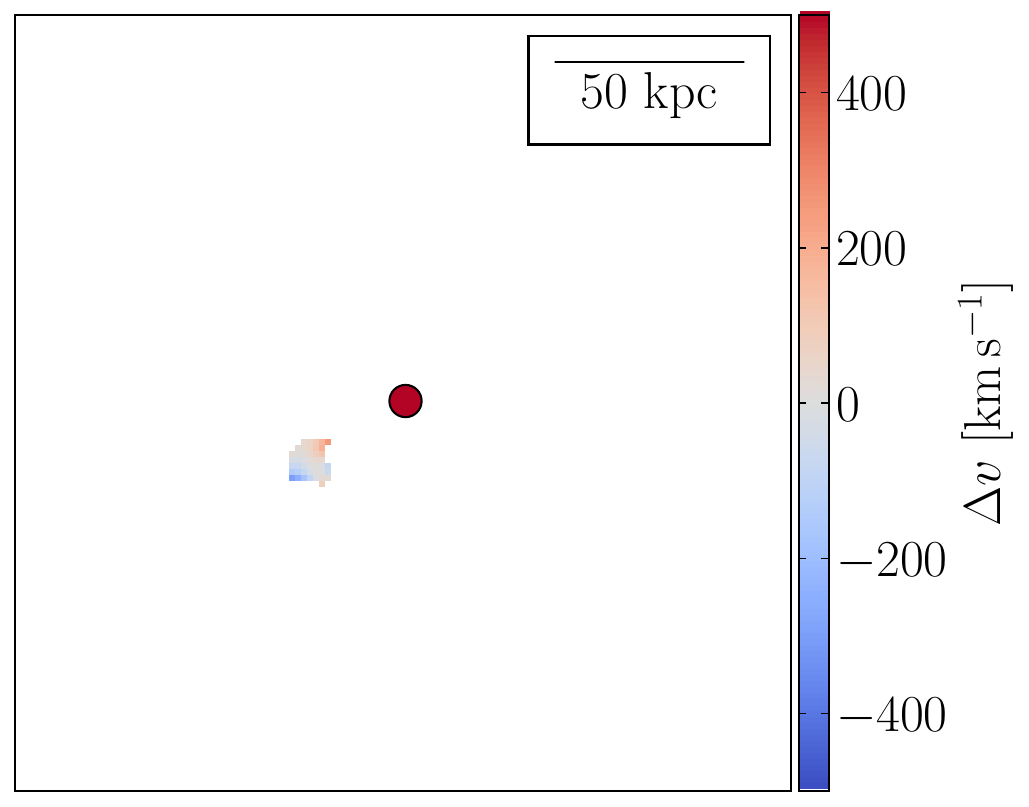}
   \includegraphics[height=4cm]{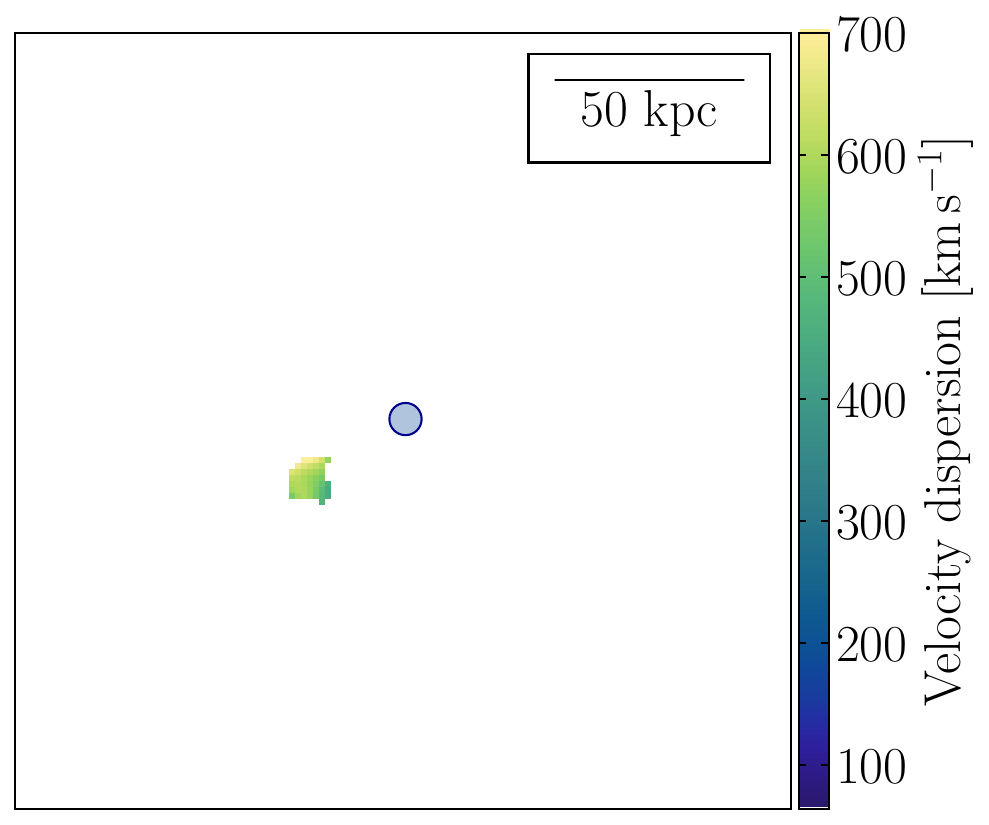}
   \includegraphics[height=5cm]{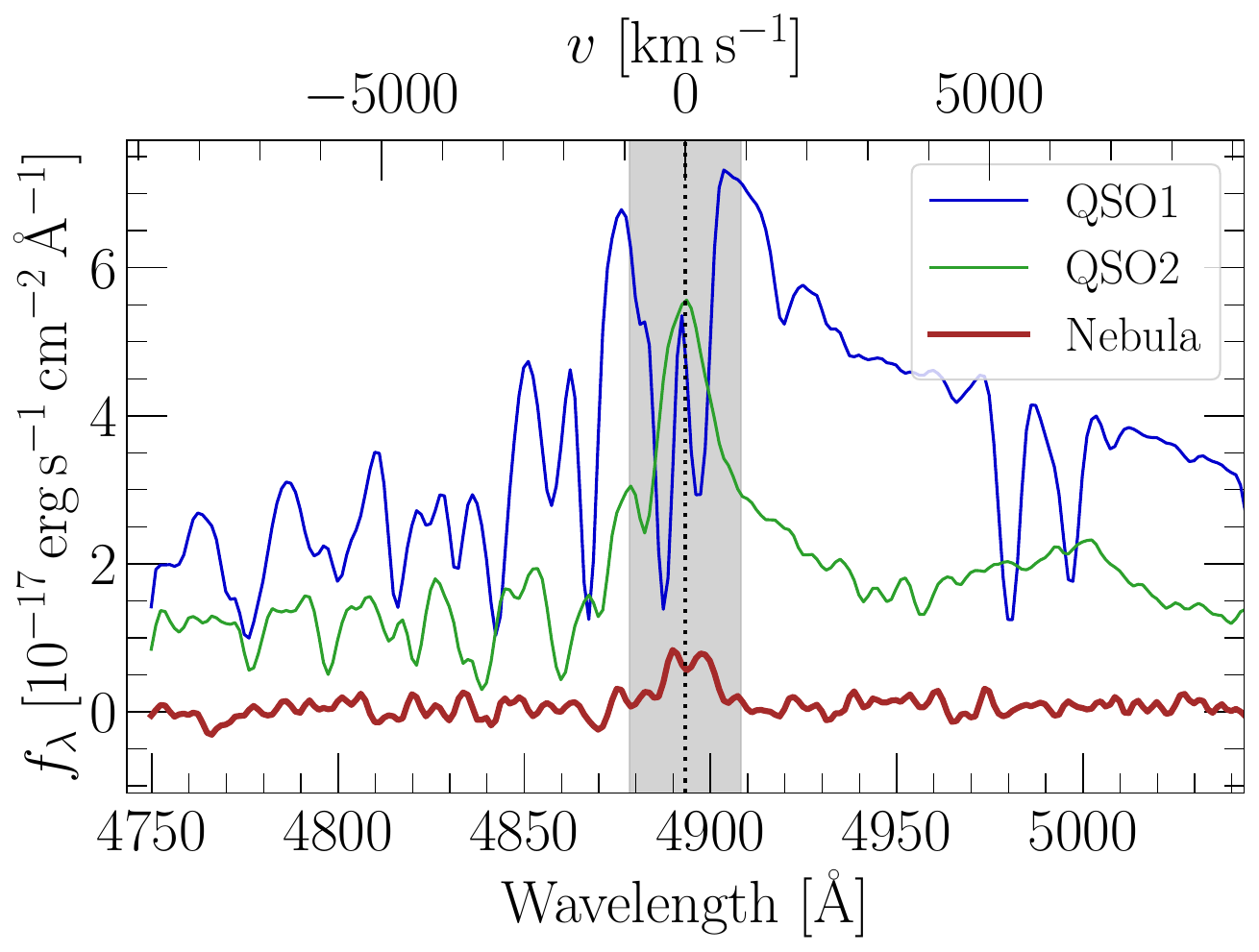}
   \includegraphics[height=5cm]{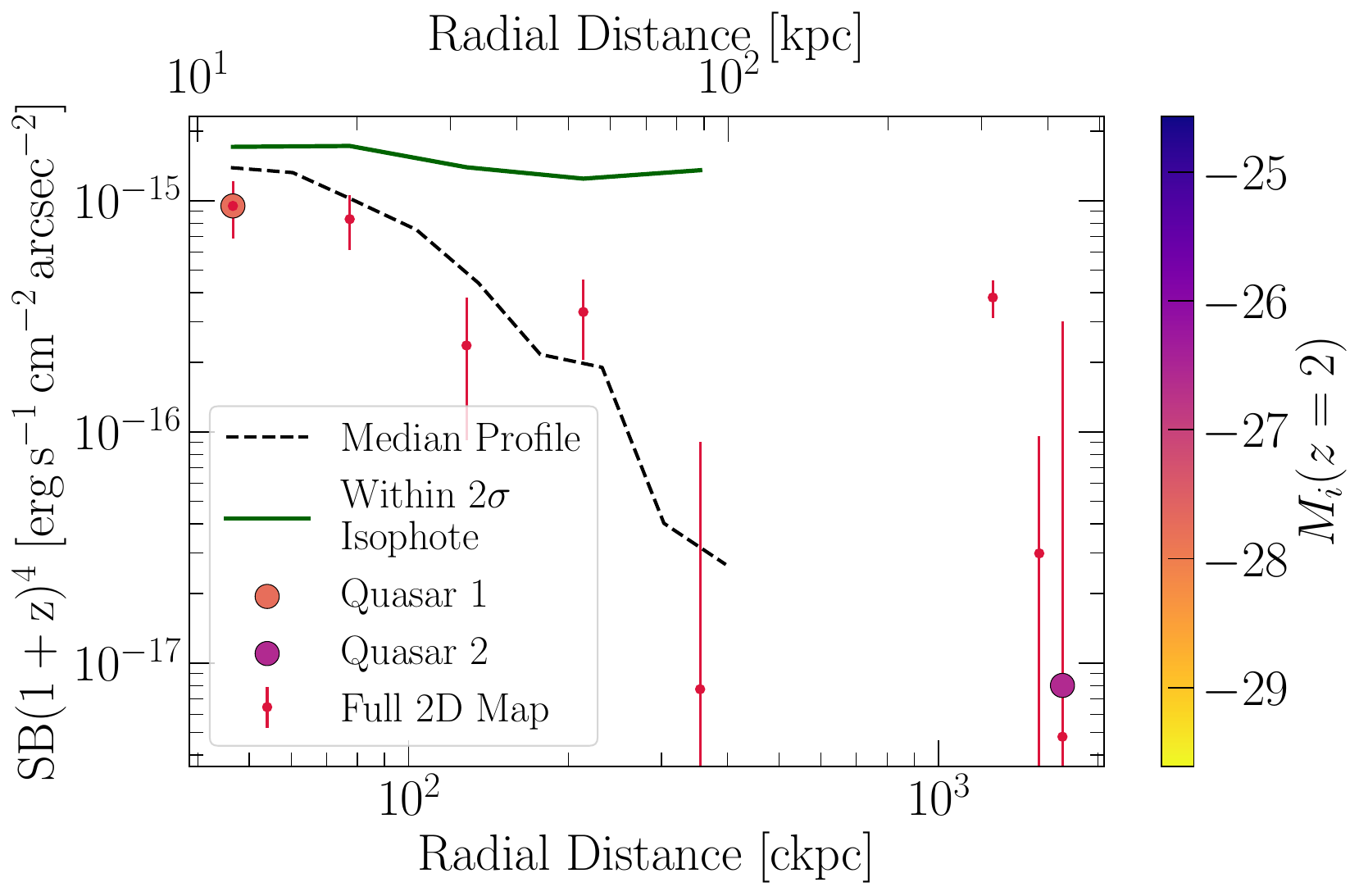}
      \caption{Same as Fig.~\ref{fig:qso1} but for quasar pair 7. As this is a wide pair, maps are presented for each quasar individually. \textbf{Top row:} Ly$\alpha$ surface brightness map, line velocity and velocity dispersion for quasar 7.1. \textbf{Middle row:} Ly$\alpha$ surface brightness map, line velocity and velocity dispersion for quasar 7.2.}
         \label{fig:qso7}
   \end{figure*}

\begin{figure*}
   \centering
   \includegraphics[height=4cm]{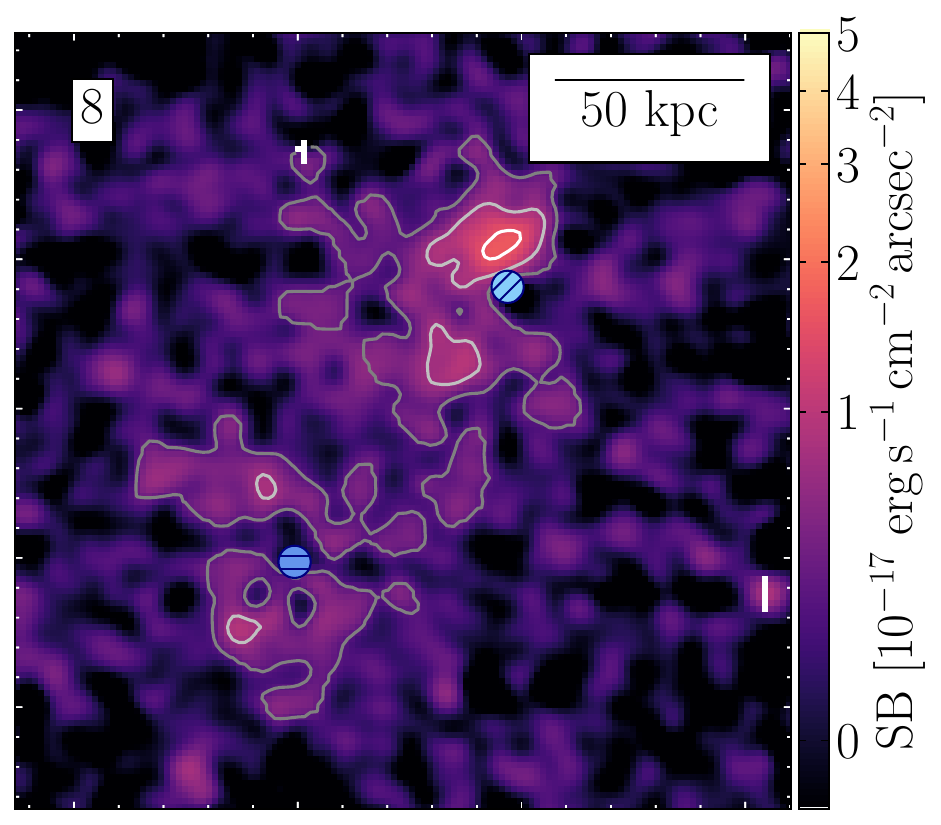}
   \includegraphics[height=4cm]{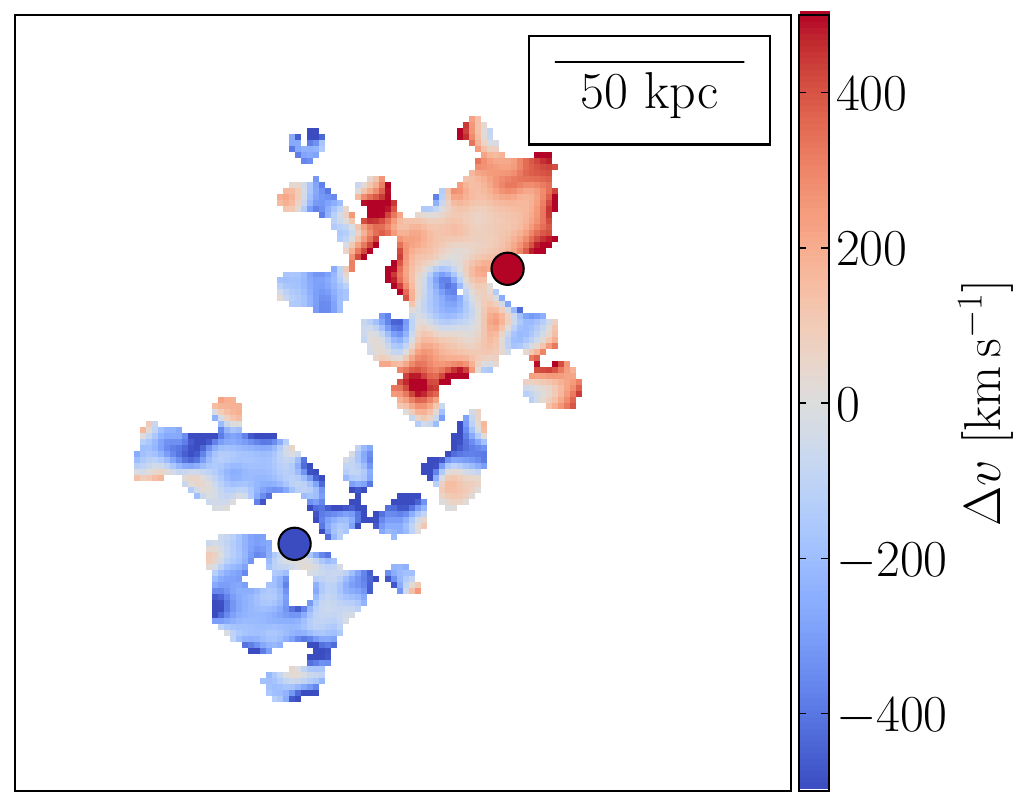}
   \includegraphics[height=4cm]{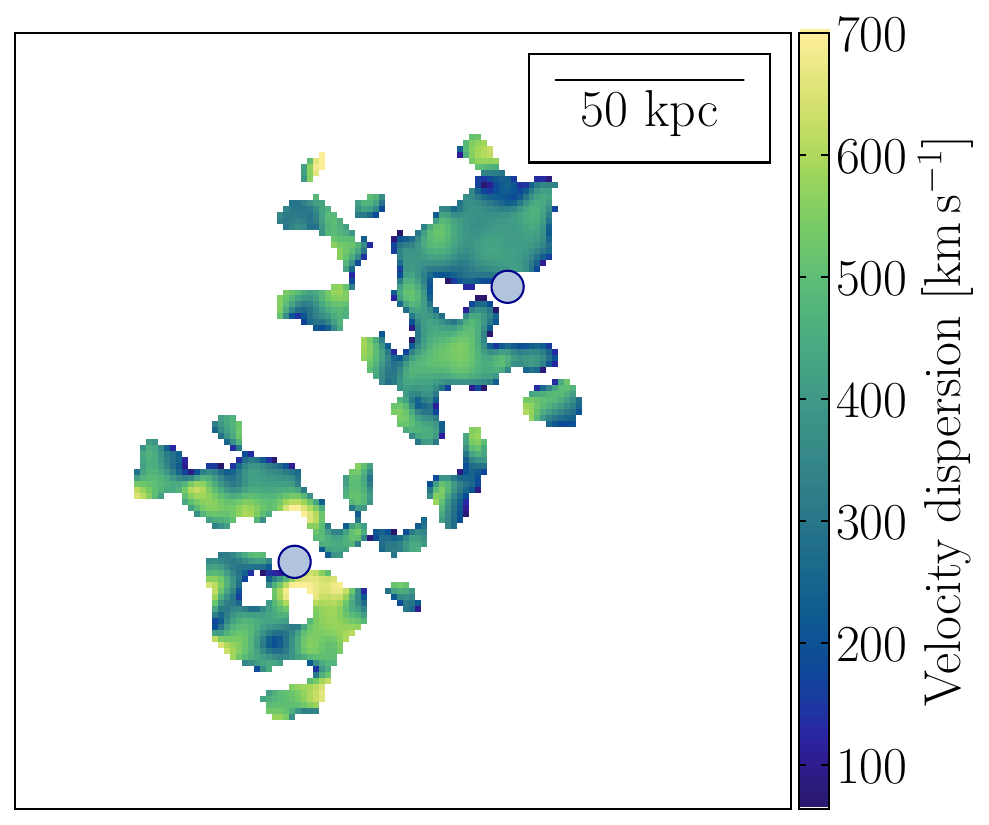}
   \includegraphics[height=5cm]{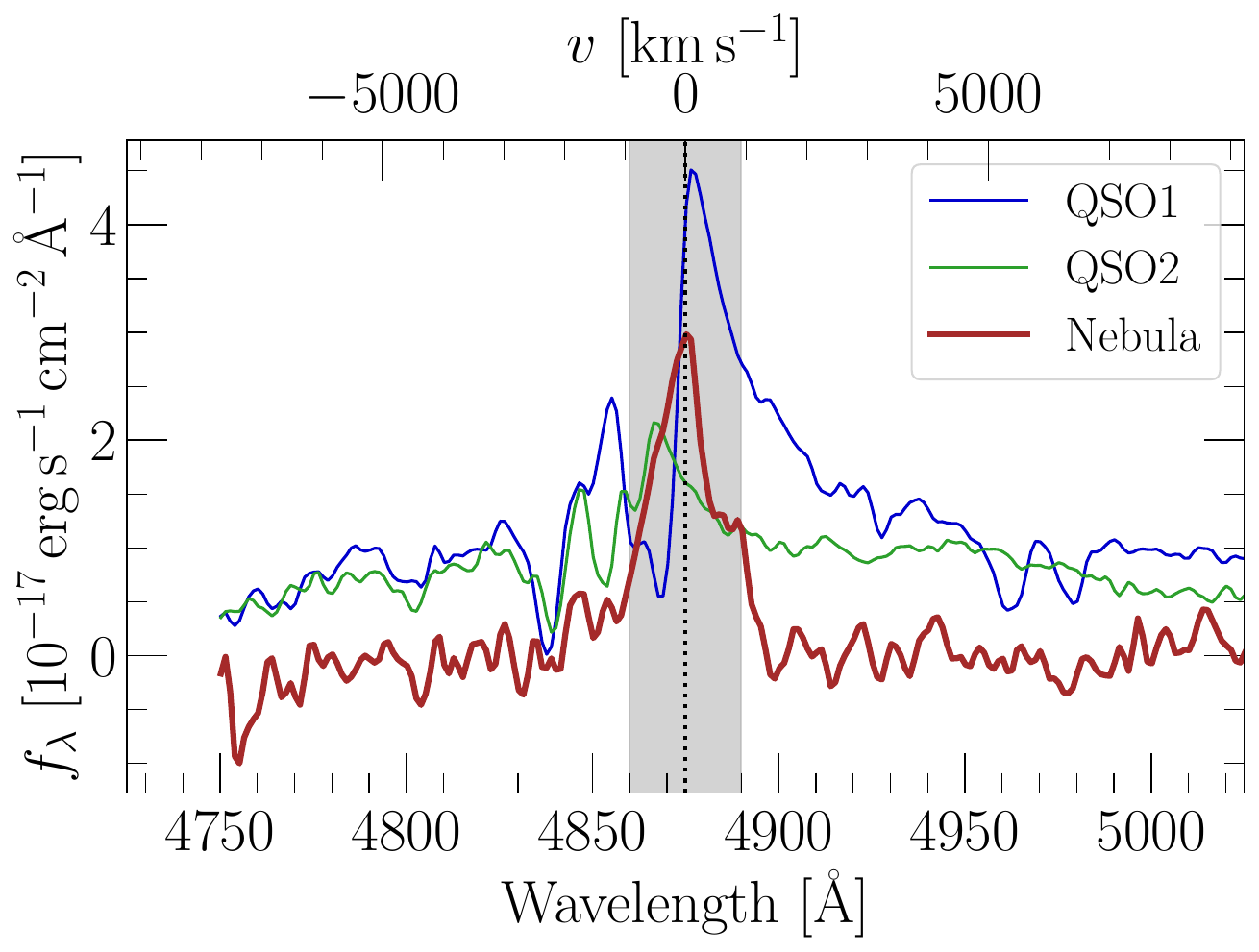}
   \includegraphics[height=5cm]{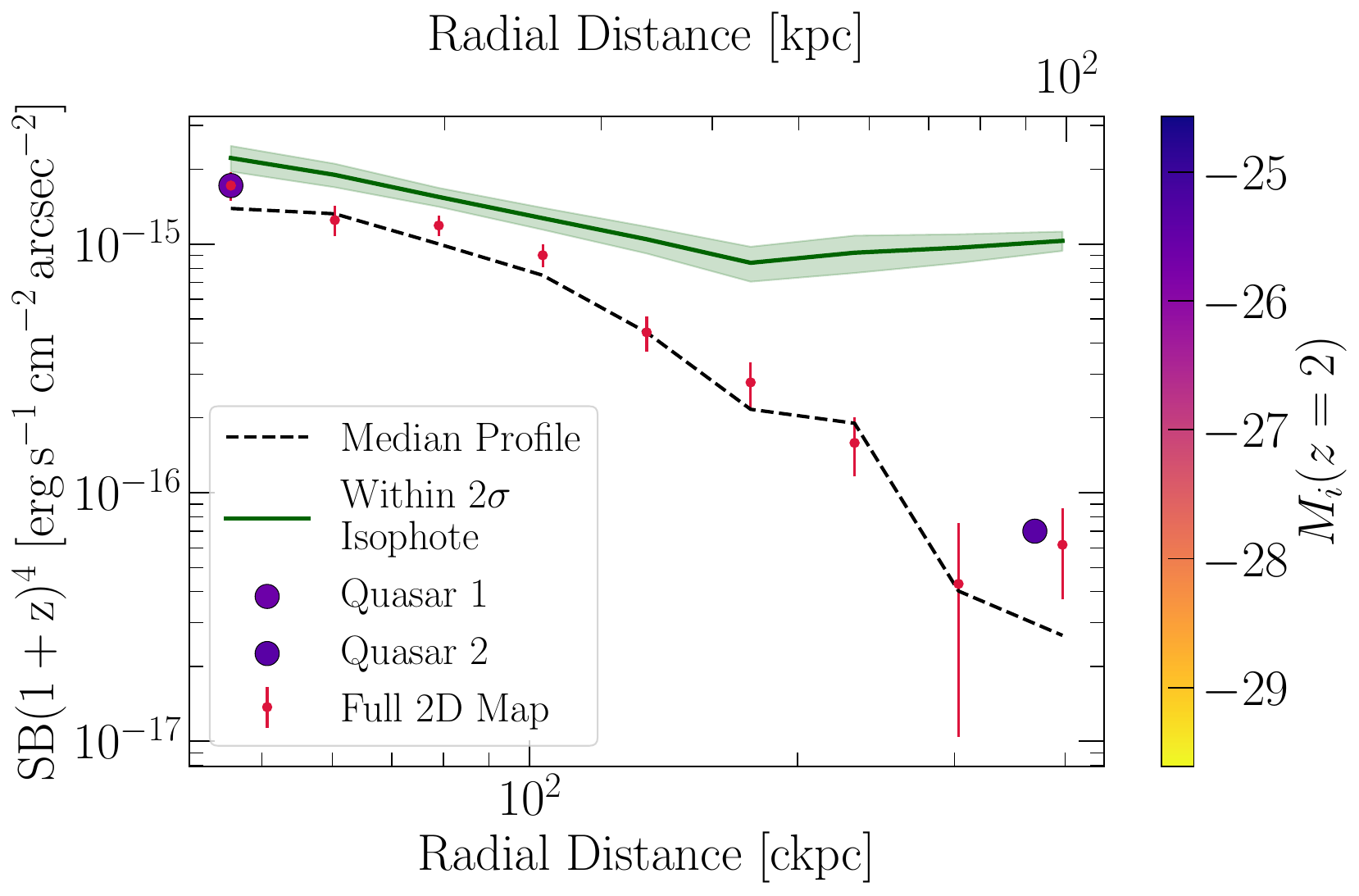}
      \caption{Same as Fig.~\ref{fig:qso1} but for quasar pair 8 (ID 8.1 and ID 8.2).}
         \label{fig:qso8}
   \end{figure*}

In this appendix, we summarize important information about the MUSE observations and the quasar properties in Table~\ref{tab:1} and show in Figs. \ref{fig:qso2} to \ref{fig:qso8}   Ly$\alpha$ maps and profiles for the remaining seven quasar pairs not shown in Fig.~\ref{fig:qso1}.

\begin{table*}[b]
\caption{Observations log and quasars properties.}             
\label{tab:1}      
\centering                          
\footnotesize
\begin{tabular}{c c c c c c c c c c}        
\hline\hline                 
ID & RA (J2000) & Dec (J2000) & Observation Date & Seeing \tablefootmark{a} & Weather \tablefootmark{b} & $i$ [mag] \tablefootmark{c} & $M_i(z=2)$ \tablefootmark{d} & $d$ [kpc/$\arcsec$] \tablefootmark{e} & $z$ \tablefootmark{f}\\    
\hline                        
1.1 & 00:01:40.6000 & 07:09:53.9999 & 20/11/2017 & 1.17$\arcsec$ & clear & 18.06 & -29.61 & 47.7/ 6.2 & 3.234\\
1.2 & 00:01:40.5984     & 07:09:47.8149 & & & & 22.75 & -24.93 & & 3.238\\ \hline
2.1 & 00:18:07.3685 & 16:12:57.5711 & 20/11/2017 & 1.59$\arcsec$ & clear &  21.31 & -26.26 & 97.2/ 12.5 & 3.138\\
2.2 & 00:18:08.0959 & 16:12:50.8420 & & & & 22.13 & -25.42 & & 3.126\\ \hline
3.1 & 02:40:05.2332 &-00:39:09.8433 & 15/12/2017 & 1.45$\arcsec$ & thin &  22.38 & -25.16 & 68.6/ 8.8 & 3.116\\
3.2 & 02:40:05.7477     &-00:39:14.0616 & & & & 22.39 & -25.16 & & 3.120\\\hline
4.1 & 02:44:42.6004     &-00:23:20.4000 & 20/12/2017 & 1.02$\arcsec$ & thin &  19.63 & -27.83 & 482.0/ 61.4 & 3.044\\
4.2 & 02:44:41.6638     &-00:24:20.1600 & --- & & &  22.54 \tablefootmark{g} & -24.57 & & 3.038\\ \hline
5.1 & 10:12:54.7449 & 03:35:48.8400     & 14/03/2018 & 0.88$\arcsec$ & thin & 21.45 & -26.15 & 477.6/ 61.6 & 3.162\\
5.2 & 10:12:51.0718 & 03:36:16.5598 & --- & & &  21.49 \tablefootmark{g} & -25.73 & & 3.168\\ \hline
6.1 & 10:21:16.4685 & 11:12:27.9389 & 14/03/2018 & 0.77$\arcsec$ & clear &  20.43 & -27.75 & 55.0/ 7.6 & 3.829\\
6.2 & 10:21:16.9849     & 11:12:27.5716 & & & & 20.28 & -28.01 & & 3.815\\ \hline
7.1 & 10:40:46.4499 & 00:59:50.9200 & 21/03/2018 & 0.87$\arcsec$ & photometric & 19.71 & -27.75 & 455.6/ 58.0 & 3.044\\
7.2 & 10:40:49.1050 & 01:00:33.0900 & & & & 20.85 & -26.59 & & 3.026\\ \hline
8.1 & 11:35:02.0325 &-02:21:10.9311 & 19/02/2018 & 1.50$\arcsec$ & clear & 21.83 & -25.60 & 92.1/ 11.7 & 3.019\\
8.2 & 11:35:02.5085 &-02:21:20.1432 & & & & 22.06 & -25.36 & & 3.009\\ \hline
\hline                                   
\end{tabular}
\tablefoot{
\tablefoottext{a}{Seeing at Ly$\alpha$, determined by extrapolating a Moffat fit to the continuum of the brighter quasar (see Sect.~\ref{sec:data_analysis} for details).}
\tablefoottext{b}{Weather conditions as reported in the ESO observing log.}
\tablefoottext{c}{Calculated in the MUSE data within an aperture of 3\arcsec.}
\tablefoottext{d}{Absolute $i$-band magnitude normalized to $z=2$ following \citet{Ross13}.}
\tablefoottext{e}{Projected physical distance between the quasars.}
\tablefoottext{f}{Redshift value taken from SDSS. The intrinsic uncertainties on quasar systemic redshifts evaluated from the available broad lines in this sample are known to be $>200$~km~s$^{-1}$ (up to $\sim400$~km~s$^{-1}$ when using only CIV; \citealt{Shen2016})
}
\tablefoottext{g}{SDSS $i$-band magnitude.}}
\end{table*}

\FloatBarrier


\FloatBarrier
\newpage
\section{Example of a field before quasar PSF subtraction}
\label{sec:appPSF}

\begin{figure}[h]
   \centering
   \includegraphics[width=\hsize]{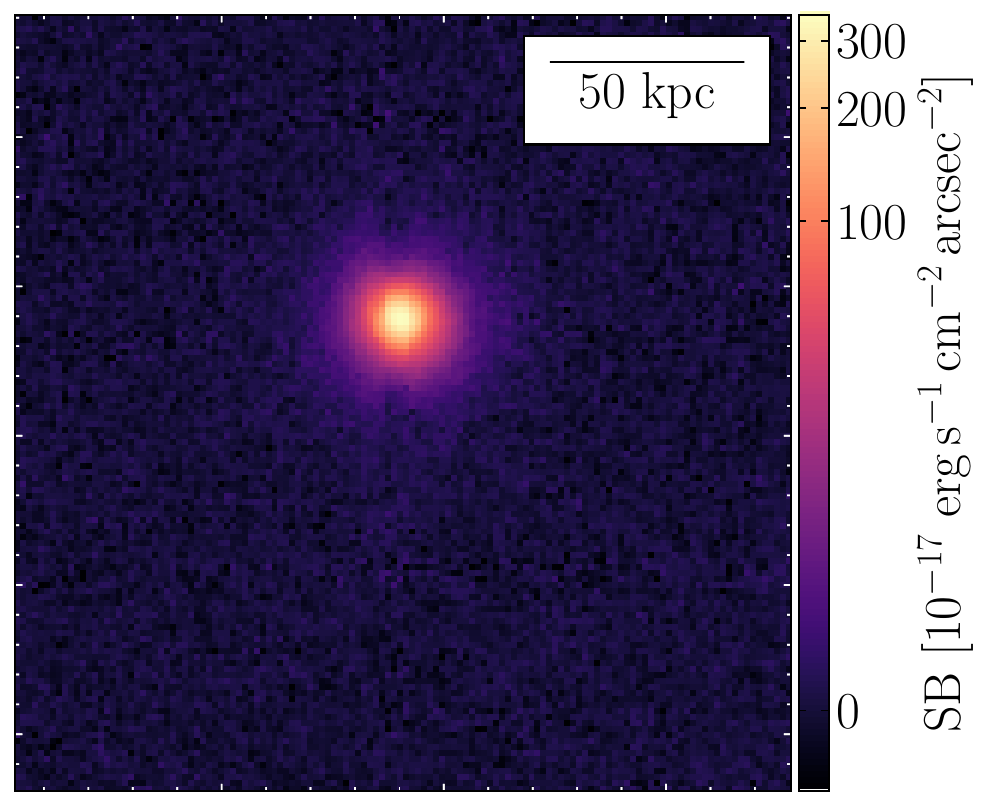}
      \caption{Surface brightness in the PNB of quasar pair 1 before PSF subtraction and spatial smoothing. Shown is the same image cutout as in Fig.~\ref{fig:qso1} but with a different color bar.}
         \label{fig:psfdemo}
   \end{figure}

As explained in Sect.~\ref{sec:data_analysis}, we subtracted the unresolved emission
from each quasar to search for the extended Ly$\alpha$ emission around
the pairs. For completeness, we show in Fig.~\ref{fig:psfdemo} one example of one field (quasar pair 1) before subtraction of the quasar PSF. As can be seen from the colorbar, the quasar PSF is much brighter than the unveiled extended emission shown on the same image cut-out in Fig.~\ref{fig:qso1}.

\begin{figure}
   \centering
   \includegraphics[width=\hsize]{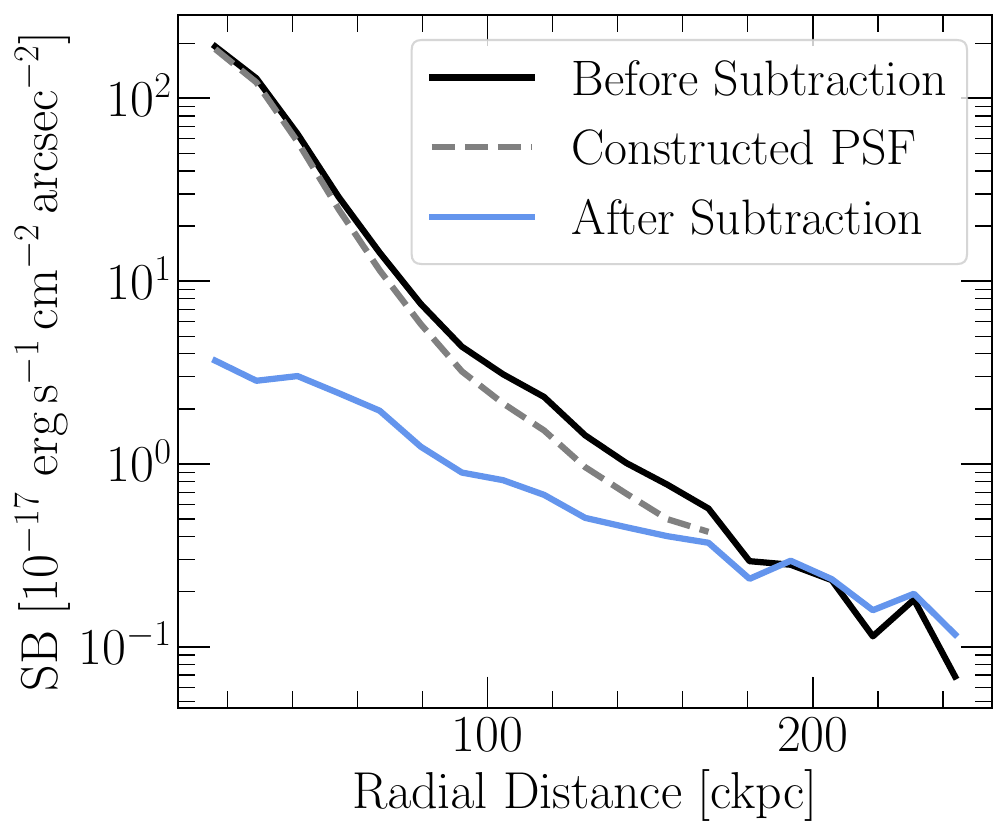}
      \caption{Surface brightness profile of quasar pair 1 determined in linearly spaced annuli in the PNB before PSF subtraction (black), after PSF subtraction (blue), and the empirical PSF constructed (gray dashed line).}
         \label{fig:psfdemoprofile}
   \end{figure}

This becomes even more apparent when plotting the surface brightness profile of the data before subtraction, the constructed PSF and the data after subtraction in linearly spaced annuli as shown in Fig.~\ref{fig:psfdemoprofile} for PNBs of quasar pair 1. At the peak of the PSF, the unveiled extended emission is roughly 1.8 orders of magnitude dimmer than the quasar emission and it only starts dominating the emission in the PNB at a distance from the quasar of roughly 200~ckpc.
We emphasize that during the analysis, we subtracted the empirical PSF channel-by-channel, and Fig.~\ref{fig:psfdemoprofile} therefore only functions as a simplified visualization of the subtraction.

\FloatBarrier

\section{Reanalyzed data of faint single quasar nebulae}
\label{sec:appMackenzie}

   \begin{figure}[h]
   \centering
   \includegraphics[width=\hsize]{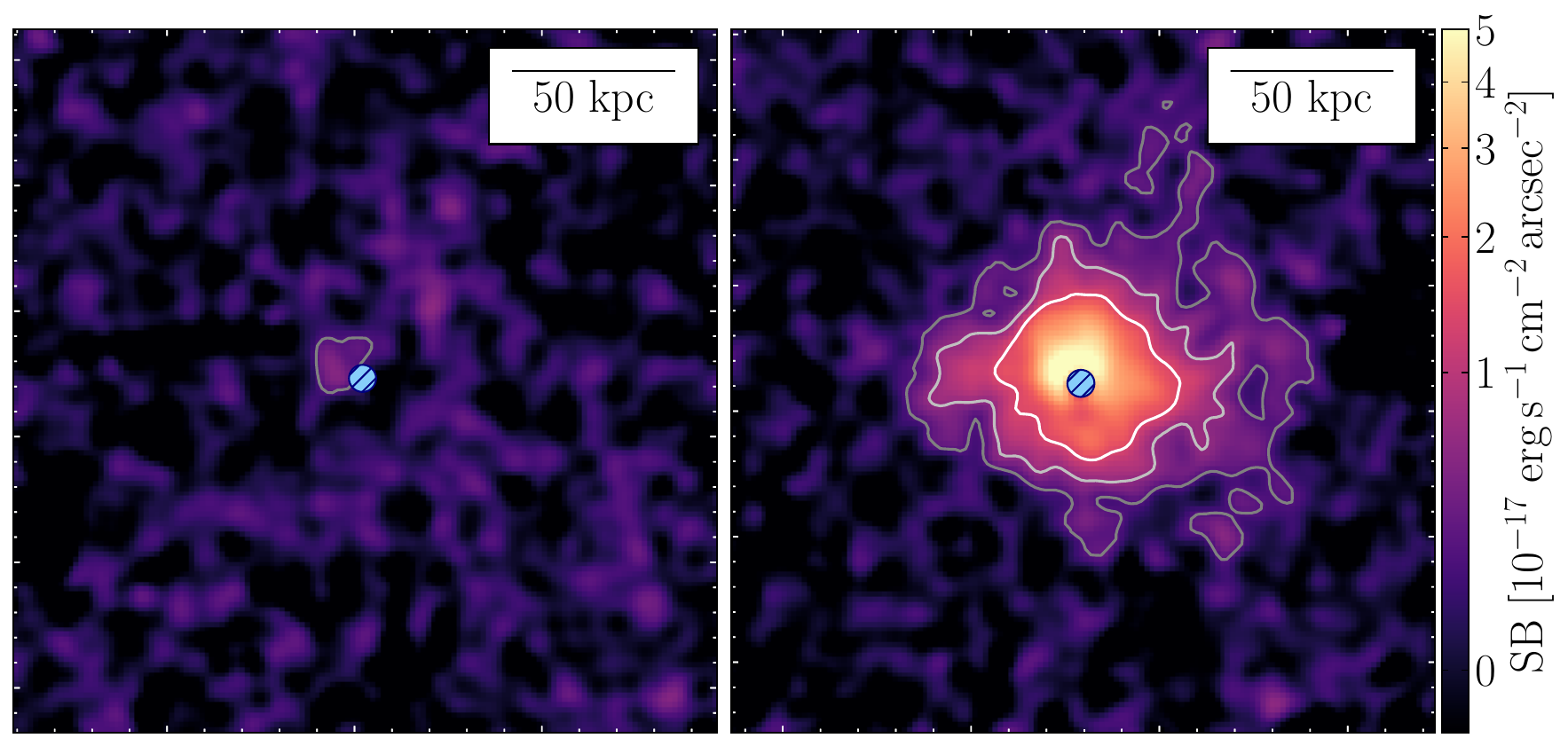}
      \caption{Ly$\alpha$ SB maps of the faintest (left) and brightest (right) nebula first reported in \cite{Mackenzie21} (ID4 and ID6 respectively) and reanalyzed using the method presented in this work. The maps are plotted as described in Fig.~\ref{fig:qso1} within a $28\arcsec \times 28\arcsec$ window.}
         \label{fig:mackenzie}
   \end{figure}

In this work, we employ a conservative approach of using PNBs to detect extended Ly$\alpha$ emission. Previously published samples of nebulae have often been optimally extracted within a 3D mask containing all pixels above a signal-to-noise ratio of 2 that are associated with the nebula. That method uses a less conservative detection criteria, and it could therefore result in more extended detections. 
To facilitate comparison, we reanalyze the two most extreme nebulae published in \cite{Mackenzie21} using 30~$\AA$ PNBs: ID4 is reported to host the faintest and smallest nebula (Fig.~\ref{fig:mackenzie}, left); ID6 is associated with the brightest and most extended Ly$\alpha$ emission (Fig.~\ref{fig:mackenzie}, right).
We find that our method is able to detect extended Ly$\alpha$ emission as shown in \citet{Mackenzie21}, with the only exception of thin structures at low S/N and at the edges of the nebulae in their optimally extracted maps.
For this reason, we find that our measurements for the extents (20~kpc for the small nebula and 134~kpc for the large nebula) are smaller than those reported in that work (60~kpc and 190~kpc). These differences are clearly due to the different extraction criteria for the nebulae but do not affect our results.

\end{appendix}

\end{document}